\documentclass[11pt]{article} 
\usepackage[utf8]{inputenc}
\usepackage[top=1in, left=0.95in, bottom=1.1in, right=0.95in]{geometry}
\usepackage[colorlinks=true,
linkcolor=blue,
urlcolor=red,
citecolor=red]{hyperref}
\usepackage[vcentermath,enableskew]{youngtab}

\usepackage{ytableau}
\usepackage{tikz}
\usepackage{lscape}
\usepackage{tabularx}
\usepackage[toc]{appendix}
\usepackage{longtable}
\usepackage{enumerate}
\usepackage{float}
\usepackage{subcaption}
\usepackage{color}
\usepackage{xcolor}
\usepackage{multirow}
\usepackage{cite}

\let\counterwithin\relax
\usepackage{chngcntr}
\usepackage{amssymb, amsmath,mathrsfs}
\usepackage{multicol,multirow}
\usepackage{ulem}
\usepackage{graphics}
\usepackage{graphicx}
\usepackage{epsf}
\usepackage{epsfig}
\usepackage{float}
\usepackage{makecell}
\usepackage{multirow}
\usepackage{color}
\usepackage{xcolor}
\usepackage{simplewick}
\usepackage{amsmath}
\usepackage{amsfonts}
\usepackage{makeidx} 
\usepackage{slashed} 
\usepackage{booktabs}
\usepackage{multirow}

\allowdisplaybreaks

\def\bar#1{\overline{#1}}

\newcommand{\nn}{\nonumber}

\newcommand{\C}{{\cal C}}

\newcommand{\todo}[1]{{\color{red} \ifmmode\else[todo]\fi #1}}

\begin{document}
\begin{center}


{\Large \textbf  {Comprehensive Effective Field Theory Analysis for Baryon Number Violating Processes}}\\[10mm]

Chuan-Qiang Song$^{a,b,c}$\footnote{songchuanqiang21@mails.ucas.ac.cn}, Jiang-Hao Yu$^{a, b, c}$\footnote{jhyu@itp.ac.cn}\\[10mm]

\noindent 
$^a${ \small School of Fundamental Physics and Mathematical Sciences, Hangzhou Institute for Advanced Study, UCAS, Hangzhou 310024, China}  \\
$^b${ \small School of Physical Sciences, University of Chinese Academy of Sciences,   Beijing 100049, P.R. China}   \\
$^c${ \small Institute of Theoretical Physics, Chinese Academy of Sciences,   Beijing 100190, P. R. China} \\[10mm]

\date{\today}   

\end{center}
\begin{abstract}

Baryon number–violating processes arise generically in many extensions of the Standard Model, with Grand Unified Theories providing the most compelling realizations. Ongoing experimental searches at JUNO, Hyper-K, and DUNE motivate a more precise and model-independent analysis utilizing the effective field theory (EFT) framework capable of connecting ultraviolet dynamics to low-energy hadronic observables. In this work, we perform a pipeline analysis that connects baryon number–violating (BNV) new physics to its low-energy description through the Standard Model EFT (SMEFT) up to dimension nine and the Low-Energy EFT (LEFT) up to dimension eight, and subsequently matches onto chiral perturbation theory using a systematic spurion method. We show that the complete set of dimension-eight LEFT operators enables the inclusion of the full set of chiral representations for three-quark operators with derivatives, which are also embedded in the chiral Lagrangian after quark–hadron matching. By contrast, the conventional dimension-six operators generate only the $(\bar{\mathbf{3}},\mathbf{3})$ and $(\mathbf{8},\mathbf{1})$ representations. The higher dimensional LEFT operators should be matched from higher order SMEFT operators, and thus a wider class of UV completions are allowed. We consider the complete tree-level UV resonances for dimension 6 and 7 SMEFT operators, and representative UVs for higher dimension operators.

\end{abstract}
\newpage
\setcounter{tocdepth}{3}
\setcounter{secnumdepth}{3}


\setcounter{footnote}{0}

\def\baselinestretch{1.5}
\counterwithin{equation}{section}

\newpage

\section{Introduction}

The Lepton number ($L$) and baryon number ($B$) are accidental global symmetries of the Standard Model (SM) Lagrangian and are remarkably well conserved in all current experimental tests. Nevertheless, many well-motivated theories beyond the SM include new physics (NP) that explicitly violates lepton or baryon number. For instance, Weinberg proposed the lepton number violating (LNV) interaction~\cite{Weinberg:1979sa} that generates Majorana neutrino masses, while Grand Unified Theories (GUTs)~\cite{Georgi:1974sy} predict baryon number violating (BNV) proton decay. Moreover, baryon number violation also plays an important role in explaining the observed matter–antimatter asymmetry~\cite{Sakharov:1967dj} of the universe, further underscoring its theoretical significance.

Motivated by these compelling theoretical predictions, experimental searches for BNV nucleon decay have been a central endeavor in particle physics for over half a century. Large-scale detectors with ultra-low backgrounds—most notably Super-K~\cite{Takhistov:2016eqm}—have probed a wide array of decay channels, ranging from canonical modes like $(p\rightarrow e^++\pi^0)$ to more exotic final states. To date, no signal has been observed, and the resulting null results have established increasingly stringent lower limits on nucleon lifetimes. The next generation of experiments, including JUNO~\cite{JUNO:2015zny}, Hyper-K~\cite{Hyper-Kamiokande:2018ofw}, and DUNE~\cite{DUNE:2020ypp}, is poised to improve sensitivity by up to an order of magnitude. Realizing the full discovery potential of these facilities demands a comprehensive understanding of all viable BNV decay topologies, which requires consistent theoretical input at both the effective and fundamental levels.

Early efforts to interpret proton decay were rooted in GUTs, which not only predicted baryon-number violation but also enabled the first quantitative estimates of nucleon lifetimes. Over the decades, detailed calculations within specific GUT frameworks have yielded a rich body of predictions for BNV decay rates~\cite{Lazarides:1980nt,Inoue:1982pi,Barger:1992ac,Nath:2006ut,Hisano:1992jj,Carena:1993ag,Ellis:1978xg,Chang:1984qr,Babu:2016bmy}, which have been continually refined in light of experimental constraints. Notably, baryon number violation typically arises from very high-scale, while observables such as proton decay occur at much lower energy scales, leading to a large energy gap between the NP scale and that of BNV observables. This makes the EFT framework particularly suitable for analyzing such processes, as it is more efficient than model-specific calculations and is model-independent, allowing applications to a wide class of models. By organizing all possible BNV interactions into a basis of effective operators ordered by mass dimension, EFT provides a systematic and model-independent classification of BNV decay channels. In this way, EFT unifies disparate UV scenarios under a single low-energy language, offering both generality and a clear connection to fundamental theory.

Within the EFT framework, the dimension 6 BNV operators were first classified in Refs.~\cite{Weinberg:1979sa,Wilczek:1979hc,Abbott:1980zj}. These operators are invariant under the SM gauge group $SU(3)_C\times SU(2)_L\times U(1)_{Y}$ and are now incorporated into the Standard Model EFT (SMEFT)~\cite{Buchmuller:1985jz,Grzadkowski:2010es,Lehman:2014jma,Li:2020gnx,Murphy:2020rsh,Li:2020xlh,Liao:2020jmn}. After electroweak symmetry breaking, they are matched onto operators in the Low-Energy EFT (LEFT)~\cite{Jenkins:2017jig,Jenkins:2017dyc,Liao:2020zyx,Li:2020tsi,Murphy:2020cly}. In this regime, the light quark sector exhibits an approximate $SU(3)_L\times SU(3)_R$ chiral symmetry~\cite{Weinberg:1968de,Weinberg:1978kz,Gasser:1983yg,Gasser:1984gg,Coleman:1969sm,Callan:1969sn}, which guides the construction of the low-energy chiral Lagrangian for BNV nucleon decays first derived in Ref.~\cite{Claudson:1981gh}. The low-energy constants (LECs) entering this Lagrangian have since been determined from lattice QCD calculations~\cite{JLQCD:1999dld,Aoki:2006ib,Aoki:2008ku,QCDSF:2008qtn,Aoki:2017puj,Yoo:2021gql}. Based on this framework, many calculations of BNV nucleon decay rates have been carried out~\cite{Hambye:2017qix,Heeck:2019kgr,He:2021mrt,He:2021sbl,Fan:2024gzc,Gargalionis:2024nij,Beneito:2023xbk}, and extensions incorporating additional NP degrees of freedom have also been explored~\cite{Li:2024liy,Li:2025slp,Fridell:2023tpb,Fan:2025xhi,Ma:2025mjy}. While previous studies have predominantly focused on dimension 6 operators, a systematic treatment of higher-dimensional contributions in the chiral Lagrangian framework has remained less explored. Recent work in Ref.~\cite{Liao:2025vlj} examined new chiral structures emerging from higher-dimensional operators, which motivates a more complete and systematic analysis.

\begin{figure}
    \centering
    \includegraphics[width=0.7\linewidth]{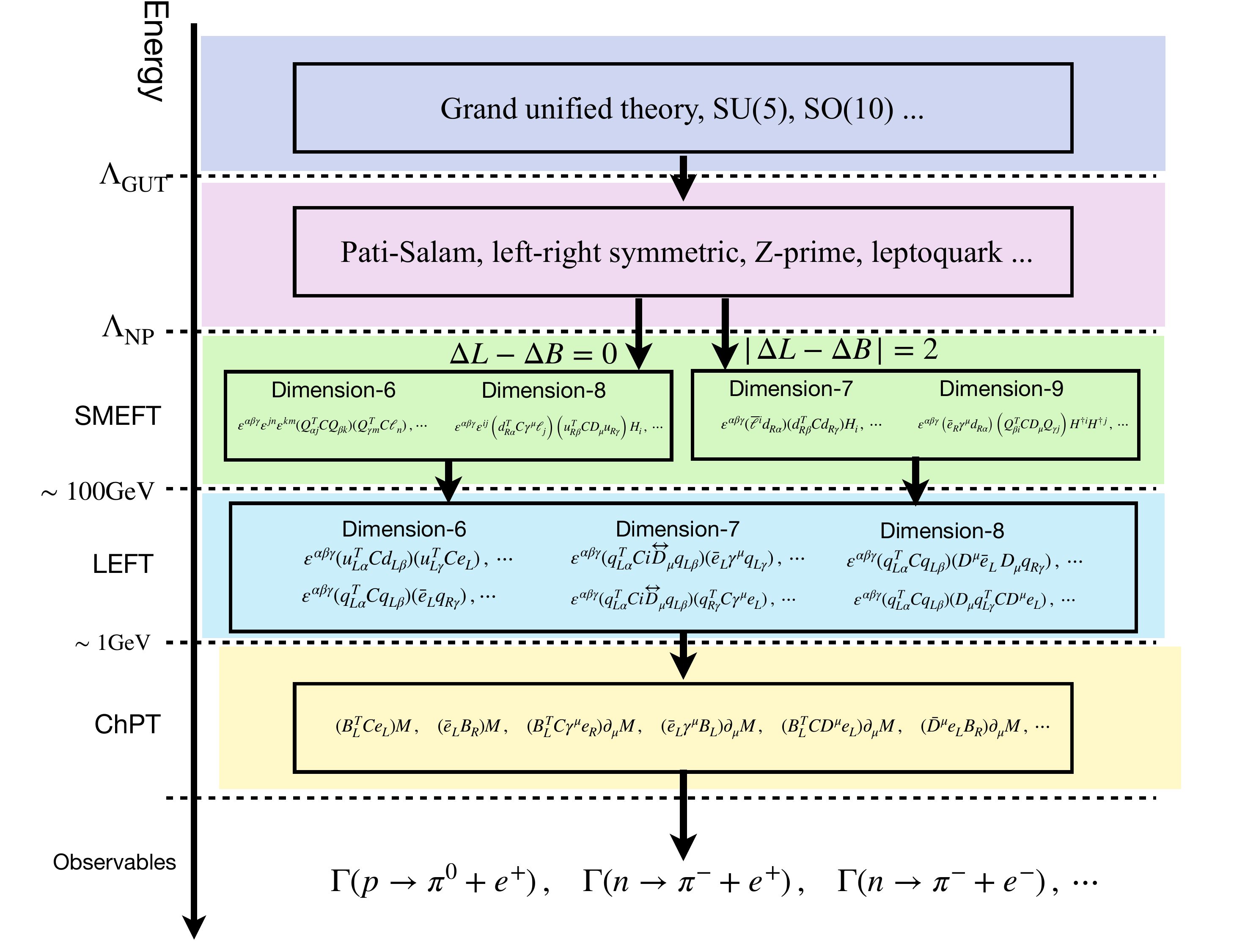}
    \caption{EFT approach to evaluate the BNV baryon decay rate.}
    \label{fig:BNVpro}
\end{figure}

In this work, we adopt the EFT framework to carry out a complete analysis of BNV nucleon decays—from their UV origins to low-energy hadronic observables, as illustrated in Fig.~\ref{fig:BNVpro}. Starting from the low-energy side, we work within the LEFT by considering higher dimensional operators up to dimension 8, and from which we construct the chiral Lagrangian for BNV interactions using a systematic spurion method~\cite{Li:2025xmq,Song:2025snz}. The observables can be calculated from the chiral Lagrangian operators, where the unknown LECs are estimated via naive dimensional analysis (NDA)~\cite{Manohar:1983md,Gavela:2016bzc,Jenkins:2013sda}. Then based on these, we consider the corresponding SMEFT operators matched to the LEFT ones, and representative UV resonances. Our analysis incorporates several distinctive features:
\begin{itemize}
    \item We systematically include higher-dimensional LEFT operators up to dimension 8; notably, these BNV operators yield a variety of chiral representations, and we account for the complete set of such representations;
    \item This extended LEFT operators in complete chiral representation would lead to a more comprehensive chiral Lagrangian, enabling us consider more broader range of nucleon decay observables;
    \item By considering SMEFT operators up to dimension 9, a wider class of UV completions that generate dimension 7, 8, and 9 BNV operators are allowed. We consider the complete tree-level UV resonances for dimension 6 and 7 SMEFT operators, and representative UVs for higher dimension operators. 
    
\end{itemize}
Together, these new features establish a comprehensive and systematic connection between Wilson coefficients at the NP scale and low-energy nucleon decay observables, providing a consistent framework for probing BNV dynamics from both bottom-up and top-down perspectives.

The paper is organized as follows. In Sec.~\ref{sec:BNVOP}, we present the BNV LEFT operator up to dimension 8, and the QCD renormalization group equations (RGEs) is considered. In Sec.~\ref{sec:CL}, we briefly review the chiral perturbation theory and give the matching between BNV LEFT operators and chiral Lagrangian systematically. The relevant effective Lagrangian and the BNV baryon decays rate are discussed in Sec.~\ref{sec:BD}. In Sec.~\ref{sec:SMUV}, we provide the BNV SMEFT operators up to dimension 9, give the matching between SMEFT and LEFT, and obtained the UV models for SMEFT. In Sec.~\ref{sec:con}, the conclusion is drawn.

\section{Low Energy Effective Field Theory Operators}
\label{sec:BNVOP}

The LEFT is formulated by integrating out all heavy degrees of freedom with masses above the electroweak scale ($\Lambda_{\text{EW}} \sim 100$ GeV), leaving only 5 quark flavors and 6 lepton flavors in its effective operator basis. This effective basis is constructed to capture the low-energy dynamics of the SM and its beyond-SM extensions, ensuring that the theory remains consistent with experimental observations at energies below $\Lambda_{\text{EW}}$. The LEFT Lagrangian admits a systematic expansion in terms of the operator mass dimension, which is a key feature of effective field theories:
\begin{equation}
\mathcal{L}_{\text{LEFT}}=\sum_{d,a}\hat{\mathcal{C}}_a^{d}\mathcal{O}_a^{d}\,, 
\label{coefficient}
\end{equation}
where $\mathcal{O}_a^{d}$ denotes the $a$-th independent, gauge-invariant operator of mass dimension $d$, and $\hat{\mathcal{C}}_a^{d}$ represents the corresponding Wilson coefficient with mass dimension $4-d$. Notably, the Wilson coefficients inherently incorporate the $1/\Lambda_{\text{EW}}^{\,d-4}$ suppression factor, which ensures the convergence of the EFT expansion—higher-dimensional operators (larger $d$) are suppressed at low energies, as expected for an effective theory. The LEFT inherently encompasses both QCD and QED, which is explicitly reflected in the covariant derivative $D_\mu=\partial_\mu+ig_3T^AG_\mu^A+ieQA_\mu$, where $g_3$ is the QCD coupling, $T^A$ are the SU(3) color generators, $G_\mu^A$ are the gluon fields, $e$ is the electric charge, $Q$ is the electric charge operator, and $A_\mu$ is the photon field.

In addition, when considering the matching procedure between the LEFT and ChPT, only the three light quark flavors ($u, d, s$) are necessary, as the heavier quarks ($c, b$) are already integrated out in the LEFT construction. For the purpose of simplifying the chiral expansion in ChPT, the light quarks are collectively redefined as a 3-component flavor vector:
\begin{align}
    q=(u,d,s)^T\,.
\end{align}
Furthermore, the LEFT operators can be classified by their baryon number ($B$) and lepton number ($L$), which are conserved in the SM and many beyond-SM scenarios. Specifically, each quark carries a baryon number $B=1/3$, while leptons have $B=0$ and lepton number $L=1$. In our analysis, we focus on classifying the LEFT operators by the baryon number and lepton number violations, specifically considering the cases of $\Delta B = \Delta L = 1$ and $\Delta B = -\Delta L = 1$. The hermitian conjugation parts of the operators are analogous in structure and physical interpretation to their counterparts, and thus are not explicitly presented.

Then the dimension 6 LEFT BNV operators are given by~\cite{Jenkins:2017jig}:
\begin{align}
\Delta B &= \Delta L = 1 && \Delta B = -\Delta L = 1 \notag \\
\label{flavorex}
{\cal O}_{1}^{(6)} &= \varepsilon^{\alpha\beta\gamma}(q_{L\alpha}^{T}C q_{L\beta})(q_{L\gamma}^{T}C e_L)\,, && {\cal O}_{2}^{(6)} = \varepsilon^{\alpha\beta\gamma}(q_{L\alpha}^{T}C q_{L\beta})(\bar{e}_L q_{R\gamma})\,, \\
{\cal O}_{3}^{(6)} &= \varepsilon^{\alpha\beta\gamma}(q_{R\alpha}^{T}C q_{R\beta})(q_{L\gamma}^{T}C e_L)\,, && {\cal O}_{4}^{(6)} = \varepsilon^{\alpha\beta\gamma}(q_{L\alpha}^{T}C q_{L\beta})(\bar{e}_R q_{L\gamma})\,, \\
{\cal O}_{5}^{(6)} &= \varepsilon^{\alpha\beta\gamma}(q_{L\alpha}^{T}C q_{L\beta})(q_{R\gamma}^{T}C e_R)\,, && {\cal O}_{6}^{(6)} = \varepsilon^{\alpha\beta\gamma}(q_{R\alpha}^{T}C q_{R\beta})(\bar{e}_L q_{R\gamma})\,, \\
{\cal O}_{7}^{(6)} &= \varepsilon^{\alpha\beta\gamma}(q_{R\alpha}^{T}C q_{R\beta})(q_{R\gamma}^{T}C e_R)\,, && {\cal O}_{8}^{(6)} = \varepsilon^{\alpha\beta\gamma}(q_{R\alpha}^{T}C q_{R\beta})(\bar{e}_R q_{L\gamma})\,, \\
{\cal O}_{9}^{(6)} &= \varepsilon^{\alpha\beta\gamma}(q_{L\alpha}^{T}C q_{L\beta})(q_{L\gamma}^{T}C \nu_L)\,, && {\cal O}_{10}^{(6)} = \varepsilon^{\alpha\beta\gamma}(q_{L\alpha}^{T}C q_{L\beta})(\bar{\nu}_L q_{R\gamma})\,, \\
{\cal O}_{11}^{(6)} &= \varepsilon^{\alpha\beta\gamma}(q_{R\alpha}^{T}C q_{R\beta})(q_{L\gamma}^{T}C \nu_L)\,, && {\cal O}_{12}^{(6)} = \varepsilon^{\alpha\beta\gamma}(q_{R\alpha}^{T}C q_{R\beta})(\bar{\nu}_L q_{R\gamma})\,,
\end{align}
where the fermions have chiral properties and $\alpha\,,\beta\,,\gamma$ are color index, $\epsilon_{\alpha\beta\gamma}$ is the Levi-Civita tensor with $\varepsilon_{123}=1$. In addition, the $C$ is the charge-conjugation matrix $C = i\gamma^{0}\gamma^{2}$ and $T$ denotes the transpose of the fermion. Similarly, the dimension 7 LEFT BNV operators are given by~\cite{Liao:2020zyx}:
\begin{align}
\Delta B&= \Delta L = 1 && \Delta B = -\Delta L = 1 \notag \\
{\cal O}_{1}^{(7)} &= \varepsilon^{\alpha\beta\gamma}(q_{L\alpha}^{T}Ci {\overleftrightarrow D_\mu} q_{L\beta})(q_{R\gamma}^{T}C\gamma^\mu e_L)\,, && {\cal O}_{2}^{(7)} = \varepsilon^{\alpha\beta\gamma}(q_{L\alpha}^{T}Ci {\overleftrightarrow D_\mu} q_{L\beta})(\bar{e}_L\gamma^\mu q_{L\gamma})\,, \\
{\cal O}_{3}^{(7)} &= \varepsilon^{\alpha\beta\gamma}(q_{R\alpha}^{T}Ci {\overleftrightarrow D_\mu} q_{R\beta})(q_{R\gamma}^{T}C\gamma^\mu e_L)\,, && {\cal O}_{4}^{(7)} = \varepsilon^{\alpha\beta\gamma}(q_{R\alpha}^{T}Ci {\overleftrightarrow D_\mu} q_{R\beta})(\bar{e}_L\gamma^\mu q_{L\gamma})\,, \\
{\cal O}_{5}^{(7)} &= \varepsilon^{\alpha\beta\gamma}(q_{L\alpha}^{T}Ci {\overleftrightarrow D_\mu} q_{L\beta})(q_{L\gamma}^{T}C\gamma^\mu e_R)\,, && {\cal O}_{6}^{(7)} = \varepsilon^{\alpha\beta\gamma}(q_{L\alpha}^{T}Ci {\overleftrightarrow D_\mu} q_{L\beta})(\bar{e}_R\gamma^\mu q_{R\gamma})\,, \\
{\cal O}_{7}^{(7)} &= \varepsilon^{\alpha\beta\gamma}(q_{R\alpha}^{T}Ci {\overleftrightarrow D_\mu} q_{R\beta})(q_{L\gamma}^{T}C\gamma^\mu e_R)\,, && {\cal O}_{8}^{(7)} = \varepsilon^{\alpha\beta\gamma}(q_{R\alpha}^{T}Ci {\overleftrightarrow D_\mu} q_{R\beta})(\bar{e}_R\gamma^\mu q_{R\gamma})\,, \\
{\cal O}_{9}^{(7)} &= \varepsilon^{\alpha\beta\gamma}(q_{L\alpha}^{T}Ci {\overleftrightarrow D_\mu} q_{L\beta})(q_{R\gamma}^{T}C\gamma^\mu \nu_L)\,, && {\cal O}_{10}^{(7)} = \varepsilon^{\alpha\beta\gamma}(q_{L\alpha}^{T}Ci {\overleftrightarrow D_\mu} q_{L\beta})(\bar{\nu}_L\gamma^\mu q_{L\gamma})\,, \\
{\cal O}_{11}^{(7)} &= \varepsilon^{\alpha\beta\gamma}(q_{R\alpha}^{T}Ci {\overleftrightarrow D_\mu} q_{R\beta})(q_{R\gamma}^{T}C\gamma^\mu \nu_L)\,, && {\cal O}_{12}^{(7)} = \varepsilon^{\alpha\beta\gamma}(q_{R\alpha}^{T}Ci {\overleftrightarrow D_\mu} q_{R\beta})(\bar{\nu}_L\gamma^\mu q_{L\gamma})\,,
\end{align}

Moreover, the dimension 8 LEFT BNV operators are given by\cite{Li:2020tsi,Murphy:2020cly}:
\begin{align}
\Delta B &= \Delta L = 1 && \Delta B = -\Delta L = 1 \notag \\
{\cal O}_{1}^{(8)}   &= \varepsilon^{\alpha\beta\gamma}(q_{L\alpha}^{T}C q_{L\beta})(D_\mu q_{L\gamma}^{T} C D^\mu e_L)\,, && {\cal O}_{2}^{(8)}   = \varepsilon^{\alpha\beta\gamma}(q_{L\alpha}^{T}C q_{L\beta})(D^\mu \bar{e}_L\, D_\mu q_{R\gamma})\,, \\
{\cal O}_{3}^{(8)}   &= \varepsilon^{\alpha\beta\gamma}(q_{R\alpha}^{T}C q_{R\beta})(D_\mu q_{L\gamma}^{T} C D^\mu e_L)\,, && {\cal O}_{4}^{(8)}   = \varepsilon^{\alpha\beta\gamma}(q_{R\alpha}^{T}C q_{R\beta})(D^\mu \bar{e}_L\, D_\mu q_{R\gamma})\,, \\
{\cal O}_{5}^{(8)}   &= \varepsilon^{\alpha\beta\gamma}(q_{L\alpha}^{T}C q_{L\beta})(D_\mu q_{R\gamma}^{T} C D^\mu e_R)\,, && {\cal O}_{6}^{(8)}   = \varepsilon^{\alpha\beta\gamma}(q_{L\alpha}^{T}C q_{L\beta})(D^\mu \bar{e}_R\, D_\mu q_{L\gamma})\,, \\
{\cal O}_{7}^{(8)}   &= \varepsilon^{\alpha\beta\gamma}(q_{R\alpha}^{T}C q_{R\beta})(D_\mu q_{R\gamma}^{T} C D^\mu e_R)\,, && {\cal O}_{8}^{(8)}   = \varepsilon^{\alpha\beta\gamma}(q_{R\alpha}^{T}C q_{R\beta})(D^\mu \bar{e}_R\, D_\mu q_{L\gamma})\,, \\
{\cal O}_{9}^{(8)}   &= \varepsilon^{\alpha\beta\gamma}(q_{L\alpha}^{T}C q_{L\beta})(D_\mu q_{L\gamma}^{T} C D^\mu \nu_L)\,, && {\cal O}_{10}^{(8)}  = \varepsilon^{\alpha\beta\gamma}(q_{L\alpha}^{T}C q_{L\beta})(D^\mu \bar{\nu}_L\, D_\mu q_{R\gamma})\,, \\
{\cal O}_{11}^{(8)}  &= \varepsilon^{\alpha\beta\gamma}(q_{R\alpha}^{T}C q_{R\beta})(D_\mu q_{L\gamma}^{T} C D^\mu \nu_L)\,, && {\cal O}_{12}^{(8)}  = \varepsilon^{\alpha\beta\gamma}(q_{R\alpha}^{T}C q_{R\beta})(D^\mu \bar{\nu}_L\, D_\mu q_{R\gamma})\,, \\
{\cal O}_{13}^{(8)}  &= \varepsilon^{\alpha\beta\gamma}(q_{L\alpha}^{T}C\sigma^{\mu\nu} q_{L\beta})(D_\mu q_{L\gamma}^{T} C D_\nu e_L)\,, && {\cal O}_{14}^{(8)}  = \varepsilon^{\alpha\beta\gamma}(q_{L\alpha}^{T}C\sigma^{\mu\nu} q_{L\beta})(D_\mu \bar{e}_L\, D_\nu q_{R\gamma})\,, \\
{\cal O}_{15}^{(8)}  &= \varepsilon^{\alpha\beta\gamma}(q_{R\alpha}^{T}C\sigma^{\mu\nu} q_{R\beta})(D_\mu q_{L\gamma}^{T} C D_\nu e_L)\,, && {\cal O}_{16}^{(8)}  = \varepsilon^{\alpha\beta\gamma}(q_{R\alpha}^{T}C\sigma^{\mu\nu} q_{R\beta})(D_\mu \bar{e}_L\, D_\nu q_{R\gamma})\,, \\
{\cal O}_{17}^{(8)}  &= \varepsilon^{\alpha\beta\gamma}(q_{L\alpha}^{T}C\sigma^{\mu\nu} q_{L\beta})(D_\mu q_{R\gamma}^{T} C D_\nu e_R)\,, && {\cal O}_{18}^{(8)}  = \varepsilon^{\alpha\beta\gamma}(q_{L\alpha}^{T}C\sigma^{\mu\nu} q_{L\beta})(D_\mu \bar{e}_R\, D_\nu q_{L\gamma})\,, \\
{\cal O}_{19}^{(8)}  &= \varepsilon^{\alpha\beta\gamma}(q_{R\alpha}^{T}C\sigma^{\mu\nu} q_{R\beta})(D_\mu q_{R\gamma}^{T} C D_\nu e_R)\,, && {\cal O}_{20}^{(8)}  = \varepsilon^{\alpha\beta\gamma}(q_{R\alpha}^{T}C\sigma^{\mu\nu} q_{R\beta})(D_\mu \bar{e}_R\, D_\nu q_{L\gamma})\,, \\
{\cal O}_{21}^{(8)}  &= \varepsilon^{\alpha\beta\gamma}(q_{L\alpha}^{T}C\sigma^{\mu\nu} q_{L\beta})(D_\mu q_{L\gamma}^{T} C D_\nu \nu_L)\,, && {\cal O}_{22}^{(8)}  = \varepsilon^{\alpha\beta\gamma}(q_{L\alpha}^{T}C\sigma^{\mu\nu} q_{L\beta})(D_\mu \bar{\nu}_L\, D_\nu q_{R\gamma})\,, \\
{\cal O}_{23}^{(8)}  &= \varepsilon^{\alpha\beta\gamma}(q_{R\alpha}^{T}C\sigma^{\mu\nu} q_{R\beta})(D_\mu q_{L\gamma}^{T} C D_\nu \nu_L)\,, && {\cal O}_{24}^{(8)}  = \varepsilon^{\alpha\beta\gamma}(q_{R\alpha}^{T}C\sigma^{\mu\nu} q_{R\beta})(D_\mu \bar{\nu}_L\, D_\nu q_{R\gamma})\,, \\
{\cal O}_{25}^{(8)}  &= \varepsilon^{\alpha\beta\gamma}(q_{L\alpha}^{T}C\sigma^{\mu\nu} q_{L\beta})(q_{L\gamma}^{T}C e_L) F_{\mu\nu}\,, && {\cal O}_{26}^{(8)}  = \varepsilon^{\alpha\beta\gamma}(q_{L\alpha}^{T}C\sigma^{\mu\nu} q_{L\beta})(\bar{e}_L q_{R\gamma}) F_{\mu\nu}\,, \\
{\cal O}_{27}^{(8)}  &= \varepsilon^{\alpha\beta\gamma}(q_{R\alpha}^{T}C\sigma^{\mu\nu} q_{R\beta})(q_{L\gamma}^{T}C e_L) F_{\mu\nu}\,, && {\cal O}_{28}^{(8)}  = \varepsilon^{\alpha\beta\gamma}(q_{R\alpha}^{T}C\sigma^{\mu\nu} q_{R\beta})(\bar{e}_L q_{R\gamma}) F_{\mu\nu}\,, \\
{\cal O}_{29}^{(8)}  &= \varepsilon^{\alpha\beta\gamma}(q_{L\alpha}^{T}C\sigma^{\mu\nu} q_{L\beta})(q_{R\gamma}^{T}C e_R) F_{\mu\nu}\,, && {\cal O}_{30}^{(8)}  = \varepsilon^{\alpha\beta\gamma}(q_{L\alpha}^{T}C\sigma^{\mu\nu} q_{L\beta})(\bar{e}_R q_{L\gamma}) F_{\mu\nu}\,, \\
{\cal O}_{31}^{(8)}  &= \varepsilon^{\alpha\beta\gamma}(q_{R\alpha}^{T}C\sigma^{\mu\nu} q_{R\beta})(q_{R\gamma}^{T}C e_R) F_{\mu\nu}\,, && {\cal O}_{32}^{(8)}  = \varepsilon^{\alpha\beta\gamma}(q_{R\alpha}^{T}C\sigma^{\mu\nu} q_{R\beta})(\bar{e}_R q_{L\gamma}) F_{\mu\nu}\,, \\
{\cal O}_{33}^{(8)}  &= \varepsilon^{\alpha\beta\gamma}(q_{L\alpha}^{T}C\sigma^{\mu\nu} q_{L\beta})(q_{L\gamma}^{T}C \nu_L) F_{\mu\nu}\,, && {\cal O}_{34}^{(8)}  = \varepsilon^{\alpha\beta\gamma}(q_{L\alpha}^{T}C\sigma^{\mu\nu} q_{L\beta})(\bar{\nu}_L q_{R\gamma}) F_{\mu\nu}\,, \\
{\cal O}_{35}^{(8)}  &= \varepsilon^{\alpha\beta\gamma}(q_{R\alpha}^{T}C\sigma^{\mu\nu} q_{R\beta})(q_{L\gamma}^{T}C \nu_L) F_{\mu\nu}\,, && {\cal O}_{36}^{(8)}  = \varepsilon^{\alpha\beta\gamma}(q_{R\alpha}^{T}C\sigma^{\mu\nu} q_{R\beta})(\bar{\nu}_L q_{R\gamma}) F_{\mu\nu}\,,\\
{\cal O}_{37}^{(8)}  &= \varepsilon^{\alpha\beta\gamma}(q_{L\alpha}^{T}C q_{L\beta})(q_{L\gamma}^{T}C \sigma^{\mu\nu}e_L) F_{\mu\nu}\,, && {\cal O}_{38}^{(8)}  = \varepsilon^{\alpha\beta\gamma}(q_{L\alpha}^{T}C q_{L\beta})(\bar{e}_L \sigma^{\mu\nu}q_{R\gamma}) F_{\mu\nu}\,, \\
{\cal O}_{39}^{(8)}  &= \varepsilon^{\alpha\beta\gamma}(q_{R\alpha}^{T}C q_{R\beta})(q_{L\gamma}^{T}C \sigma^{\mu\nu}e_L) F_{\mu\nu}\,, && {\cal O}_{40}^{(8)}  = \varepsilon^{\alpha\beta\gamma}(q_{R\alpha}^{T}C q_{R\beta})(\bar{e}_L \sigma^{\mu\nu}q_{R\gamma}) F_{\mu\nu}\,, \\
{\cal O}_{41}^{(8)}  &= \varepsilon^{\alpha\beta\gamma}(q_{L\alpha}^{T}C q_{L\beta})(q_{R\gamma}^{T}C \sigma^{\mu\nu}e_R) F_{\mu\nu}\,, && {\cal O}_{42}^{(8)}  = \varepsilon^{\alpha\beta\gamma}(q_{L\alpha}^{T}C q_{L\beta})(\bar{e}_R \sigma^{\mu\nu}q_{L\gamma}) F_{\mu\nu}\,, \\
{\cal O}_{43}^{(8)}  &= \varepsilon^{\alpha\beta\gamma}(q_{R\alpha}^{T}C q_{R\beta})(q_{R\gamma}^{T}C \sigma^{\mu\nu}e_R) F_{\mu\nu}\,, && {\cal O}_{44}^{(8)}  = \varepsilon^{\alpha\beta\gamma}(q_{R\alpha}^{T}C q_{R\beta})(\bar{e}_R \sigma^{\mu\nu}q_{L\gamma}) F_{\mu\nu}\,, \\
{\cal O}_{45}^{(8)}  &= \varepsilon^{\alpha\beta\gamma}(q_{L\alpha}^{T}C q_{L\beta})(q_{L\gamma}^{T}C \sigma^{\mu\nu}\nu_L) F_{\mu\nu}\,, && {\cal O}_{46}^{(8)}  = \varepsilon^{\alpha\beta\gamma}(q_{L\alpha}^{T}C q_{L\beta})(\bar{\nu}_L \sigma^{\mu\nu}q_{R\gamma}) F_{\mu\nu}\,, \\
{\cal O}_{47}^{(8)}  &= \varepsilon^{\alpha\beta\gamma}(q_{R\alpha}^{T}C q_{R\beta})(q_{L\gamma}^{T}C \sigma^{\mu\nu}\nu_L) F_{\mu\nu}\,, && {\cal O}_{48}^{(8)}  = \varepsilon^{\alpha\beta\gamma}(q_{R\alpha}^{T}C q_{R\beta})(\bar{\nu}_L \sigma^{\mu\nu}q_{R\gamma}) F_{\mu\nu}\,,
\end{align}
where $F_{\mu\nu}$ denotes the electromagnetic field-strength tensor, and the operators involving the gluon field-strength tensor cannot be classified during the hadronic matching and are thus neglected in this work. The RGEs are dominated by strong interactions; the QCD RGEs of the dimension 6 BNV LEFT operators are given by~\cite{Jenkins:2017dyc}:
\begin{align} \label{eq:RGEC}
  \frac{d}{d\ln\mu}\hat{\mathcal{C}}_n^{(6)} = -\frac{\alpha_s}{\pi}\hat{\mathcal{C}}_n^{(6)}\,,
\end{align}
where $\alpha_s = g_s^2/(4\pi)$ with $g_s$ being the strong coupling constant. Furthermore, leptons and quarks carry distinct electric charges. The conservation of electric charge is a necessary constraint, which restricts the allowed combinations of the quark field $q$ as follows:
\begin{align}
    qqqe^- &\rightarrow {\rm span}\{uud,~uus\}e^-\,,\\
    qqq\nu &\rightarrow {\rm span}\{udd,~uds,~uss\}\nu\,,\\
    qqqe^+ &\rightarrow {\rm span}\{ddd,~dds,~dss,~sss\}e^+\,.
\end{align}
In addition, we classify the flavor of quark fields in the next section upon the inclusion of spurions.

\section{Chiral Lagrangian}
\label{sec:CL}

Having obtained the LEFT operators and considered the QCD RG running, the next step is to match the quark-level interactions onto a low-energy hadronic description at the scale $\Lambda_\chi\sim 1~\mathrm{GeV}$ in ChPT. The ChPT~\cite{Weinberg:1968de,Weinberg:1978kz,Gasser:1983yg,Gasser:1984gg} is a low-energy theory of the LEFT below $\Lambda_\chi\sim 1\text{ GeV}$, where the chiral symmetry $SU(3)_{L}\times SU(3)_{R}$ is spontaneously broken down to the subgroup $SU(3)_V$. According to the Goldstone theorem~\cite{Goldstone:1961eq, Goldstone:1962es}, the broken symmetries generate 8 pseudo-Nambu-Goldstone bosons (NGBs) $\phi^A$, and these NGBs are pseudoscalar bosons with negative parity, which can be parameterized by the $SU(3)_V$ as
\begin{equation}
\label{eq:NGBPi}
    \Pi(x)=\sum_{a=1}^8\phi^A(x)\lambda^A = \left(\begin{array}{ccc}
    \pi^0 + \frac{1}{\sqrt{3}}\eta & \sqrt{2}\pi^+ & \sqrt{2}K^+ \\
    \sqrt{2}\pi^- & -\pi^0+\frac{1}{\sqrt{3}}\eta & \sqrt{2}K^0 \\
    \sqrt{2}K^- & \sqrt{2}~\Bar{K}^0 & -\frac{2}{\sqrt{3}}\eta
    \end{array}\right)\,.
\end{equation}

The NGBs of the ChPT can be described under Coleman-Callan-Wess-Zumino (CCWZ) coset construction~\cite{Coleman:1969sm,Callan:1969sn}, by which they are characterized non-linearly, and are collected in a unitary matrix
\begin{equation}
\label{eq:CCWZ}
    u(x)= \exp\left(\frac{i\phi(x)^A \lambda^A
    }{2f}\right)\,,
\end{equation}
where $f\simeq 92\ \mathrm{MeV}$ is the pion decay constant at leading order (LO) in the chiral expansion.
The unitary matrix transforms under the chiral symmetry as $u(x)\rightarrow Ru(x)V^\dagger=Vu(x)L^\dagger$, where $L\in SU(3)_L\,,R\in SU(3)_R$ and $V\in SU(3)_V$. In terms of $u$, we define the building blocks and the covariant derivative as~\cite{Li:2025xmq,Li:2026mco}
\begin{align}
    u_\mu &= i(u^\dagger D_\mu u-uD_\mu u^\dagger ) \,,\label{building1} \\
    \chi^{\pm} &= u^\dagger \chi u^\dagger \pm u \chi^\dagger u\label{building2} \,,\\
    \nabla_\mu&=\partial_\mu+\frac{1}{2}(u^\dagger D_\mu u+uD_\mu u^\dagger )\,,
    \label{eq:u_without_lr}
\end{align}
where the covariant derivative including the photon field is given by $D_\mu = \partial_\mu + i eQ A_\mu$, the quark mass term is absorbed in $\chi = 2 B_0 \mathcal{M}_q$ with the quark condensate parameter $B_0\simeq 2.8~\mathrm{GeV}$ and the building blocks become the adjoint representation of $SU(3)_V$ group.

Moreover, the baryons are also included in the ChPT~\cite{Gasser:1987rb,Bernard:1992qa,Becher:1999he,Jenkins:1990jv,Krause:1990xc,Oller:2006yh,Song:2024fae} and compose a matrix $B$ of the adjoint representation of the $SU(3)_V$, 
\begin{equation}
    B = \frac{B^A\lambda^A}{\sqrt{2}} = \left(\begin{array}{ccc}
\frac{1}{\sqrt{2}}\Sigma^0+\frac{1}{\sqrt{6}}\Lambda & \Sigma^+ & p \\
\Sigma^- & -\frac{1}{\sqrt{2}}\Sigma^0 + \frac{1}{\sqrt{6}}\Lambda & n \\
\Xi^- & \Xi^0 & -\frac{2}{\sqrt{6}}\Lambda 
    \end{array}\right)\,,
    \label{eq:Bdefine}
\end{equation}
which transforms as $B\rightarrow VBV^{\dagger}$ under the chiral symmetry. The standard LO chiral Lagrangian for the interactions of mesons and baryons are given by:
\begin{align}
\label{eq:BC}
    \mathcal{L}_B=\frac{f^2}{4}\langle u_\mu u^\mu+\chi_+\rangle+\langle\bar B(i\slashed \nabla-m_B)B\rangle-D\langle\bar B\gamma^5\gamma^\mu [B,u_\mu]\rangle-F\langle\bar B\gamma^5\gamma^\mu \{B,u_\mu\}\rangle\,,
\end{align}
where $D=0.730(11)$ and $F=0.730(11)$~\cite{Bali:2022qja} are LECs.

\subsection{Complete Chiral Representation and Building Blocks}

The matching between LEFT and $\chi$PT has been systematically performed in Refs.~\cite{Song:2025snz,Li:2025xmq}. However, the extended spurion must be considered for the BNV chiral Lagrangian. The quark part for LEFT operators can be classified as
\begin{align}
    (q_Lq_Lq_L)\,,\quad(q_Rq_Rq_R)\,,\quad(q_Lq_Rq_R)\,,\quad(q_Lq_Lq_R)\,,
\end{align}
which need to be invariant under the $SU(3)_L\times SU(3)_R$ group and the hermitian conjugate parts are similar with these types. Here, we add spurions into quark sector as
\begin{align}
    (T_Lq_LT_Lq_LT_Lq_L)\,,\quad (T_Rq_RT_Rq_RT_Rq_R)\,,\quad (T_Lq_LT_Rq_RT_Rq_R)\,,\quad (T_Lq_LT_Lq_LT_Rq_R)\,,
\end{align}
in which $T_L$ belongs to the $(\bar{\mathbf{3}}_L,\mathbf{1}_R)$ and $T_R$ belongs to the $(\mathbf{1}_L,\bar{\mathbf{3}}_R)$ representation of the $SU(3)_L\times SU(3)_R$ group. Thus, the whole quark sectors are invariant under the chiral symmetry. Using this rotation, the flavors of the LEFT operators in Sec.~\ref{sec:BNVOP} are classified by the spurion $T_{L\,,R}$. The LEFT operators are labeled by this additional $SU(3)$ index. For example, one type of the flavor for LEFT operator $\mathcal{O}_{1}^{(6)}$ in Eq.~\eqref{flavorex} becomes
\begin{align}
\label{eq:buildblock}
    [{\cal O}_{1}^{(6)}]_{121} &= \varepsilon^{\alpha\beta\gamma}(u_{L\alpha}^{T}C d_{L\beta})(u_{L\gamma}^{T}C e_L)\,.
\end{align}

At the hadron level, the spurions $T_L$ and $T_R$ will not change and the quarks becomes hadrons which also satisfy the properties of chiral symmetry. Since these two types of spurions are representations of the $SU(3)_L\times SU(3)_R$ group, we employ the $SU(3)_L\times SU(3)_R$ group, and the corresponding building blocks differ from those in Eqs.~\eqref{building1} and \eqref{building2}. In the baryon number conserving case, the $SU(3)_L$ and $SU(3)_R$ symmetries can be related via the properties of parity ($P$). However, $P$ is not used for classification here; instead, the representations of the $SU(3)_L\times SU(3)_R$ chiral symmetry must be classified. Accordingly, the building blocks of the BNV ChPT dressed with $u$ are
\begin{equation}
\label{eq:building_blocks}
\left(uu_\mu u^\dagger\,, \, u^\dagger u_\mu u\,, \,uu_\mu u\,, \, u^\dagger u_\mu u^\dagger\,, \,\chi\,,\,\chi^\dagger\,,\, uBu \,,\, u^\dagger Bu^\dagger \,,\, uBu^\dagger \,,\, u^\dagger Bu \,,\, e_L\,,\, e_R \,,\, \nu_L \,,\, F_{\mu\nu}\,,\, T_L\,,\, T_R\right)^T\,,
\end{equation}
which transform under the $SU(3)_L\times SU(3)_R$ symmetry as
\begin{equation}
    \left(\begin{array}{c}
          u u_\mu u^\dagger\\
          u^\dagger u_\mu u\\
          u u_\mu u\\
          u^\dagger u_\mu u^\dagger\\
          \chi\\
          \chi^\dagger\\
          u Bu\\
          u^\dagger Bu^\dagger\\
          u Bu^\dagger\\
          u^\dagger Bu\\
          e_R\\
          e_L\\
          \nu_L\\
          F_{\mu\nu}\\
          T_L\\
          T_R
    \end{array}\right)\rightarrow\left(\begin{array}{c}
          R(u u_\mu u^\dagger)R^\dagger\\
          L(u^\dagger u_\mu u) L^\dagger\\
          R(u u_\mu u)L^\dagger\\
          L(u^\dagger u_\mu u^\dagger) R^\dagger\\
          R\,\chi\, L^{\dagger}\\
          L\,\chi^\dagger\,R^\dagger\\
          R(uBu)L^{\dagger}\\
          L(u^\dagger Bu^\dagger) R^{\dagger}\\
          R(u Bu^\dagger) R^{\dagger}\\
          L(u^\dagger Bu)L^{\dagger}\\
          e_R\\
          e_L\\
          \nu_L\\
          F_{\mu\nu}\\
          T_LL^\dagger\\
          T_RR^\dagger
    \end{array}\right)\,,\quad L\in SU(3)_L\,{\rm and}\,R\in SU(3)_R\,.
\end{equation}
The chiral dimension of the hadronic building blocks are listed in Tab.~\ref{tab:building_blocks_mesons} and the covariant derivative $\nabla_\mu$ has chiral dimension one. Then the BNV chiral Lagrangian can be constructed through the Young tensor technique~\cite{Li:2020gnx,Li:2020xlh,Li:2022tec} which has been used to construct standard chiral Lagrangian~\cite{Low:2022iim,Song:2024fae,Li:2024ghg,Li:2025xmq,Song:2025snz}.

\begin{table}
    \centering
    \begin{tabular}{|l|cccccc|}
    \hline
           & $uu_\mu u^\dagger$ & $u^\dagger u_\mu u$ & $uu_\mu u$ & $u^\dagger u_\mu u^\dagger$ & $\chi$ & $\chi^\dagger$ \\
    \hline
     $d_\chi$ & $1$ & $1$ & $1$ & $1$ & $2$ & $2$ \\
     \hline
           & $uBu $ & $u^\dagger B u^\dagger$ & $uB u^\dagger$ & $u^\dagger B u$ & $T_L$ & $T_R$ \\
    \hline
     $d_\chi$ & $0$ & $0$ & $0$ & $0$ & $0$ & $0$ \\
    \hline
    \end{tabular}
    \caption{The chiral dimension $d_\chi$ of the building blocks in BNV ChPT.}
    \label{tab:building_blocks_mesons}
\end{table}

The next step is matching the LEFT operators to the BNV chiral Lagrangian. The BNV LEFT operators have their corresponding representation of the $SU(3)_L\times SU(3)_R$ chiral symmetry. For example, the dimension 6 BNV LEFT operators with three left-handed quarks are the $\mathbf{8}$ representation of $SU(3)_L$ symmetry, the $\mathbf{1}$ and $\mathbf{10}$ representations are vanish due to the Fierz identity~\cite{Nieves:2003in,Liao:2012uj}. The corresponding representation of chiral symmetry for BNV LEFT operators are listed in Tab.~\ref{tab:chiral_re}. Here, we give the matching rules between BNV LEFT operators and chiral operators:
\begin{itemize}
    \item The LEFT and chiral operators must share the same lepton and photon part.
    \item The spurions of the LEFT and chiral operators are identical, which ensures the invariance of these two type operators under the chiral symmetry.
    \item In the matching procedure, the chiral representations of the LEFT operators must be consistent with those of the chiral operators.
\end{itemize}
In order to show the corresponding chiral representation of the hadronic operators, the symmetries of spurions in chiral operators are defined as  
\begin{align}
    (T_X)_{[a}(T_X)_{b]}(T_X)_{c}=&\frac{1}{2}\bigg[(T_X)_{a}(T_X)_{b}(T_X)_{c}-(T_X)_{b}(T_X)_{a}(T_X)_{c}\bigg]\,,\\
    (T_X)_{\{a}(T_X)_{b\}}(T_X)_{c}=&\frac{1}{2}\bigg[(T_X)_{a}(T_X)_{b}(T_X)_{c}+(T_X)_{b}(T_X)_{a}(T_X)_{c}\bigg]\,,\\
    (T_X)_{\{a}(T_X)_b(T_X)_{c\}}=&\frac{1}{6}\bigg[(T_X)_{a}(T_X)_b(T_X)_{c}+(T_X)_{a}(T_X)_c(T_X)_{b}+(T_X)_{b}(T_X)_a(T_X)_{c}\notag\\
    &+(T_X)_{b}(T_X)_c(T_X)_{a}+(T_X)_{c}(T_X)_a(T_X)_{b}+(T_X)_{c}(T_X)_b(T_X)_{a}\bigg]\,,
\end{align}
and Levi-Civita tensor can also constrain the symmetry of spurion indices.

\begin{table}[H]
    \centering
    \begin{tabular}{|c|c|c|c|}
    \hline
         Operators& $SU(3)L\times SU(3)_R$&Operators& $SU(3)L\times SU(3)_R$ \\
         \hline
         $\mathcal{O}_{1,4,9}^{(6)}$& $(\mathbf{8}_L,\mathbf{1}_R)$&$\mathcal{O}_{6,7,12}^{(6)}$& $(\mathbf{1}_L,\mathbf{8}_R)$\\
         $\mathcal{O}_{2,5,10}^{(6)}$& $(\bar{\mathbf{3}}_L,\mathbf{3}_R)$&$\mathcal{O}_{3,8,11}^{(6)}$& $(\mathbf{3}_L,\bar{\mathbf{3}}_R)$\\
         $\mathcal{O}_{2,5,10}^{(7)}$&$(\mathbf{10}_L,\mathbf{1}_R)$&$\mathcal{O}_{3,8,11}^{(7)}$&$(\mathbf{1}_L,\mathbf{10}_R)$\\
         $\mathcal{O}_{1,6,9}^{(7)}$&$(\mathbf{6}_L,\mathbf{3}_R)$&$\mathcal{O}_{4,7,12}^{(7)}$&$(\mathbf{3}_L,\mathbf{6}_R)$\\
         $\mathcal{O}_{1,6,9}^{(8)}$&$(\mathbf{1}_L,\mathbf{1}_R)$ or $(\mathbf{8}_L,\mathbf{1}_R)$&$\mathcal{O}_{4,7,12}^{(8)}$&$(\mathbf{1}_L,\mathbf{1}_R)$ or $(\mathbf{1}_L,\mathbf{8}_R)$\\
         $\mathcal{O}_{2,5,10}^{(8)}$&$(\bar{\mathbf{3}}_L,\mathbf{3}_R)$&$\mathcal{O}_{3,8,11}^{(8)}$&$(\mathbf{3}_L,\bar{\mathbf{3}}_R)$\\
         $\mathcal{O}_{13,18,21}^{(8)}$&$(\mathbf{8}_L,\mathbf{1}_R)$ or  $(\mathbf{10}_L,\mathbf{1}_R)$&$\mathcal{O}_{16,19,24}^{(8)}$&$(\mathbf{1}_L,\mathbf{8}_R)$ or $(\mathbf{1}_L,\mathbf{10}_R)$\\
         $\mathcal{O}_{14,17,22}^{(8)}$&$(\mathbf{6}_L,\mathbf{3}_R)$&$\mathcal{O}_{15,20,23}^{(8)}$&$(\mathbf{3}_L,\mathbf{6}_R)$\\
         $\mathcal{O}_{25,30,33}^{(8)}$&$(\mathbf{8}_L,\mathbf{1}_R)$ or $(\mathbf{10}_L,\mathbf{1}_R)$&$\mathcal{O}_{28,31,36}^{(8)}$&$(\mathbf{1}_L,\mathbf{8}_R)$ or $(\mathbf{1}_L,\mathbf{10}_R)$\\
         $\mathcal{O}_{26,29,34}^{(8)}$&$(\mathbf{6}_L,\mathbf{3}_R)$&$\mathcal{O}_{27,32,35}^{(8)}$& $(\mathbf{3}_L,\mathbf{6}_R)$\\
         $\mathcal{O}_{37,42,45}^{(8)}$&$(\mathbf{1}_L,\mathbf{1}_R)$ or $(\mathbf{8}_L,\mathbf{1}_R)$&$\mathcal{O}_{40,43,48}^{(8)}$&$(\mathbf{1}_L,\mathbf{1}_R)$ or $(\mathbf{1}_L,\mathbf{8}_R)$\\
         $\mathcal{O}_{38,41,46}^{(8)}$&$(\bar{\mathbf{3}}_L,\mathbf{3}_R)$&$\mathcal{O}_{39,44,47}^{(8)}$& $(\mathbf{3}_L,\bar{\mathbf{3}}_R)$\\
         \hline
    \end{tabular}
    \caption{Chiral representation of the BNV LEFT operators.}
    \label{tab:chiral_re}
\end{table}

\subsection{Matching Results}

 Based on the construction of the building blocks and the matching rules, the full matching results can be derived. Here, we take the type $B_LT_L^3e$ as the example. There are three $T_L$ spurions in this type, thus, the baryon sector need only involve the $SU(3)_L$ index. Then the BNV chiral operators at LO are
\begin{align}
\label{exmatch}
 {\rm LO}:\quad &\varepsilon^{abd} (\lambda^A)^c_d(T_L)_{a}(T_L)_{\{b}(T_L)_{c\}}[\bar e_R (u^\dagger B_Lu)^A]\,,\notag\\
 &\varepsilon^{abd}(\lambda^A)^c_d(T_L)_{a}(T_L)_{\{b}(T_L)_{c\}}[(u^\dagger{B_L^TC}u)^Ae_L]\,,
\end{align}
where $\varepsilon_{acd}$ is the Levi-Civita tensor with $\varepsilon_{123}=1$ and $(abcd...)$ are the $SU(3)$ flavor index. In addition, the left-handed baryon is generated from the properties of lepton part. The $\varepsilon_{acd}$ tensor and $\{..\}$ label make the product of three spurions become the $\mathbf{8}$ representation of $SU(3)_L$ symmetry. Thus, the former chiral operators of Eq.~\eqref{exmatch} corresponds to operator $\mathcal{O}^{(6)}_4$ and the latter one corresponds to operator $\mathcal{O}^{(6)}_1$. As discussed in Ref.~\cite{Li:2025xmq}, these two chiral operators have the same chiral representation and they will share the same LECs which has been calculated as $\beta=0.01269(107)~{\rm GeV}^3$~\cite{Yoo:2021gql} through lattice QCD and the three $T_R$ spurion of this type also have the $\alpha$ constant due to parity conservation of strong interactions. Moreover, the LEC $\alpha=-0.01257(111)~{\rm GeV}^3$~\cite{Yoo:2021gql} for the LO $T_L^2T_R$ and $T_R^2T_L$ type has also been calculated through lattice QCD.

Thus, the BNV chiral Lagrangian of dimension 6 can be obtained
\begin{align}
\label{eq:dimension6}
    \mathcal{L}^{(6)}_{\slashed B}=&\beta\bigg\{[\hat{\mathcal{C}}^{(6)}_1]_{abc}\varepsilon^{abd}(\lambda^A)_d^c(T_L)_a(T_L)_{\{b}(T_L)_{c\}}[(u^\dagger{B_L^TC}u)^Ae_L]\notag\\
    &\quad+[\hat{\mathcal{C}}^{(6)}_4]_{abc}\varepsilon^{abd}(\lambda^A)_d^c(T_L)_a(T_L)_{\{b}(T_L)_{c\}}[\bar e_R (u^\dagger B_Lu)^A]\notag\\
    &\quad+[\hat{\mathcal{C}}^{(6)}_9]_{abc}\varepsilon^{abd}(\lambda^A)_d^c(T_L)_a(T_L)_{\{b}(T_L)_{c\}}[(u^\dagger{B_L^TC}u)^A\nu_L]\notag\\
    &\quad-[\hat{\mathcal{C}}^{(6)}_6]_{abc}\varepsilon^{abd}(\lambda^A)_d^c(T_R)_a(T_R)_{\{b}(T_R)_{c\}}[\bar e_L (u B_Ru^\dagger)^A]\notag\\
    &\quad-[\hat{\mathcal{C}}^{(6)}_7]_{abc}\varepsilon^{abd}(\lambda^A)_d^c(T_R)_a(T_R)_{\{b}(T_R)_{c\}}[(u{B_R^TC}u^\dagger)^Ae_R]\notag\\
    &\quad-[\hat{\mathcal{C}}^{(6)}_{12}]_{abc}\varepsilon^{abd}(\lambda^A)_d^c(T_R)_a(T_R)_{\{b}(T_R)_{c\}}[\bar \nu_L (u B_Ru^\dagger)^A]\bigg\}\notag\\
    &+\alpha\bigg\{[\hat{\mathcal{C}}^{(6)}_2]_{abc}\varepsilon^{abd} (T_L)_a(T_L)_b(T_R)_c[\bar e_L (uB_Ru)^c_d]\notag\\
    &\quad\quad+[\hat{\mathcal{C}}^{(6)}_5]_{acb}\varepsilon^{abd} (T_L)_a(T_L)_b(T_R)_c[ (uB_R^TCu)^c_de_R]\notag\\
    &\quad\quad+[\hat{\mathcal{C}}^{(6)}_{10}]_{acb}\varepsilon^{abd} (T_L)_a(T_L)_b(T_R)_c[\bar \nu_L (uB_Ru)^c_d]\notag\\
    &\quad\quad-[\hat{\mathcal{C}}^{(6)}_3]_{acb}\varepsilon^{abd} (T_R)_a(T_R)_b(T_L)_c[ (u^\dagger B_L^TCu^\dagger)^c_de_L]\notag\\
    &\quad\quad-[\hat{\mathcal{C}}^{(6)}_8]_{acb}\varepsilon^{abd} (T_R)_a(T_R)_b(T_L)_c[\bar e_R (u^\dagger B_Lu^\dagger)^c_d]\notag\\
    &\quad\quad-[\hat{\mathcal{C}}^{(6)}_{11}]_{acb}\varepsilon^{abd} (T_R)_a(T_R)_b(T_L)_c[ (u^\dagger B_L^TCu^\dagger)^c_d\nu_L]\bigg\}+\rm h.c.\,,
\end{align}
where the ($abcd$) are the flavor index. Here, the Levi-Civita tensor $\epsilon_{acd}$ constrain the chiral representation of the operators which listed in Tab.~\ref{tab:chiral_re}.

For the dimension 7 BNV LEFT operators, the LO chiral operators do not have the corresponding chiral representation and the next leading order (NLO) chiral operators are necessary. The additional derivative should be considered and there are two types of the derivative as $\nabla_\mu$ and $u_\mu$. The additional covariant derivative for type $BT^3l$ can be removed through the integration by parts (IBP) and equation of motion (EOM). Here, we take the operator $\mathcal{O}_2^{(7)}$ as the example.  We then construct the complete and independent chiral operators of of type $B_LT_L^3e_L$ as
\begin{align}
    \label{exmatch2}
 {\rm NLO}:\quad &\varepsilon^{abc} (T_L)_a(T_L)_b(T_L)_c(\bar e_L\gamma^\mu B_L^A)u_\mu^A\,, \notag\\
 &\varepsilon^{cde}(\lambda^A)^a_d(\lambda^B)^b_e (T_L)_{\{a}(T_L)_b(T_L)_{c\}}[\bar e_L\gamma_\mu (u^\dagger B_Lu)^A](u^\dagger u_\mu u)^B\,,
\end{align}
where the combination $B_L^A u_\mu^A$ becomes $\langle B_L u_\mu\rangle$. Furthermore, the first chiral operator in Eq.~\eqref{exmatch2} becomes the $\mathbf{1}$ representation of $SU(3)_L$, which cannot match to the LEFT operator $\mathcal{O}_2^{(7)}$. Moreover, the chiral representation of $\mathcal{O}_2^{(7)}$ is presented through $\{...\}$. Additionally, the totally symmetric representation $\mathbf{10}$ can generate the decuplet baryons, which will be considered in future work.

Then we present the BNV chiral Lagrangian for dimension 7 LEFT operators 
\begin{align}
    \mathcal{L}^{(7)}_{\slashed B}=C_1\bigg\{&[\hat{\mathcal{C}}^{(7)}_2]_{abc}\varepsilon^{cde} (\lambda^A)_d^a(\lambda^B)_e^b(T_L)_{\{a}(T_L)_b(T_L)_{c\}}[\bar e_L\gamma_\mu (u^\dagger B_Lu)^A](u^\dagger u^\mu u)^B\notag\\
    &+[\hat{\mathcal{C}}^{(7)}_5]_{abc}\varepsilon^{cde} (\lambda^A)_d^a(\lambda^B)_e^b(T_L)_{\{a}(T_L)_b(T_L)_{c\}}[(u ^\dagger B_L^TCu)^A\gamma^\mu e_R ](u^\dagger u^\mu u)^B\notag\\
    &+[\hat{\mathcal{C}}^{(7)}_{10}]_{abc}\varepsilon^{cde} (\lambda^A)_d^a(\lambda^B)_e^b(T_L)_{\{a}(T_L)_b(T_L)_{c\}}[\bar \nu_L\gamma_\mu (u^\dagger B_Lu)^A](u^\dagger u^\mu u)^B\notag\\
    &-[\hat{\mathcal{C}}^{(7)}_3]_{abc}\varepsilon^{cde} (\lambda^A)_d^a(\lambda^B)_e^b(T_R)_{\{a}(T_R)_b(T_R)_{c\}}[(u B_R^TCu^\dagger)^A\gamma^\mu e_L ](u u^\mu u^\dagger)^B\notag\\
    &-[\hat{\mathcal{C}}^{(7)}_8]_{abc}\varepsilon^{cde} (\lambda^A)_d^a(\lambda^B)_e^b(T_R)_{\{a}(T_R)_b(T_R)_{c\}}[\bar e_R\gamma_\mu (u B_Ru^\dagger)^A](u u^\mu u^\dagger)^B\notag\\
    &-[\hat{\mathcal{C}}^{(7)}_{11}]_{abc}\varepsilon^{cde} (\lambda^A)_d^a(\lambda^B)_e^b(T_R)_{\{a}(T_R)_b(T_R)_{c\}}[(u B_R^TCu^\dagger)^A\gamma^\mu \nu_L ](u u^\mu u^\dagger)^B\bigg\}\notag\\
    +C_2&\bigg\{[\hat{\mathcal{C}}^{(7)}_1]_{abc}\varepsilon^{cde} (T_L)_{\{a}(T_L)_{b\}}(T_R)_c[ (u^\dagger B_R^TCu^\dagger)^a_d\gamma^\mu e_L](u^\dagger u^\mu u^\dagger)^b_e\notag\\
    &\quad+[\hat{\mathcal{C}}^{(7)}_6]_{abc}\varepsilon^{cde} (T_L)_{\{a}(T_L)_{b\}}(T_R)_c[ \bar e_R\gamma^\mu(u^\dagger B_R^TCu^\dagger)^a_d ](u^\dagger u^\mu u^\dagger)^b_e\notag\\
    &\quad+[\hat{\mathcal{C}}^{(7)}_9]_{abc}\varepsilon^{cde} (T_L)_{\{a}(T_L)_{b\}}(T_R)_c[ (u^\dagger B_R^TCu^\dagger)^a_d\gamma^\mu \nu_L](u^\dagger u^\mu u^\dagger)^b_e\notag\\
    &\quad-[\hat{\mathcal{C}}^{(7)}_4]_{abc}\varepsilon^{cde} (T_R)_{\{a}(T_R)_{b\}}(T_L)_c[ \bar e_L\gamma^\mu(u B_L^TCu)^a_d ](u u^\mu u)^b_e\notag\\
    &\quad-[\hat{\mathcal{C}}^{(7)}_7]_{abc}\varepsilon^{cde} (T_R)_{\{a}(T_R)_{b\}}(T_L)_c[ (u B_L^TCu)^a_d\gamma^\mu e_R](u u^\mu u)^b_e\notag\\
    &\quad-[\hat{\mathcal{C}}^{(7)}_{12}]_{abc}\varepsilon^{cde} (T_R)_{\{a}(T_R)_{b\}}(T_L)_c[ \bar \nu_L\gamma^\mu(u B_L^TCu)^a_d ](u u^\mu u)^b_e\bigg\}+\rm h.c.\,,
\end{align}
where $C_1$ and $C_2$ are two LECs of the chiral Lagrangian.

Similarly, the corresponding chiral operators for dimension 8 LEFT operators become more complicated. We classify the different chiral representations of the chiral operators. The trivial representation $(\mathbf{1}_L,\mathbf{1}_R)$ chiral operators become
\begin{align}
   \mathcal{L}^{(8)}_{\slashed B\,,(\mathbf{1},\mathbf{1})}=& D_1\bigg\{[\hat{\mathcal{C}}^{(8)}_{1}]_{abc}\varepsilon^{abc} (T_L)_a(T_L)_b(T_L)_c( B_L^{AT}CD^\mu e_L)u_{\mu A}\notag\\
   &\quad\quad+[\hat{\mathcal{C}}^{(8)}_{6}]_{abc}\varepsilon^{abc} (T_L)_a(T_L)_b(T_L)_c(D^\mu\bar e_R B_L^A)u_{\mu A}\notag\\
   &\quad\quad+[\hat{\mathcal{C}}^{(8)}_{9}]_{abc}\varepsilon^{abc} (T_L)_a(T_L)_b(T_L)_c( B_L^{AT}CD^\mu \nu_L)u_{\mu A}\notag\\
   &\quad\quad-[\hat{\mathcal{C}}^{(8)}_{4}]_{abc}\varepsilon^{abc} (T_R)_a(T_R)_b(T_R)_c(D^\mu\bar e_L B_R^A)u_{\mu A}\notag\\
   &\quad\quad-[\hat{\mathcal{C}}^{(8)}_{7}]_{abc}\varepsilon^{abc} (T_R)_a(T_R)_b(T_R)_c( B_R^{AT}CD^\mu e_R)u_{\mu A}\notag\\
   &\quad\quad-[\hat{\mathcal{C}}^{(8)}_{12}]_{abc}\varepsilon^{abc} (T_R)_a(T_R)_b(T_R)_c(D^\mu\bar \nu_L B_R^A)u_{\mu A}\bigg\}\notag\\
   &+D_2\bigg\{[\hat{\mathcal{C}}^{(8)}_{37}]_{abc}\varepsilon^{abc} (T_L)_a(T_L)_b(T_L)_c {\rm Tr}(U^\dagger\chi U B_L^{T})C\sigma_{\mu\nu}e_LF_{\mu\nu}\notag\\
   &\quad\quad+[\hat{\mathcal{C}}^{(8)}_{42}]_{abc}\varepsilon^{abc} (T_L)_a(T_L)_b(T_L)_c\bar e_R\sigma_{\mu\nu}{\rm Tr}(U^\dagger\chi UB_L)F^{\mu\nu}\notag\\
   &\quad\quad+[\hat{\mathcal{C}}^{(8)}_{45}]_{abc}\varepsilon^{abc} (T_L)_a(T_L)_b(T_L)_c{\rm Tr}(U^\dagger\chi UB_L^{T})C\sigma_{\mu\nu}\nu_LF^{\mu\nu}\notag\\
   &\quad\quad-[\hat{\mathcal{C}}^{(8)}_{40}]_{abc}\varepsilon^{abc} (T_R)_a(T_R)_b(T_R)_c\bar e_L\sigma_{\mu\nu}{\rm Tr}(U^\dagger\chi UB_R)F^{\mu\nu}\notag\\
   &\quad\quad-[\hat{\mathcal{C}}^{(8)}_{43}]_{abc}\varepsilon^{abc} (T_R)_a(T_R)_b(T_R)_c{\rm Tr}(U^\dagger\chi UB_R^{T})C\sigma_{\mu\nu}e_RF^{\mu\nu}\notag\\
   &\quad\quad-[\hat{\mathcal{C}}^{(8)}_{48}]_{abc}\varepsilon^{abc} (T_R)_a(T_R)_b(T_R)_c\bar \nu_L\sigma_{\mu\nu}{\rm Tr}(U^\dagger\chi UB_R)F^{\mu\nu}\bigg\}\notag\\
   &+D_3\bigg\{[\hat{\mathcal{C}}^{(8)}_{37}]_{abc}d_{ABC}\varepsilon^{abc} (T_L)_a(T_L)_b(T_L)_c(B_L^{AT}C\sigma^{\mu\nu}e_L)u_\rho^Bu^{\rho C}F_{\mu\nu}\notag\\
   &\quad\quad+[\hat{\mathcal{C}}^{(8)}_{42}]_{abc}d_{ABC}\varepsilon^{abc} (T_L)_a(T_L)_b(T_L)_c(\bar e_R\sigma^{\mu\nu}B_L^{A})u_\rho^Bu^{\rho C}F_{\mu\nu}\notag\\
   &\quad\quad+[\hat{\mathcal{C}}^{(8)}_{45}]_{abc}d_{ABC}\varepsilon^{abc} (T_L)_a(T_L)_b(T_L)_c(B_L^{AT}C\sigma^{\mu\nu}\nu_L)u_\rho^Bu^{\rho C}F_{\mu\nu}\notag\\
   &\quad\quad-[\hat{\mathcal{C}}^{(8)}_{40}]_{abc}d_{ABC}\varepsilon^{abc} (T_R)_a(T_R)_b(T_R)_c(\bar e_L\sigma^{\mu\nu}B_R^{A})u_\rho^Bu^{\rho C}F_{\mu\nu}\notag\\
   &\quad\quad-[\hat{\mathcal{C}}^{(8)}_{43}]_{abc}d_{ABC}\varepsilon^{abc} (T_R)_a(T_R)_b(T_R)_c(B_R^{AT}C\sigma^{\mu\nu}e_R)u_\rho^Bu^{\rho C}F_{\mu\nu}\notag\\
   &\quad\quad-[\hat{\mathcal{C}}^{(8)}_{48}]_{abc}d_{ABC}\varepsilon^{abc} (T_R)_a(T_R)_b(T_R)_c(\bar \nu_L\sigma^{\mu\nu}B_R^{A})u_\rho^Bu^{\rho C}F_{\mu\nu}\bigg\}\,,
\end{align}
where $d_{ABC}=\frac{1}{4i}{\rm Tr}(\{\lambda_A,\lambda_B\}\lambda_C)$ is the $SU(3)$ structure constant, $U$ denotes $u^2$ and $\chi$ is the quark mass term. In addition, the combinations of the $u_\mu$ and $B$ become
\begin{align}
    B^Au_{\mu A}={\rm Tr}[u^\dagger Bu u^\dagger u_\mu u]={\rm Tr}[uBu^\dagger uu_\mu u^\dagger]\,.
\end{align}.

The representations $(\bar{\mathbf{3}}_L,\mathbf{3}_R)$, $(\mathbf{3}_L,\bar{\mathbf{3}}_R)$, $(\mathbf{8}_L,\mathbf{1}_R)$ and $(\mathbf{1}_L,\mathbf{8}_R)$ for dimension 8 BNV LEFT operators can also be classified. The corresponding hadronic structure of the operators $\mathcal{O}_{37-48}^{(8)}$ has the same properties with dimension 6 operators. Thus, their chiral operators are given directly
\begin{align}
    \mathcal{L}^{(8)}_{\slashed B,\rm Standard}=&\beta\bigg\{[\hat{\mathcal{C}}^{(8)}_{37}]_{abc}\varepsilon^{abd}(\lambda^A)_d^c(T_L)_a(T_L)_{\{b}(T_L)_{c\}}[(u^\dagger{B_L^TC}u)^A\sigma^{\mu\nu}e_L]F_{\mu\nu}\notag\\
    &\quad+[\hat{\mathcal{C}}^{(8)}_{42}]_{abc}\varepsilon^{abd}(\lambda^A)_d^c(T_L)_a(T_L)_{\{b}(T_L)_{c\}}[\bar e_R \sigma^{\mu\nu}(u^\dagger B_Lu)^A]F_{\mu\nu}\notag\\
    &\quad+[\hat{\mathcal{C}}^{(8)}_{45}]_{abc}\varepsilon^{abd}(\lambda^A)_d^c(T_L)_a(T_L)_{\{b}(T_L)_{c\}}[(u^\dagger{B_L^TC}u)^A\sigma^{\mu\nu}\nu_L]F_{\mu\nu}\notag\\
    &\quad-[\hat{\mathcal{C}}^{(8)}_{43}]_{abc}\varepsilon^{abd}(\lambda^A)_d^c(T_R)_a(T_R)_{\{b}(T_R)_{c\}}[(u{B_R^TC}u^\dagger)^A\sigma^{\mu\nu}e_R]F_{\mu\nu}\notag\\
    &\quad-[\hat{\mathcal{C}}^{(8)}_{40}]_{abc}\varepsilon^{abd}(\lambda^A)_d^c(T_R)_a(T_R)_{\{b}(T_R)_{c\}}[\bar e_L \sigma^{\mu\nu}(u B_Ru^\dagger)^A]F_{\mu\nu}\notag\\
    &\quad-[\hat{\mathcal{C}}^{(8)}_{48}]_{abc}\varepsilon^{abd}(\lambda^A)_d^c(T_R)_a(T_R)_{\{b}(T_R)_{c\}}[\bar \nu_L \sigma^{\mu\nu}(u B_Ru^\dagger)^A]F_{\mu\nu}\bigg\}\notag\\
    &+\alpha\bigg\{[\hat{\mathcal{C}}^{(8)}_{38}]_{abc}\varepsilon^{abd} (T_L)_a(T_L)_b(T_R)_c[\bar e_L \sigma^{\mu\nu}(uB_Ru)^c_d]F_{\mu\nu}\notag\\
    &\quad\quad+[\hat{\mathcal{C}}^{(8)}_{41}]_{acb}\varepsilon^{abd} (T_L)_a(T_L)_b(T_R)_c[ (uB_R^TCu)^c_d\sigma^{\mu\nu}e_R]F_{\mu\nu}\notag\\
    &\quad\quad+[\hat{\mathcal{C}}^{(8)}_{46}]_{acb}\varepsilon^{abd} (T_L)_a(T_L)_b(T_R)_c[\bar \nu_L \sigma^{\mu\nu}(uB_Ru)^c_d]F_{\mu\nu}\notag\\
    &\quad\quad-[\hat{\mathcal{C}}^{(8)}_{39}]_{acb}\varepsilon^{abd} (T_R)_a(T_R)_b(T_L)_c[ (u^\dagger B_L^TCu^\dagger)^c_d\sigma^{\mu\nu}e_L]F_{\mu\nu}\notag\\
    &\quad\quad-[\hat{\mathcal{C}}^{(8)}_{44}]_{acb}\varepsilon^{abd} (T_R)_a(T_R)_b(T_L)_c[\bar e_R \sigma^{\mu\nu}(u^\dagger B_Lu^\dagger)^c_d]F_{\mu\nu}\notag\\
    &\quad\quad-[\hat{\mathcal{C}}^{(8)}_{47}]_{acb}\varepsilon^{abd} (T_R)_a(T_R)_b(T_L)_c[ (u^\dagger B_L^TCu^\dagger)^c_d\sigma^{\mu\nu}\nu_L]F_{\mu\nu}\bigg\}+\rm h.c.\,,
\end{align}
where the LECs share the same values with dimension 6 operators.

The other $(\bar{\mathbf{3}}_L,\mathbf{3}_R)$ and  $(\mathbf{3}_L,\bar{\mathbf{3}}_R)$ chiral operators are
\begin{align}
    \mathcal{L}^{(8)}_{\slashed B,\rm (\bar{\mathbf{3}},\mathbf{3})}=&E_1\bigg\{[\hat{\mathcal{C}}^{(8)}_{2}]_{abc} \varepsilon^{cde} (T_L)_{[a}(T_L)_{b]}(T_R)_{c}[D^\mu\bar e_L(u^\dagger B_Ru^\dagger)^a_d](u^\dagger u^\mu u^\dagger)^b_e\notag\\
    &\quad\quad+[\hat{\mathcal{C}}^{(8)}_{5}]_{abc} \varepsilon^{cde} (T_L)_{[a}(T_L)_{b]}(T_R)_{c}[(u^\dagger B_R^TCu^\dagger)^a_dD^\mu e_R](u^\dagger u^\mu u^\dagger)^b_e\notag\\
    &\quad\quad+[\hat{\mathcal{C}}^{(8)}_{10}]_{abc} \varepsilon^{cde} (T_L)_{[a}(T_L)_{b]}(T_R)_{c}[D^\mu\bar \nu_L(u^\dagger B_Ru^\dagger)^a_d](u^\dagger u^\mu u^\dagger)^b_e\notag\\
    &\quad\quad-[\hat{\mathcal{C}}^{(8)}_{3}]_{abc}\varepsilon^{cde} (T_R)_{[a}(T_R)_{b]}(T_L)_{c}[(u B_L^TCu)^a_dD^\mu e_L](uu^\mu u)^b_e\notag\\
    &\quad\quad-[\hat{\mathcal{C}}^{(8)}_{8}]_{abc} \varepsilon^{cde} (T_R)_{[a}(T_R)_{b]}(T_L)_{c}[D^\mu\bar e_R(u B_Lu)^a_d](uu^\mu u)^b_e\notag\\
    &\quad\quad-[\hat{\mathcal{C}}^{(8)}_{11}]_{abc} \varepsilon^{cde} (T_R)_{[a}(T_R)_{b]}(T_L)_{c}[(u B_L^TCu)^a_dD^\mu \nu_L](uu^\mu u)^b_e\bigg\}+\rm h.c.\,,
\end{align}
where the asymmetry parts of the index $a,b$ are selected. Moreover, the other $(\mathbf{8}_L,\mathbf{1}_R)$ and $(\mathbf{1}_L,\mathbf{8}_R)$ chiral operators without photon fields are
\begin{align}
    \mathcal{L}^{(8)}_{\slashed B,\rm (\mathbf{8},\mathbf{1})}=&E_2\bigg\{[\hat{\mathcal{C}}^{(8)}_{1}]_{abc}f_{ABC} \varepsilon^{abd} (\lambda^A)_d^c(T_L)_a(T_L)_{\{b}(T_L)_{c\}}[(u^\dagger B_L^TCu)^B D^\mu e_L]u^\mu_C\notag\\
    &\quad\quad+[\hat{\mathcal{C}}^{(8)}_{6}]_{abc}f_{ABC} \varepsilon^{abd} (\lambda^A)_d^c(T_L)_a(T_L)_{\{b}(T_L)_{c\}}[D^\mu\bar e_R(u^\dagger B_Lu)^B]u^\mu_C\notag\\
    &\quad\quad+[\hat{\mathcal{C}}^{(8)}_{9}]_{abc}f_{ABC} \varepsilon^{abd} (\lambda^A)_d^c(T_L)_a(T_L)_{\{b}(T_L)_{c\}}[(u^\dagger B_L^TCu)^B D^\mu \nu_L]u^\mu_C\notag\\
    &\quad\quad+[\hat{\mathcal{C}}^{(8)}_{13}]_{abc}f_{ABC} \varepsilon^{acd} (\lambda^A)_d^b(T_L)_{\{a}(T_L)_{b\}}(T_L)_{c}[(u^\dagger B_L^TCu)^B D^\mu e_L]u^\mu_C\notag\\
    &\quad\quad+[\hat{\mathcal{C}}^{(8)}_{18}]_{abc}f_{ABC} \varepsilon^{acd} (\lambda^A)_d^b(T_L)_{\{a}(T_L)_{b\}}(T_L)_{c}[D^\mu\bar e_R(u^\dagger B_Lu)^B]u^\mu_C\notag\\
    &\quad\quad+[\hat{\mathcal{C}}^{(8)}_{21}]_{abc}f_{ABC} \varepsilon^{acd} (\lambda^A)_d^b(T_L)_{\{a}(T_L)_{b\}}(T_L)_{c}[(u^\dagger B_L^TCu)^B D^\mu \nu_L]u^\mu_C\notag\\
    &\quad\quad-[\hat{\mathcal{C}}^{(8)}_{4}]_{abc}f_{ABC} \varepsilon^{abd} (\lambda^A)_d^c(T_R)_a(T_R)_{\{b}(T_R)_{c\}}[D^\mu\bar e_L(uB_Ru^\dagger)^B]u^\mu_C\notag\\
    &\quad\quad-[\hat{\mathcal{C}}^{(8)}_{7}]_{abc}f_{ABC} \varepsilon^{abd} (\lambda^A)_d^c(T_R)_a(T_R)_{\{b}(T_R)_{c\}}[(u B_R^TCu^\dagger)^B D^\mu e_R]u^\mu_C\notag\\
    &\quad\quad-[\hat{\mathcal{C}}^{(8)}_{12}]_{abc}f_{ABC} \varepsilon^{abd} (\lambda^A)_d^c(T_R)_a(T_R)_{\{b}(T_R)_{c\}}[D^\mu\bar \nu_L(uB_Ru^\dagger)^B]u^\mu_C\notag\\
    &\quad\quad-[\hat{\mathcal{C}}^{(8)}_{16}]_{abc}f_{ABC} \varepsilon^{acd} (\lambda^A)_d^b(T_R)_{\{a}(T_R)_{b\}}(T_R)_{c}[D^\mu\bar e_L(u^\dagger B_Ru)^B]u^\mu_C\notag\\
    &\quad\quad-[\hat{\mathcal{C}}^{(8)}_{19}]_{abc}f_{ABC} \varepsilon^{acd} (\lambda^A)_d^b(T_R)_{\{a}(T_R)_{b\}}(T_R)_{c}[(u^\dagger B_L^TCu)^B D^\mu e_R]u^\mu_C\notag\\
    &\quad\quad-[\hat{\mathcal{C}}^{(8)}_{24}]_{abc}f_{ABC} \varepsilon^{acd} (\lambda^A)_d^b(T_R)_{\{a}(T_R)_{b\}}(T_R)_{c}[D^\mu\bar \nu_L(u^\dagger B_Ru)^B]u^\mu_C\bigg\}\notag\\
    &+E_3\bigg\{[\hat{\mathcal{C}}^{(8)}_{1}]_{abc}d_{ABC} \varepsilon^{abd} (\lambda^A)_d^c(T_L)_a(T_L)_{\{b}(T_L)_{c\}}[(u^\dagger B_L^TCu)^B D^\mu e_L]u^\mu_C\notag\\
    &\quad\quad\quad+[\hat{\mathcal{C}}^{(8)}_{6}]_{abc}d_{ABC} \varepsilon^{abd} (\lambda^A)_d^c(T_L)_a(T_L)_{\{b}(T_L)_{c\}}[D^\mu\bar e_R(u^\dagger B_Lu)^B]u^\mu_C\notag\\
    &\quad\quad\quad+[\hat{\mathcal{C}}^{(8)}_{9}]_{abc}d_{ABC} \varepsilon^{abd} (\lambda^A)_d^c(T_L)_a(T_L)_{\{b}(T_L)_{c\}}[(u^\dagger B_L^TCu)^B D^\mu \nu_L]u^\mu_C\notag\\
    &\quad\quad\quad+[\hat{\mathcal{C}}^{(8)}_{13}]_{abc}d_{ABC} \varepsilon^{acd} (\lambda^A)_d^b(T_L)_{\{a}(T_L)_{b\}}(T_L)_{c}[(u^\dagger B_L^TCu)^B D^\mu e_L]u^\mu_C\notag\\
    &\quad\quad\quad+[\hat{\mathcal{C}}^{(8)}_{18}]_{abc}d_{ABC} \varepsilon^{acd} (\lambda^A)_d^b(T_L)_{\{a}(T_L)_{b\}}(T_L)_{c}[D^\mu\bar e_R(u^\dagger B_Lu)^B]u^\mu_C\notag\\
    &\quad\quad\quad+[\hat{\mathcal{C}}^{(8)}_{21}]_{abc}d_{ABC} \varepsilon^{acd} (\lambda^A)_d^b(T_L)_{\{a}(T_L)_{b\}}(T_L)_{c}[(u^\dagger B_L^TCu)^B D^\mu \nu_L]u^\mu_C\notag\\
    &\quad\quad\quad-[\hat{\mathcal{C}}^{(8)}_{4}]_{abc}d_{ABC} \varepsilon^{abd} (\lambda^A)_d^c(T_R)_a(T_R)_{\{b}(T_R)_{c\}}[D^\mu\bar e_L(uB_Ru^\dagger)^B]u^\mu_C\notag\\
    &\quad\quad\quad-[\hat{\mathcal{C}}^{(8)}_{7}]_{abc}d_{ABC} \varepsilon^{abd} (\lambda^A)_d^c(T_R)_a(T_R)_{\{b}(T_R)_{c\}}[(u B_R^TCu^\dagger)^B D^\mu e_R]u^\mu_C\notag\\
    &\quad\quad\quad-[\hat{\mathcal{C}}^{(8)}_{12}]_{abc}d_{ABC} \varepsilon^{abd} (\lambda^A)_d^c(T_R)_a(T_R)_{\{b}(T_R)_{c\}}[D^\mu\bar \nu_L(uB_Ru^\dagger)^B]u^\mu_C\notag\\
    &\quad\quad\quad-[\hat{\mathcal{C}}^{(8)}_{16}]_{abc}d_{ABC} \varepsilon^{acd} (\lambda^A)_d^b(T_R)_{\{a}(T_R)_{b\}}(T_R)_{c}[D^\mu\bar e_L(u^\dagger B_Ru)^B]u^\mu_C\notag\\
    &\quad\quad\quad-[\hat{\mathcal{C}}^{(8)}_{19}]_{abc}d_{ABC} \varepsilon^{acd} (\lambda^A)_d^b(T_R)_{\{a}(T_R)_{b\}}(T_R)_{c}[(u^\dagger B_L^TCu)^B D^\mu e_R]u^\mu_C\notag\\
    &\quad\quad\quad-[\hat{\mathcal{C}}^{(8)}_{24}]_{abc}d_{ABC} \varepsilon^{acd} (\lambda^A)_d^b(T_R)_{\{a}(T_R)_{b\}}(T_R)_{c}[D^\mu\bar \nu_L(u^\dagger B_Ru)^B]u^\mu_C\bigg\}+\rm h.c.\,,
\end{align}
where $f_{ABC}=\frac{1}{4i}{\rm Tr}([\lambda_A,\lambda_B]\lambda_C)$ and both $f_{ABC}$ and $d_{ABC}$ appear in this case, whereas only $d_{ABC}$ was present in the earlier expression.

When considering the photon tensor dimension 8 operators, there are more building blocks and derivatives. Here, we only present the $(\mathbf{8}_L,\mathbf{1}_R)$ and $(\mathbf{1}_L,\mathbf{8}_R)$ chiral operators of LEFT operator $\mathcal{O}^{(8)}_{25}$
\begin{align}
    \mathcal{L}^{(8)\prime}_{\slashed B,(\mathbf{8},\mathbf{1})}=&\bigg\{F_1[\hat{\mathcal{C}}^{(8)}_{25}]_{abc}f_{ABC} \varepsilon^{abd} (\lambda^A)_d^c(T_L)_a(T_L)_{\{b}(T_L)_{c\}}[\nabla^\mu(u^\dagger B_L^Tu)^BC e_L]u^{\nu C}F_{\mu\nu}\notag\\
    &\quad+F_2[\hat{\mathcal{C}}^{(8)}_{25}]_{abc} \varepsilon^{abd} (\lambda^A)_d^c(T_L)_a(T_L)_{\{b}(T_L)_{c\}}[(u^\dagger B_L^Tu)^AC e_L]u^\mu_Bu^\nu_BF_{\mu\nu}\notag\\
    &\quad+F_3[\hat{\mathcal{C}}^{(8)}_{25}]_{abc} \varepsilon^{abd} (\lambda^A)_d^c(T_L)_a(T_L)_{\{b}(T_L)_{c\}}[(u^\dagger B_L^Tu)_BC e_L]u^\mu_Au^\nu_BF_{\mu\nu}\notag\\
    &\quad+F_4[\hat{\mathcal{C}}^{(8)}_{25}]_{abc}f_{ABD}f^{CDE} \varepsilon^{abd} (\lambda^E)_d^c(T_L)_a(T_L)_{\{b}(T_L)_{c\}}[(u^\dagger B_L^Tu)_CC e_L]u^\mu_Au^\nu_BF_{\mu\nu}\notag\\
    &\quad+F_5[\hat{\mathcal{C}}^{(8)}_{25}]_{abc}f_{ABD}f^{CDE} \varepsilon^{abd} (\lambda^E)_d^c(T_L)_a(T_L)_{\{b}(T_L)_{c\}}[(u^\dagger B_L^Tu)^AC e_L]u^\mu_Cu^\nu_BF_{\mu\nu}\notag\\
    &\quad+F_6[\hat{\mathcal{C}}^{(8)}_{25}]_{abc}f_{ABD}d^{CDE} \varepsilon^{abd} (\lambda^E)_d^c(T_L)_a(T_L)_{\{b}(T_L)_{c\}}[(u^\dagger B_L^Tu)_CC e_L]u^\mu_Au^\nu_BF_{\mu\nu}\notag\\
    &\quad+F_7[\hat{\mathcal{C}}^{(8)}_{25}]_{abc}f_{ABD}d^{CDE} \varepsilon^{abd} (\lambda^E)_d^c(T_L)_a(T_L)_{\{b}(T_L)_{c\}}[(u^\dagger B_L^Tu)^AC e_L]u^\mu_Cu^\nu_BF_{\mu\nu}\notag\\
    &\quad+F_8[\hat{\mathcal{C}}^{(8)}_{25}]_{abc}d_{ABD}f^{CDE} \varepsilon^{abd} (\lambda^E)_d^c(T_L)_a(T_L)_{\{b}(T_L)_{c\}}[(u^\dagger B_L^Tu)^AC e_L]u^\mu_Cu^\nu_BF_{\mu\nu}\notag\\
    &\quad+F_9[\hat{\mathcal{C}}^{(8)}_{25}]_{abc}d_{ABD}d^{CDE} \varepsilon^{abd} (\lambda^E)_d^c(T_L)_a(T_L)_{\{b}(T_L)_{c\}}[(u^\dagger B_L^Tu)^AC e_L]u^\mu_Cu^\nu_BF_{\mu\nu}\notag\\
    &\quad\quad+\cdots\bigg\}\,,
\end{align}
where the $(\cdots)$ denotes the similar $(\mathbf{8}_L,\mathbf{1}_R)$ and $(\mathbf{1}_L,\mathbf{8}_R)$ chiral operators for $\mathcal{O}_{28}^{(8)}$, $\mathcal{O}_{30}^{(8)}$, $\mathcal{O}_{31}^{(8)}$, $\mathcal{O}_{33}^{(8)}$ and $\mathcal{O}_{36}^{(8)}$, they share the same LECs $F_{1-9}$ with operator $\mathcal{O}^{(8)}_{25}$.

For representation $(\mathbf{10}_L,\mathbf{1}_R)$, $(\mathbf{1}_L,\mathbf{10}_R)$, $(\mathbf{6}_L,\mathbf{3}_R)$ and $(\mathbf{3}_L,\mathbf{6}_R)$ types of dimension 8 BNV LEFT operators, the quark part of the operators $\mathcal{O}^{(8)}_{13-24}$ have the same chiral representation and Lorentz index with the dimension 7 operators. Thus, they will share the same LECs $C_1$ and $C_2$. The corresponding chiral operators of $\mathcal{O}^{(8)}_{13-24}$ in these representations are given by
\begin{align}
   \mathcal{L}^{(8)}_{\slashed B\,,13-24}=C_1\bigg\{&[\hat{\mathcal{C}}^{(8)}_{13}]_{abc}\varepsilon^{cde} (\lambda^A)_d^a(\lambda^B)_e^b(T_L)_{\{a}(T_L)_b(T_L)_{c\}}[(u ^\dagger B_L^TCu)^AD_\mu e_L ](u^\dagger u_\mu u)^B\notag\\
   &+[\hat{\mathcal{C}}^{(8)}_{18}]_{abc}\varepsilon^{cde} (\lambda^A)_d^a(\lambda^B)_e^b(T_L)_{\{a}(T_L)_b(T_L)_{c\}}[D_\mu\bar e_R (u^\dagger B_Lu)^A](u^\dagger u_\mu u)^B\notag\\
   &+[\hat{\mathcal{C}}^{(8)}_{21}]_{abc}\varepsilon^{cde} (\lambda^A)_d^a(\lambda^B)_e^b(T_L)_{\{a}(T_L)_b(T_L)_{c\}}[(u ^\dagger B_L^TCu)^AD_\mu \nu_L ](u^\dagger u_\mu u)^B\notag\\
   &-[\hat{\mathcal{C}}^{(8)}_{16}]_{abc}\varepsilon^{cde} (\lambda^A)_d^a(\lambda^B)_e^b(T_R)_{\{a}(T_R)_b(T_R)_{c\}}[D_\mu \bar e_L (u B_Ru^\dagger)^A](u u_\mu u^\dagger)^B\notag\\
    &-[\hat{\mathcal{C}}^{(8)}_{19}]_{abc}\varepsilon^{cde} (\lambda^A)_d^a(\lambda^B)_e^b(T_R)_{\{a}(T_R)_b(T_R)_{c\}}[(u B_R^TCu^\dagger)^AD_\mu e_R](u u_\mu u^\dagger)^B\notag\\
    &-[\hat{\mathcal{C}}^{(8)}_{24}]_{abc}\varepsilon^{cde} (\lambda^A)_d^a(\lambda^B)_e^b(T_R)_{\{a}(T_R)_b(T_R)_{c\}}[D_\mu \bar \nu_L (u B_Ru^\dagger)^A](u u_\mu u^\dagger)^B\bigg\}\notag\\
    +C_2&\bigg\{[\hat{\mathcal{C}}^{(8)}_{14}]_{abc}\varepsilon^{cde} (T_L)_{\{a}(T_L)_{b\}}(T_R)_c[ D_\mu\bar e_L(u^\dagger B_Ru^\dagger)^a_d ](u^\dagger u_\mu u^\dagger)^b_e\notag\\
    &\quad+[\hat{\mathcal{C}}^{(8)}_{17}]_{abc}\varepsilon^{cde} (T_L)_{\{a}(T_L)_{b\}}(T_R)_c[ (u^\dagger B_R^TCu^\dagger)^a_dD_\mu e_R](u^\dagger u_\mu u^\dagger)^b_e\notag\\
    &\quad+[\hat{\mathcal{C}}^{(8)}_{22}]_{abc}\varepsilon^{cde} (T_L)_{\{a}(T_L)_{b\}}(T_R)_c[ D_\mu\bar \nu_L(u^\dagger B_Ru^\dagger)^a_d ](u^\dagger u_\mu u^\dagger)^b_e\notag\\
    &\quad-[\hat{\mathcal{C}}^{(8)}_{15}]_{abc}\varepsilon^{cde} (T_R)_{\{a}(T_R)_{b\}}(T_L)_c[ (u B_L^TCu)^a_dD_\mu e_L](u u_\mu u)^b_e\notag\\
    &\quad-[\hat{\mathcal{C}}^{(8)}_{20}]_{abc}\varepsilon^{cde} (T_R)_{\{a}(T_R)_{b\}}(T_L)_c[ D_\mu\bar e_R(u B_Lu)^a_d ](u u_\mu u)^b_e\notag\\
    &\quad-[\hat{\mathcal{C}}^{(8)}_{23}]_{abc}\varepsilon^{cde} (T_R)_{\{a}(T_R)_{b\}}(T_L)_c[ (u B_L^TCu)^a_dD_\mu \nu_L](u u_\mu u)^b_e\bigg\}+\rm h.c.\,.
\end{align}

Furthermore, chiral operators of operators $\mathcal{O}^{(8)}_{25-36}$  with the representation $(\mathbf{10}_L,\mathbf{1}_R)$, $(\mathbf{1}_L,\mathbf{10}_R)$, $(\mathbf{6}_L,\mathbf{3}_R)$ and $(\mathbf{3}_L,\mathbf{6}_R)$ are given by
\begin{align}
    \mathcal{L}^{(8)}_{\slashed B,25-36}=&\bigg\{ F_{10}[\hat{\mathcal{C}}^{(8)}_{25}]_{abc}\varepsilon^{cde} (\lambda^A)_d^a(\lambda^B)^b_e(T_L)_{\{a}(T_L)_{b}(T_L)_{c\}}[\nabla^\mu(u^\dagger B_L^Tu)^AC e_L]u^{\nu B}F_{\mu\nu}\notag\\
    &\quad+F_{11}[\hat{\mathcal{C}}^{(8)}_{25}]_{abc}f_{ABD} \varepsilon^{cde} (\lambda^C)^a_d(\lambda^D)^b_e(T_L)_a(T_L)_{\{b}(T_L)_{c\}}[(u^\dagger B_L^Tu)^CC e_L](u^\dagger u_\mu u)^A(u^\dagger u_\nu u)^BF_{\mu\nu}\notag\\
    &\quad+F_{12}[\hat{\mathcal{C}}^{(8)}_{25}]_{abc}f_{ABD} \varepsilon^{cde} (\lambda^C)^a_d(\lambda^D)^b_e(T_L)_a(T_L)_{\{b}(T_L)_{c\}}[(u^\dagger B_L^Tu)^AC e_L](u^\dagger u_\mu u)^C(u^\dagger u_\nu u)^BF_{\mu\nu}\notag\\
    &\quad+F_{13}[\hat{\mathcal{C}}^{(8)}_{25}]_{abc}d_{ABD} \varepsilon^{cde} (\lambda^C)^a_d(\lambda^D)^b_e(T_L)_a(T_L)_{\{b}(T_L)_{c\}}[(u^\dagger B_L^Tu)^AC e_L](u^\dagger u_\mu u)^C(u^\dagger u_\nu u)^BF_{\mu\nu}\notag\\
    &\quad\quad+\cdots\bigg\}\notag\\
    &+\bigg\{F_{14}[\hat{\mathcal{C}}^{(8)}_{26}]_{abc}\varepsilon^{cde}(T_L)_{\{a}(T_L)_{b\}}(T_R)_c[\bar e_L(u^\dagger B_Ru^\dagger)_d^a](u^\dagger u_\mu u_\nu u^\dagger)^b_eF^{\mu\nu}\notag\\
    &\quad\quad+F_{15}[\hat{\mathcal{C}}^{(8)}_{26}]_{abc}\varepsilon^{ade}(T_L)_{\{a}(T_L)_{b\}}(T_R)_c[\bar e_L(u^\dagger B_Ru^\dagger)_f^b](u u_\mu u)_d^c( u u_\nu u)_e^fF^{\mu\nu}\notag\\
    &\quad\quad\quad+\cdots\bigg\}+\rm h.c.\,,
\end{align}
where the $(\cdots)$ denotes the similar chiral operators for $\mathcal{O}_{27-36}^{(8)}$. We estimate all unknown LECs through the NDA~\cite{Manohar:1983md,Gavela:2016bzc,Jenkins:2013sda}, and the resulting estimates are summarized in Tab.~\ref{tab:NDA}. A full account of the methodology is provided in App.~\ref{app:nda-matching}.

\begin{table}[H]
    \centering
    \begin{tabular}{|c|c|c|c|c|c|}
    \hline
        $C_1$ & $\mathcal{O}(\Lambda_\chi^3)$ &$C_2$&$\mathcal{O}(\Lambda_\chi^3)$&$D_1$&$\mathcal{O}(\Lambda_\chi^3)$\\
        $D_2$ & $\mathcal{O}(\Lambda_\chi^3)$&$D_3$&$\mathcal{O}(\Lambda_\chi)$&$E_1$&$\mathcal{O}(\Lambda_\chi^3)$\\
        $E_2$ & $\mathcal{O}(\Lambda_\chi^3)$&$E_3$&$\mathcal{O}(\Lambda_\chi^3)$&$F_1$&$\mathcal{O}(\Lambda_\chi)$\\
        $F_2$ & $\mathcal{O}(\Lambda_\chi)$&$F_3$&$\mathcal{O}(\Lambda_\chi)$&$F_4$&$\mathcal{O}(\Lambda_\chi)$\\
        $F_5$ & $\mathcal{O}(\Lambda_\chi)$&$F_6$&$\mathcal{O}(\Lambda_\chi)$&$F_{7}$&$\mathcal{O}(\Lambda_\chi)$\\
        $F_{8}$ & $\mathcal{O}(\Lambda_\chi)$&$F_{9}$&$\mathcal{O}(\Lambda_\chi)$&$F_{10}$&$\mathcal{O}(\Lambda_\chi)$\\
         $F_{11}$& $\mathcal{O}(\Lambda_\chi)$&$F_{12}$&$\mathcal{O}(\Lambda_\chi)$&$F_{13}$&$\mathcal{O}(\Lambda_\chi)$\\
         $F_{14}$& $\mathcal{O}(\Lambda_\chi)$&$F_{15}$&$\mathcal{O}(\Lambda_\chi)$&&\\
        \hline
    \end{tabular}
    \caption{The estimation of the unknown LECs via NDA.}
    \label{tab:NDA}
\end{table}

\section{Effective Lagrangian and Nucleon Decays}
\label{sec:BD}

After deriving the corresponding chiral Lagrangian via the systematic spurion method, we proceed to identify the effective Lagrangians that contribute to the BNV nucleon decay process. The expansions of the building blocks are presented in App.~\ref{app:chiral-expansion}, and we list the complete flavor structures of the BNV LEFT operators up to dimension 8 in Tab.~\ref{tab:flavor} and \ref{tab:flavord8}. Based on the chiral operators in Sec.~\ref{sec:CL} and the expansion rules, the complete decay channels of the LEFT operators can be directly determined. Specifically, the dimension 6 operators can contribute to BNV nucleon decay through two mechanisms: baryon pole contributions and direct contributions. In contrast, dimension 7 operators only contribute to BNV nucleon decay via the direct channel.

\begin{table}[H]
    \centering
    \resizebox{0.7\textwidth}{!}{\begin{tabular}{|c|c|c|c|c|c|}
    \hline
    \multicolumn{6}{|c|}{Dimension 6}\\
    \hline
        $[\mathcal{O}^{(6)}_1]_{121}$ &$[\mathcal{O}^{(6)}_1]_{131}$&$[\mathcal{O}^{(6)}_2]_{232}$&$[\mathcal{O}^{(6)}_2]_{233}$&$[\mathcal{O}^{(6)}_3]_{121}$&$[\mathcal{O}^{(6)}_3]_{131}$  \\
        $[\mathcal{O}^{(6)}_4]_{232}$ &$[\mathcal{O}^{(6)}_4]_{233}$&$[\mathcal{O}^{(6)}_5]_{121}$&$[\mathcal{O}^{(6)}_5]_{131}$&$[\mathcal{O}^{(6)}_6]_{232}$&$[\mathcal{O}^{(6)}_6]_{233}$  \\
        $[\mathcal{O}^{(6)}_7]_{121}$ &$[\mathcal{O}^{(6)}_7]_{131}$&$[\mathcal{O}^{(6)}_8]_{232}$&$[\mathcal{O}^{(6)}_8]_{233}$&$[\mathcal{O}^{(6)}_9]_{122}$&$[\mathcal{O}^{(6)}_9]_{123}$  \\
        $[\mathcal{O}^{(6)}_{9}]_{132}$ &$[\mathcal{O}^{(6)}_{9}]_{133}$&$[\mathcal{O}^{(6)}_{10}]_{122}$&$[\mathcal{O}^{(6)}_{10}]_{123}$&$[\mathcal{O}^{(6)}_{10}]_{132}$&$[\mathcal{O}^{(6)}_{10}]_{133}$  \\
        $[\mathcal{O}^{(6)}_{10}]_{231}$&$[\mathcal{O}^{(6)}_{11}]_{122}$&$[\mathcal{O}^{(6)}_{11}]_{123}$&$[\mathcal{O}^{(6)}_{11}]_{132}$&$[\mathcal{O}^{(6)}_{11}]_{133}$&$[\mathcal{O}^{(6)}_{11}]_{231}$ \\
        $[\mathcal{O}^{(6)}_{12}]_{122}$&$[\mathcal{O}^{(6)}_{12}]_{123}$&$[\mathcal{O}^{(6)}_{12}]_{132}$&$[\mathcal{O}^{(6)}_{12}]_{133}$& & \\  
        \hline
        \multicolumn{6}{|c|}{Dimension 7}\\
        \hline
        $[\mathcal{O}^{(7)}_1]_{112}$ &$[\mathcal{O}^{(7)}_1]_{113}$&$[\mathcal{O}^{(7)}_1]_{121}$&$[\mathcal{O}^{(7)}_1]_{131}$&$[\mathcal{O}^{(7)}_2]_{222}$&$[\mathcal{O}^{(7)}_2]_{223}$ \\
        $[\mathcal{O}^{(7)}_2]_{233}$ &$[\mathcal{O}^{(7)}_2]_{333}$ &$[\mathcal{O}^{(7)}_3]_{112}$ &$[\mathcal{O}^{(7)}_3]_{113}$&$[\mathcal{O}^{(7)}_4]_{222}$&$[\mathcal{O}^{(7)}_4]_{223}$\\
        $[\mathcal{O}^{(7)}_4]_{232}$&$[\mathcal{O}^{(7)}_4]_{233}$&$[\mathcal{O}^{(7)}_4]_{332}$&$[\mathcal{O}^{(7)}_4]_{333}$& $[\mathcal{O}^{(7)}_5]_{112}$ &$[\mathcal{O}^{(7)}_5]_{113}$ \\
        $[\mathcal{O}^{(7)}_6]_{222}$&$[\mathcal{O}^{(7)}_6]_{223}$&$[\mathcal{O}^{(7)}_6]_{232}$&$[\mathcal{O}^{(7)}_6]_{233}$&$[\mathcal{O}^{(7)}_6]_{332}$ &$[\mathcal{O}^{(7)}_6]_{333}$ \\
        $[\mathcal{O}^{(7)}_7]_{112}$ &$[\mathcal{O}^{(7)}_7]_{113}$&$[\mathcal{O}^{(7)}_7]_{121}$&$[\mathcal{O}^{(7)}_7]_{131}$&$[\mathcal{O}^{(7)}_8]_{222}$&$[\mathcal{O}^{(7)}_8]_{223}$  \\
        $[\mathcal{O}^{(7)}_8]_{233}$&$[\mathcal{O}^{(7)}_8]_{333}$&$[\mathcal{O}^{(7)}_9]_{122}$ &$[\mathcal{O}^{(7)}_9]_{123}$&$[\mathcal{O}^{(7)}_9]_{133}$&$[\mathcal{O}^{(7)}_9]_{221}$\\
        $[\mathcal{O}^{(7)}_9]_{331}$&$[\mathcal{O}^{(7)}_9]_{132}$ &$[\mathcal{O}^{(7)}_9]_{231}$ &$[\mathcal{O}^{(7)}_{10}]_{122}$&$[\mathcal{O}^{(7)}_{10}]_{123}$&$[\mathcal{O}^{(7)}_{10}]_{133}$\\
        $[\mathcal{O}^{(7)}_{11}]_{122}$&$[\mathcal{O}^{(7)}_{11}]_{123}$&$[\mathcal{O}^{(7)}_{11}]_{133}$ &$[\mathcal{O}^{(7)}_{12}]_{122}$&$[\mathcal{O}^{(7)}_{12}]_{133}$&$[\mathcal{O}^{(7)}_{12}]_{123}$\\
        $[\mathcal{O}^{(7)}_{12}]_{221}$&$[\mathcal{O}^{(7)}_{12}]_{331}$ &
        $[\mathcal{O}^{(7)}_{12}]_{132}$&$[\mathcal{O}^{(7)}_{12}]_{231}$& & \\
        \hline
    \end{tabular}}
    \caption{Flavor structures of the dimension 6 and dimension 7 BNV LEFT operators.}
    \label{tab:flavor}
\end{table}

\begin{table}[H]
    \centering
    \resizebox{0.8\textwidth}{!}{\begin{tabular}{|c|c|c|c|c|c|}
    \hline
    \multicolumn{6}{|c|}{Dimension 8}\\
    \hline
        $[\mathcal{O}^{(8)}_{1,37}]_{121}$ &$[\mathcal{O}^{(8)}_{1,37}]_{131}$&$[\mathcal{O}^{(8)}_{2,38}]_{232}$&$[\mathcal{O}^{(8)}_{2,38}]_{233}$&$[\mathcal{O}^{(8)}_{3,39}]_{121}$&$[\mathcal{O}^{(8)}_{3,39}]_{131}$  \\
        $[\mathcal{O}^{(8)}_{4,40}]_{232}$ &$[\mathcal{O}^{(8)}_{4,40}]_{233}$&$[\mathcal{O}^{(8)}_{5,41}]_{121}$&$[\mathcal{O}^{(8)}_{5,41}]_{131}$&$[\mathcal{O}^{(8)}_{6,42}]_{232}$&$[\mathcal{O}^{(8)}_{6,42}]_{233}$  \\
        $[\mathcal{O}^{(8)}_{7,43}]_{121}$ &$[\mathcal{O}^{(8)}_{7,43}]_{131}$&$[\mathcal{O}^{(8)}_{8,44}]_{232}$&$[\mathcal{O}^{(8)}_{8,44}]_{233}$&$[\mathcal{O}^{(8)}_{9,45}]_{122}$&$[\mathcal{O}^{(8)}_{9,45}]_{123}$  \\
        $[\mathcal{O}^{(8)}_{9,45}]_{132}$ &$[\mathcal{O}^{(8)}_{9,45}]_{133}$&$[\mathcal{O}^{(8)}_{9,45}]_{231}$&$[\mathcal{O}^{(8)}_{10,46}]_{122}$&$[\mathcal{O}^{(8)}_{10,46}]_{123}$&$[\mathcal{O}^{(8)}_{10,46}]_{132}$  \\
        $[\mathcal{O}^{(8)}_{10,46}]_{133}$&$[\mathcal{O}^{(8)}_{10,46}]_{231}$&$[\mathcal{O}^{(8)}_{11,47}]_{122}$&$[\mathcal{O}^{(8)}_{11,47}]_{123}$&$[\mathcal{O}^{(8)}_{11,47}]_{132}$&$[\mathcal{O}^{(8)}_{11,47}]_{133}$ \\
        $[\mathcal{O}^{(8)}_{11,47}]_{231}$&$[\mathcal{O}^{(8)}_{12,48}]_{122}$&$[\mathcal{O}^{(8)}_{12,48}]_{123}$&$[\mathcal{O}^{(8)}_{12,48}]_{132}$& $[\mathcal{O}^{(8)}_{12,48}]_{133}$& $[\mathcal{O}^{(8)}_{12,48}]_{231}$\\ 
        $[\mathcal{O}^{(8)}_{13,25}]_{112}$ &$[\mathcal{O}^{(8)}_{13,25}]_{121}$&$[\mathcal{O}^{(8)}_{13,25}]_{211}$&$[\mathcal{O}^{(8)}_{13,25}]_{131}$&$[\mathcal{O}^{(8)}_{13,25}]_{113}$&$[\mathcal{O}^{(8)}_{13,25}]_{311}$  \\
        $[\mathcal{O}^{(8)}_{14,26}]_{222}$ &$[\mathcal{O}^{(8)}_{14,26}]_{333}$&$[\mathcal{O}^{(8)}_{14,26}]_{223}$ &$[\mathcal{O}^{(8)}_{14,26}]_{232}$&$[\mathcal{O}^{(8)}_{14,26}]_{322}$&$[\mathcal{O}^{(8)}_{14,26}]_{332}$\\
        $[\mathcal{O}^{(8)}_{14,26}]_{323}$&$[\mathcal{O}^{(8)}_{14,26}]_{233}$&
        $[\mathcal{O}^{(8)}_{15,27}]_{112}$ &$[\mathcal{O}^{(8)}_{15,27}]_{121}$&$[\mathcal{O}^{(8)}_{15,27}]_{211}$&$[\mathcal{O}^{(8)}_{15,27}]_{131}$\\
        $[\mathcal{O}^{(8)}_{15,27}]_{113}$&$[\mathcal{O}^{(8)}_{15,27}]_{311}$&$[\mathcal{O}^{(8)}_{16,28}]_{222}$ &$[\mathcal{O}^{(8)}_{16,28}]_{333}$&$[\mathcal{O}^{(8)}_{16,28}]_{223}$ &$[\mathcal{O}^{(8)}_{16,28}]_{232}$ \\
        $[\mathcal{O}^{(8)}_{16,28}]_{322}$&$[\mathcal{O}^{(8)}_{16,28}]_{332}$&$[\mathcal{O}^{(8)}_{16,28}]_{323}$&$[\mathcal{O}^{(8)}_{16,28}]_{233}$&$[\mathcal{O}^{(8)}_{17,29}]_{112}$ &$[\mathcal{O}^{(8)}_{17,29}]_{121}$\\
        $[\mathcal{O}^{(8)}_{17,29}]_{211}$&$[\mathcal{O}^{(8)}_{17,29}]_{131}$&$[\mathcal{O}^{(8)}_{17,29}]_{113}$&$[\mathcal{O}^{(8)}_{17,29}]_{311}$&$[\mathcal{O}^{(8)}_{18,30}]_{222}$ &$[\mathcal{O}^{(8)}_{18,30}]_{333}$\\
        $[\mathcal{O}^{(8)}_{18,30}]_{223}$ &$[\mathcal{O}^{(8)}_{18,30}]_{232}$&$[\mathcal{O}^{(8)}_{18,30}]_{322}$&$[\mathcal{O}^{(8)}_{18,30}]_{332}$&$[\mathcal{O}^{(8)}_{18,30}]_{323}$&$[\mathcal{O}^{(8)}_{18,30}]_{233}$  \\
        $[\mathcal{O}^{(8)}_{19,31}]_{112}$ &$[\mathcal{O}^{(8)}_{19,31}]_{121}$&$[\mathcal{O}^{(8)}_{19,31}]_{211}$&$[\mathcal{O}^{(8)}_{19,31}]_{131}$&$[\mathcal{O}^{(8)}_{19,31}]_{113}$&$[\mathcal{O}^{(8)}_{19,31}]_{311}$  \\
        $[\mathcal{O}^{(8)}_{20,32}]_{222}$ &$[\mathcal{O}^{(8)}_{20,32}]_{333}$&$[\mathcal{O}^{(8)}_{20,32}]_{223}$ &$[\mathcal{O}^{(8)}_{20,32}]_{232}$&$[\mathcal{O}^{(8)}_{20,32}]_{322}$&$[\mathcal{O}^{(8)}_{20,32}]_{332}$ \\
        $[\mathcal{O}^{(8)}_{20,32}]_{323}$&$[\mathcal{O}^{(8)}_{20,32}]_{233}$&
        $[\mathcal{O}^{(8)}_{21,33}]_{122}$ &$[\mathcal{O}^{(8)}_{21,33}]_{212}$&$[\mathcal{O}^{(8)}_{21,33}]_{221}$&$[\mathcal{O}^{(8)}_{21,33}]_{133}$\\
        $[\mathcal{O}^{(8)}_{21,33}]_{313}$&$[\mathcal{O}^{(8)}_{21,33}]_{331}$&$[\mathcal{O}^{(8)}_{21,33}]_{123}$ &$[\mathcal{O}^{(8)}_{21,33}]_{132}$&$[\mathcal{O}^{(8)}_{21,33}]_{213}$&$[\mathcal{O}^{(8)}_{21,33}]_{231}$\\
        $[\mathcal{O}^{(8)}_{21,33}]_{312}$&  $[\mathcal{O}^{(8)}_{22,34}]_{122}$&$[\mathcal{O}^{(8)}_{22,34}]_{212}$&$[\mathcal{O}^{(8)}_{22,34}]_{221}$&$[\mathcal{O}^{(8)}_{22,34}]_{133}$&$[\mathcal{O}^{(8)}_{22,34}]_{313}$\\
        $[\mathcal{O}^{(8)}_{22,34}]_{331}$  &$[\mathcal{O}^{(8)}_{22,34}]_{123}$&$[\mathcal{O}^{(8)}_{22,34}]_{132}$&$[\mathcal{O}^{(8)}_{22,34}]_{213}$&$[\mathcal{O}^{(8)}_{22,34}]_{231}$&$[\mathcal{O}^{(8)}_{22,34}]_{312}$\\
        $[\mathcal{O}^{(8)}_{22,34}]_{321}$  &$[\mathcal{O}^{(8)}_{23,35}]_{122}$&$[\mathcal{O}^{(8)}_{23,35}]_{212}$&$[\mathcal{O}^{(8)}_{23,35}]_{221}$&$[\mathcal{O}^{(8)}_{23,35}]_{133}$&$[\mathcal{O}^{(8)}_{23,35}]_{313}$\\
        $[\mathcal{O}^{(8)}_{23,35}]_{331}$  &$[\mathcal{O}^{(8)}_{23,35}]_{123}$&$[\mathcal{O}^{(8)}_{23,35}]_{132}$&$[\mathcal{O}^{(8)}_{23,35}]_{213}$&$[\mathcal{O}^{(8)}_{23,35}]_{231}$&$[\mathcal{O}^{(8)}_{23,35}]_{312}$\\
        $[\mathcal{O}^{(8)}_{23,35}]_{321}$ &$[\mathcal{O}^{(8)}_{24,36}]_{122}$&$[\mathcal{O}^{(8)}_{24,36}]_{212}$&$[\mathcal{O}^{(8)}_{24,36}]_{221}$&$[\mathcal{O}^{(8)}_{24,36}]_{133}$&$[\mathcal{O}^{(8)}_{24,36}]_{313}$\\
        $[\mathcal{O}^{(8)}_{24,36}]_{331}$ & $[\mathcal{O}^{(8)}_{24,36}]_{123}$&$[\mathcal{O}^{(8)}_{24,36}]_{132}$&$[\mathcal{O}^{(8)}_{24,36}]_{213}$&$[\mathcal{O}^{(8)}_{24,36}]_{231}$&$[\mathcal{O}^{(8)}_{24,36}]_{312}$\\
         \hline
    \end{tabular}}
    \caption{Flavor structures of the dimension 8 BNV LEFT operators.}
    \label{tab:flavord8}
\end{table}

Here, we consider the complete expansion of the LO chiral Lagrangian. According to the BNV chiral operators in Sec.~\ref{sec:CL} and the expansion in App.~\ref{app:chiral-expansion}, the relevant effective Lagrangian for BNV nucleon decays at LO are obtained 
\begin{align}
\mathcal{L}_{\slashed{B}}^{\rm LO}
&= \frac{i}{f}\bigg\{\frac{1}{\sqrt{6}}(3\beta [\hat{\mathcal{C}}^{(6)}_1]_{121}+\alpha[\hat{\mathcal{C}}^{(6)}_3]_{121})\bigg[(p^T_LCe_L)\eta-(p^T_RCe_R)\eta\bigg]\notag\\
&\quad\quad\quad+\frac{1}{\sqrt{2}}(\beta[\hat{\mathcal{C}}^{(6)}_1]_{121}+\alpha[\hat{\mathcal{C}}^{(6)}_3]_{121})\bigg[(p^T_LCe_L)\pi^0-(p^T_RCe_R)\pi^0\bigg]\notag\\
&\quad\quad\quad+(\beta[\hat{\mathcal{C}}^{(6)}_1]_{121}+\alpha[\hat{\mathcal{C}}^{(6)}_3]_{131})\bigg[(n^T_LCe_L)\pi^+-(n^T_RCe_R)\pi^+\bigg]\notag\\
&\quad\quad\quad+(\beta[\hat{\mathcal{C}}^{(6)}_1]_{131}+\alpha[\hat{\mathcal{C}}^{(6)}_3]_{131})\bigg[( p_L^T C e_L)\bar K^0-(p_R^T C e_R)\bar K^0\bigg]\notag\\
&\quad\quad\quad+(\beta[\hat{\mathcal{C}}^{(6)}_4]_{232}+\alpha[\hat{\mathcal{C}}^{(6)}_8]_{232})\bigg[(\bar e_Rn_L)K^--(\bar e_Ln_R)K^-\bigg]\notag\\
&\quad\quad\quad+\frac{1}{\sqrt{6}}(3\beta[\hat{\mathcal{C}}^{(6)}_9]_{122} + \alpha[\hat{\mathcal{C}}^{(6)}_{11}]_{122})(n^T_L C \nu_L)\eta\notag\\
&\quad\quad\quad + \frac{1}{\sqrt{6}}(\alpha[\hat{\mathcal{C}}^{(6)}_{10}]_{122} + 3\beta[\hat{\mathcal{C}}^{(6)}_{12}]_{122})(\bar\nu_L n_R)\eta\notag\\
&\quad\quad\quad+\frac{1}{\sqrt{2}}(\beta[\hat{\mathcal{C}}^{(6)}_9]_{122} + \alpha[\hat{\mathcal{C}}^{(6)}_{11}]_{122})(n^T_L C \nu_L)\pi^0\notag\\
&\quad\quad\quad  + \frac{1}{\sqrt{2}}(\alpha[\hat{\mathcal{C}}^{(6)}_{10}]_{122} + \beta[\hat{\mathcal{C}}^{(6)}_{12}]_{122})(\bar\nu_L n_R)\pi^0\notag\\
& \quad\quad\quad+ (\beta[\hat{\mathcal{C}}^{(6)}_9]_{122} - \alpha[\hat{\mathcal{C}}^{(6)}_{11}]_{122})(p^T_L C \nu_L)\pi^-
  - (\alpha[\hat{\mathcal{C}}^{(6)}_{10}]_{122} - \beta[\hat{\mathcal{C}}^{(6)}_{12}]_{122})(\bar\nu_L p_R)\pi^- \notag\\
&\quad\quad\quad+
  (\beta[\hat{\mathcal{C}}^{(6)}_9]_{123} + \beta[\hat{\mathcal{C}}^{(6)}_9]_{132} + \alpha[\hat{\mathcal{C}}^{(6)}_{11}]_{123} + \alpha[\hat{\mathcal{C}}^{(6)}_{11}]_{132})(n^T_L C \nu_L)\bar{K}^0 \notag\\
&\quad\quad\quad+
  (\alpha[\hat{\mathcal{C}}^{(6)}_{10}]_{123} + \alpha[\hat{\mathcal{C}}^{(6)}_{10}]_{132} + \beta[\hat{\mathcal{C}}^{(6)}_{12}]_{123} + \beta[\hat{\mathcal{C}}^{(6)}_{12}]_{132})(\bar\nu_L n_R)\bar{K}^0 \notag\\
&\quad\quad\quad+
  (\beta[\hat{\mathcal{C}}^{(6)}_9]_{123} + \alpha[\hat{\mathcal{C}}^{(6)}_{11}]_{123} + \alpha[\hat{\mathcal{C}}^{(6)}_{11}]_{231})(p^T_L C \nu_L)K^-\notag\\
&\quad\quad\quad+ (\alpha[\hat{\mathcal{C}}^{(6)}_{10}]_{123} + \alpha[\hat{\mathcal{C}}^{(6)}_{10}]_{231} + \beta[\hat{\mathcal{C}}^{(6)}_{12}]_{123})(\bar\nu_L p_R)K^-
\Bigg\}+\rm .h.c.\,,
\end{align}
which is the same as Refs.~\cite{Beneito:2023xbk}. Moreover, the NLO chiral operators are similar with the LO if we only consider the two-body decay, there is the derivative in the the meson as $\partial_\mu M$ with the vector fermion bilinear.

\begin{figure}[h!]
    \centering
    \begin{minipage}[t]{0.4\textwidth}
        \centering
        \includegraphics[width=\linewidth]{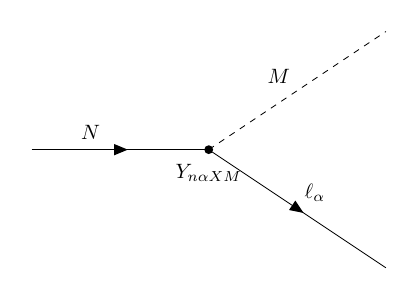}
    \end{minipage}
    \begin{minipage}[t]{0.4\textwidth}
        \centering
        \includegraphics[width=\linewidth]{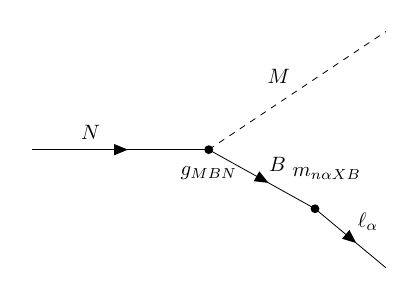}
    \end{minipage}
    \centering
    \caption{The Feynman diagrams of BNV nucleon decay processes.}
    \label{fig:feynman_both}
\end{figure}

In addition, there is also the indirect contribution through the Fig.~\ref{fig:feynman_both}. The interactions $g_{MBN}$ are generated from the Eq.~\eqref{eq:BC}
\begin{align}
    \mathcal{L}_B=g_{MBN}(\bar N\gamma^5\gamma^\mu N)\partial_\mu M\,.
\end{align}
The BNV property of the indirect contribution comes from the non meson expansion of the chiral operators
\begin{align}
    \mathcal{L}^{\rm indrect}_{\slashed B}=m_{1\alpha XB}(\bar l_{\alpha}P_X B)+m_{2\alpha XB}(B^TCP_Xl)
\end{align}
where the $P_X=\frac{1\pm\gamma^5}{2}$ is the chirality projection and the constants $m_{X}$ can be obtained through the Eq.~\eqref{eq:dimension6}. The interaction of the direct contribution can be organized 
\begin{align}
    \mathcal{L}_{\slashed B}^{\rm direct}=&Y_{1\alpha XM}(\bar l_\alpha P_XN)M+Y_{2\alpha XM}(N^TCP_Xl_\alpha)M\notag\\
    &+Y_{3\alpha XM}(\bar l_\alpha P_X\gamma^\mu N)\partial_\mu M+Y_{4\alpha XM}(N^TCP_X\gamma^\mu l_\alpha)\partial_\mu M
\end{align}

There are two types $\Delta (B-L)=0$ and $|\Delta(B-L)|=2$ BNV nucleon decay. The amplitude of decay $N\to \ell M$ with $\Delta (B-L)=2$ then becomes
\begin{align}\label{eq:ampBL2}
    i\mathcal{M} =  \bar u_\ell  P_{X}\bigg[-Y_{1\alpha XM} -Y_{3\alpha XM}\slashed p  + \sum_B m_{1\alpha XB} \frac{\slashed{k}+m_B}{k^2-m_B^2} (D+F) \slashed{p}\gamma_5\bigg]  u_N  
\end{align}
with the lepton momentum $k$ and meson momentum $p$. The summation over the baryons includes any of $p,n,\Lambda^0,\Sigma^{0,\pm}$ and the amplitude of the $\Delta (B-L)=0$ process is similar to Eq.~\eqref{eq:ampBL2}. Moreover, the two-body nucleon decay rate can be estimated directly~\cite{Kleiss:1985gy}
\begin{align}
    \Gamma(N\to M \ell_\alpha) & = \frac{1}{16\pi m_N} \overline{|M|^2}\,,
\end{align}
and the dimension 6 decay rates have been obtained systematically in Ref.~\cite{Beneito:2023xbk}. We can consider the dimension 7 LEFT operators which only contribute to the decay directly. Here, we take the dimension 7 LEFT operators $[\mathcal{O}_{3}^{(7)}]_{112}$ as the example. The chiral operators of the $[\mathcal{O}_{3}^{(7)}]_{112}$ are
\begin{align}
    \mathcal{L}_{[\mathcal{O}_{3}^{(7)}]_{112}}=&C_1[\hat{\mathcal{C}}^{(7)}_3]_{112}\varepsilon_{2de} (\lambda^A)^d_1(\lambda^B)^e_1(T_R)^{\{1}(T_R)^1(T_R)^{2\}}[(u B_R^TCu^\dagger)^A\gamma^\mu e_L ](u u_\mu u^\dagger)^B\notag\\
    =&\frac{1}{3}C_1[\hat{\mathcal{C}}^{(7)}_3]_{112}\varepsilon_{2de} (\lambda^A)^d_1(\lambda^B)^e_1(T_R)^{1}(T_R)^1(T_R)^{2}[(u B_R^TCu^\dagger)^A\gamma^\mu e_L ](u u_\mu u^\dagger)^B\notag\\
    &+\frac{1}{3}C_1[\hat{\mathcal{C}}^{(7)}_3]_{112}\varepsilon_{1de} (\lambda^A)^d_1(\lambda^B)^e_2(T_R)^{1}(T_R)^2(T_R)^{1}[(u B_R^TCu^\dagger)^A\gamma^\mu e_L ](u u_\mu u^\dagger)^B\notag\\
    &+\frac{1}{3}C_1[\hat{\mathcal{C}}^{(7)}_3]_{112}\varepsilon_{1de} (\lambda^A)^d_2(\lambda^B)^e_1(T_R)^{2}(T_R)^1(T_R)^{1}[(u B_R^TCu^\dagger)^A\gamma^\mu e_L ](u u_\mu u^\dagger)^B\notag\\
    =&\frac{1}{3}C_1[\hat{\mathcal{C}}^{(7)}_3]_{112}\varepsilon_{213} (\lambda^A)^1_1(\lambda^B)^3_1[(u B_R^TCu^\dagger)^A\gamma^\mu e_L ](u u_\mu u^\dagger)^B\notag\\
    &+\frac{1}{3}C_1[\hat{\mathcal{C}}^{(7)}_3]_{112}\varepsilon_{231} (\lambda^A)^3_1(\lambda^B)^1_1[(u B_R^TCu^\dagger)^A\gamma^\mu e_L ](u u_\mu u^\dagger)^B\notag\\
    &+\frac{1}{3}C_1[\hat{\mathcal{C}}^{(7)}_3]_{112}\varepsilon_{123} (\lambda^A)^2_1(\lambda^B)^3_2[(u B_R^TCu^\dagger)^A\gamma^\mu e_L ](u u_\mu u^\dagger)^B\notag\\
    &+\frac{1}{3}C_1[\hat{\mathcal{C}}^{(7)}_3]_{112}\varepsilon_{132} (\lambda^A)^3_1(\lambda^B)^2_2[(u B_R^TCu^\dagger)^A\gamma^\mu e_L ](u u_\mu u^\dagger)^B\notag\\
    &+\frac{1}{3}C_1[\hat{\mathcal{C}}^{(7)}_3]_{112}\varepsilon_{123} (\lambda^A)^2_2(\lambda^B)^3_1[(u B_R^TCu^\dagger)^A\gamma^\mu e_L ](u u_\mu u^\dagger)^B\notag\\
    &+\frac{1}{3}C_1[\hat{\mathcal{C}}^{(7)}_3]_{112}\varepsilon_{132} (\lambda^A)^3_2(\lambda^B)^2_1[(u B_R^TCu^\dagger)^A\gamma^\mu e_L ](u u_\mu u^\dagger)^B\notag\\
    =&\frac{2\sqrt{2}}{3}C_1[\hat{\mathcal{C}}^{(7)}_3]_{112}[p^T_RC\gamma^\mu e_L]\partial_\mu\pi^0-\frac{2}{3}C_1[\hat{\mathcal{C}}^{(7)}_3]_{112} [n^T_RC\gamma^\mu e_L ]\partial_\mu\pi^++\cdots\,,
\end{align}
where $(\cdots)$ denote the hyperon interactions. The BNV decay rates of the proton and neutron for $[\mathcal{O}_{3}^{(7)}]_{112}$ are given by
\begin{align}
    \Gamma(p\to \pi^0 e^+) & = \frac{1}{16\pi m_N} \overline{|M|^2}=\frac{C_1^2 \left[ \hat{\mathcal{C}}_3^{(7)} \right]_{112}^2}{36\pi m_p}  \left[ m_p^4 - m_p^2\left( m_\pi^2 + 2m_e^2 \right) + m_e^2\left( m_e^2 - m_\pi^2 \right) \right],\notag\\
    \Gamma(n\to \pi^- e^+) & = \frac{1}{16\pi m_N} \overline{|M|^2}=\frac{C_1^2 \left[ \hat{\mathcal{C}}_3^{(7)} \right]_{112}^2}{18\pi m_n}  \left[ m_n^4 - m_n^2\left( m_\pi^2 + 2m_e^2 \right) + m_e^2\left( m_e^2 - m_\pi^2 \right) \right].
\end{align}
For other LEFT operators (e.g., $[\mathcal{O}_{3}^{(7)}]_{113}$, which contributes to the decay channel $p\rightarrow \bar K^0+e^+$), the functional form of the decay rate remains unchanged, with only the mass terms (associated with the final-state mesons/leptons) modified. This systematic dependence on the final-state particle masses allows the complete set of dimension-7 BNV decay rates to be derived in a straightforward manner, following the same formalism as the illustrative case above.

Moreover, the BNV hyperon decay processes exhibit a similar theoretical framework to those of BNV nucleon decays. Specifically, the relevant hyperon interaction terms can be systematically identified by leveraging the chiral operators and flavor structures tabulated in Tab.~\ref{tab:flavor}, following the same methodology applied to nucleon decays. When extending the analysis to include more meson expansions of LO chiral operators and complete contributions from dimension-8 chiral operators, multi-body decay channels (involving more than two final-state particles) emerge as non-negligible components of the total BNV decay width. For the quantitative estimation of the decay rates of such multi-body BNV processes, we adopt the well-established formula from Ref.~\cite{Kleiss:1985gy}:
\begin{align}
    \Gamma(N \to k\ \text{particles}) \sim \frac{2^{5-4k} \pi^{3-2k}}{2m_p} \frac{m_p^{2k-4} |\mathcal{M}|^2}{(k-1)!(k-2)!}\,,
\end{align}
where the masses of the final-state particles are neglected for simplicity, and the integer $k$ denotes the total number of particles in the final state of the decay process.

\section{Standard Model Effective Field Theory and UV Models}
\label{sec:SMUV}

Having established the low-energy description of BNV processes in the LEFT and carried it through ChPT to the differential decay rates, we now turn to
the UV side of the same EFT chain. The SMEFT serves as the proper framework to characterize the high-energy dynamics above the electroweak scale.

\subsection{Standard Model Effective Field Theory operators}

The SMEFT is constructed by expanding the NP Lagrangian in a power series of $1/\Lambda$, retaining all gauge-invariant, Lorentz-invariant operators that can be formed using the SM field content. The general form of the SMEFT Lagrangian is given by~\cite{Grzadkowski:2010es,Lehman:2014jma,Li:2020gnx,Murphy:2020rsh,Li:2020xlh,Liao:2020jmn}:
\begin{equation}
\mathcal{L}_{\text{SMEFT}} = \mathcal{L}_{\text{SM}} + \sum_{d \geq 5} \frac{1}{\Lambda^{d-4}} \sum_{i} C_i^{(d)} \mathcal{O}_i^{(d)}\,,    
\end{equation}
where $\mathcal{L}_{\text{SM}}$ is the SM Lagrangian, $C_i^{(d)}$ are dimensionless Wilson coefficients that characterize the strength of NP effects, and $\mathcal{O}_i^{(d)}$ represent dimension-$d$ effective operators. BNV in SMEFT is encoded in those effective operators that carry a non-zero net baryon number ($\Delta B \neq 0$). 

The SM Lagrangian in unbroken phase reads:
\begin{eqnarray}
\mathcal L_{\text{SM}} &=& -\dfrac{1}{4} G^A_{\mu\nu} G^{A\mu\nu} - \dfrac{1}{4} W^I_{\mu\nu} W^{I\mu\nu}
- \dfrac{1}{4} B_{\mu\nu} B^{\mu\nu} \nn\\
&& +\sum_{\psi=Q,u_R,d_R,\ell,e_R} \overline\psi i\slashed D\psi + (D_\mu H)^\dagger (D^\mu H)
- \lambda \left(H^\dagger H-\dfrac{1}{2}v^2\right)^2 \nn\\
&& -\left[\overline \ell e_R(Y_e)H+\overline Qu_R(Y_u)\tilde H+\overline Qd_R(Y_d)H+\text{h.c.}\right]
\nn\\
&& +\dfrac{\theta_3{g_s^2}}{32\pi^2}G^A_{\mu\nu}\tilde G^{A\mu\nu}
+\dfrac{\theta_2g^2}{32\pi^2}W^I_{\mu\nu}\tilde W^{I\mu\nu}+\dfrac{\theta_1{g'}^2}{32\pi^2}B_{\mu\nu}\tilde B^{\mu\nu} ~.
\label{SMLagrangian}
\end{eqnarray}
The gauge covariant derivative is $D_\mu=\partial_\mu+ig_1YB_\mu+ig_2t^IW_\mu^I+ig_3T^AG_\mu^A$. Here, $T^A=\lambda^A/2$ and $t^I=\tau^I/2$ are the $SU(3)_C$ and $SU(2)_L$ generators, while $\lambda^A$ and $\tau^I$ are the Gell-Mann and Pauli matrices, and $Y$ is the $U(1)_Y$ generator. The complex conjugate of the Higgs field is denoted either as $H^\dagger$ or $\tilde{H}$. $\tilde{H}$ is defined by $\tilde{H}_i=\varepsilon_{ij}H^{\dagger j}$ where $\varepsilon_{ij}$ is totally antisymmetric with $\varepsilon_{12}=1$.

The fermion sector consists of left-handed doublet fields $Q = (u_L,d_L)^T$ and $\ell = (\nu_L,e_L)^T$, together with right-handed singlet fields $u_R$, $d_R$, $s_R$ and $e_R$. We only consider the up ($u$), down ($d$) and strange ($s$) quarks in our LEFT operators, so the quark flavor degrees of freedom are not taken into account in SMEFT operators. All fermion fields carry weak-eigenstate indices $r=1,2,3$, and the Yukawa couplings $Y_e$, $Y_u$, $Y_d$ are $3\times 3$ matrices in flavor space. The subscripts $\alpha,\beta,\gamma$ are $SU(3)_C$ color indices, $i,j,k$ are $SU(2)_L$ indices, and $\varepsilon$ denotes the antisymmetric tensor in the SMEFT operators shown in Tab.~\ref{tab:SMEFT}.

In the SM, quark fields (left-handed doublet $Q$ and right-handed singlets $u_R, d_R$) carry a baryon number of $B = 1/3$, while all other SM fields (leptons, Higgs boson, and gauge bosons) have $B = 0$. Consequently, a BNV operator in SMEFT must contain an odd number of quark fields to yield a non-zero $\Delta B$. The SMEFT operators are invariant under the $SU(3)_c\times SU(2)_L\times U(1)_Y$ symmetry and the Lorentz symmetry, implying that baryon number violation and lepton number violation satisfy:
\begin{align}
(\Delta B,~\Delta L)=\left\{\begin{array}{l}
  {\rm (odd,~odd)}\\
{\rm (even,~even)}\\
    \end{array}\right.\,,\quad|\Delta B-\Delta L|=\left\{\begin{array}{l}
0,~4,~8,~...,~~{\rm for}~n=\rm even \\
2,~6,~...,~~{\rm for}~n=\rm odd\\
    \end{array}\right.\,,
\end{align}
where $n$ is the dimension of corresponding SMEFT operators. In order to include the complete set of LEFT operators up to dimension 8, we consider the BNV SMEFT operators up to dimension 9. It should be emphasized that, for dimension-9 operators, the $\Delta B=2$ SMEFT operators arise,
and we only include those SMEFT operators that contribute to LEFT operators below dimension 8. Here, we list the relevant SMEFT operators in Tab.~\ref{tab:SMEFT} and give the direct matching results in App.~\ref{app:MatchingSM2L}. Similarly, the QCD RGEs of the dimension 6 SMEFT operators are given by\cite{Jenkins:2013zja,Alonso:2013hga,Jenkins:2013wua}:
\begin{align}
    \frac{d}{d\ln\mu}C_{duq}&=-\frac{\alpha_s}{\pi}C_{duq}\,,\notag\\
    \frac{d}{d\ln\mu}C_{qqu}&=-\frac{\alpha_s}{\pi}C_{qqu}\,,\notag\\
    \frac{d}{d\ln\mu}C_{qqq}&=-\frac{\alpha_s}{\pi}C_{qqq}\,,\notag\\
    \frac{d}{d\ln\mu}C_{duu}&=-\frac{\alpha_s}{\pi}C_{duu}\,.
\end{align}

\begin{table}[H]
  \centering
  \resizebox{0.8\textwidth}{!}{
\begin{tabular}{|c|c|c|c|}
\hline 
\multicolumn{2}{|c|}{Dimension-6}&\multicolumn{2}{|c|}{Dimension-7}\\
\hline
 ${\cal O}_{duq}$ & { $\varepsilon^{\alpha\beta\gamma}\varepsilon^{jk} (d_{R\alpha}^{T} C u_{R\beta}) (Q_{\gamma j}^{T} C \ell_{k})$ } &${\cal O}_{LQddD}$ & $\varepsilon^{\alpha\beta\gamma}(\overline{\ell}^{i} \gamma^\mu Q_{\alpha i}) (d_{R\beta}^{T} C iD_{\mu} d_{R\gamma}) $\\
 ${\cal O}_{qqu}$ & { $\varepsilon^{\alpha\beta\gamma}\varepsilon^{jk} (Q_{\alpha j}^{T} C Q_{\beta k}) (u_{R\gamma}^{T} C e_R)$ } &${\cal O}_{LdudH}$ & $\varepsilon^{\alpha\beta\gamma}\varepsilon^{ij} (\overline{\ell}^i d_{R\alpha}) (u_{R\beta}^{T} C d_{R\gamma}) H^{*j} $\\
${\cal O}_{qqq}$ & { $\varepsilon^{\alpha\beta\gamma}\varepsilon^{jn}\varepsilon^{km} (Q_{\alpha j}^{T} C Q_{\beta k}) (Q_{\gamma m}^{T} C \ell_{n})$ } &${\cal O}_{LdddH}$ & $\varepsilon^{\alpha\beta\gamma}(\overline{\ell}^i d_{R\alpha}) (d_{R\beta}^{T} C d_{R\gamma}) H_i $\\ 
${\cal O}_{duu}$ & { $\varepsilon^{\alpha\beta\gamma} (d_{R\alpha}^{T} C u_{R\beta}) (u_{R\gamma}^{T} C e_R)$ } &${\cal O}_{eQddH}$ & $\varepsilon^{\alpha\beta\gamma}(\overline{e}_R Q_{\alpha i}) (d_{R\beta}^{T} C d_{R\gamma}) H^{*i} $\\
&&${\cal O}_{edddD}$ & $\varepsilon^{\alpha\beta\gamma}(\overline{e}_R \gamma^\mu d_{R\alpha}) (d_{R\beta}^{T} C iD_{\mu} d_{R\gamma}) $\\
&&${\cal O}_{LdQQH}$ & $\varepsilon^{\alpha\beta\gamma}(\overline{\ell}^{k} d_{R\alpha}) (Q_{\beta k}^{T} C Q_{\gamma i}) H^{*i} $ \\
\hline
\multicolumn{4}{|c|}{Dimension-8}\\
\hline
$Q_{lq^2uHD}$  &  $\varepsilon^{\alpha\beta\gamma} \varepsilon^{ik} \left(u_{R\alpha}^TC\gamma^\mu \ell_i\right)\left(Q_{\beta j}^TCD_\mu Q_{\gamma k}\right)H^{\dagger j}$&$Q_{lq^2dHD}$  &  $\varepsilon^{\alpha\beta\gamma} \varepsilon^{ik} \varepsilon^{jm}\left(d_{R\alpha}^TC\gamma^\mu \ell_i\right)\left(Q_{\beta j}^TCD_\mu Q_{\gamma k}\right)H_{m}$\\
$Q_{lu^2dHD}$&  $\varepsilon^{\alpha\beta\gamma} \varepsilon^{ij} \left(d_{R\alpha}^TC\gamma^\mu \ell_j\right)\left(u_{R\beta }^TCD_\mu u_{R\gamma }\right)H_i$&$Q_{lud^2HD}$  &   $\varepsilon^{\alpha\beta\gamma}  \left(d_{R\alpha}^TC\gamma^\mu \ell_i\right)\left(d_{R\beta }^TCD_\mu u_{R\gamma }\right)H^{\dagger i}$\\
$Q_{eq^3HD}^{(1)}$  &   $\varepsilon^{\alpha\beta\gamma} \varepsilon^{ik}\varepsilon^{jm}\left(Q_{\alpha i}^TC\gamma^\mu e_R\right)\left(Q_{\beta j}^TCD_\mu Q_{\gamma k}\right)H_m$&$Q_{eq^3HD}^{(2)}$  &   $\varepsilon^{\alpha\beta\gamma} \varepsilon^{ij}\varepsilon^{km}\left(Q_{\alpha i}^TC\gamma^\mu e_R\right)\left(Q_{\beta j}^TCD_\mu Q_{\gamma k}\right)H_m$\\
$Q_{equ^2HD}$&$\varepsilon^{\alpha\beta\gamma} \varepsilon^{ij}\left(Q_{\alpha i}^TC\gamma^\mu e_R\right)\left(u_{R\beta }^TCD_\mu u_{R\gamma }\right)H_j$&$Q_{equdHD}$&$\varepsilon^{\alpha\beta\gamma} \varepsilon^{ij}\left(Q_{\alpha i}^TC\gamma^\mu e_R\right)\left(u_{R\beta }^TCD_\mu d_{R\gamma }\right)H^{\dagger i}$\\
$Q_{lq^3D^2}^{(1)}$ &$\varepsilon^{\alpha\beta\gamma} \varepsilon^{ij}\varepsilon^{km}\left(Q_{\alpha i}^TCQ_{\beta k}\right)\left(D_\mu Q_{\gamma j}^TCD^\mu\ell_{m}\right)$&$Q_{lq^3D^2}^{(2)}$ &$\varepsilon^{\alpha\beta\gamma} \varepsilon^{ik}\varepsilon^{jm}\left(Q_{\alpha i}^TC\sigma^{\mu\nu}Q_{\beta k}\right)\left(D_\mu Q_{\gamma j}^TCD_\nu\ell_{m}\right)$\\
$Q_{lq^3D^2}^{(3)}$ &$\varepsilon^{\alpha\beta\gamma} \varepsilon^{ij}\varepsilon^{km}\left(Q_{\alpha i}^TC\sigma^{\mu\nu}Q_{\beta k}\right)\left(D_\mu Q_{\gamma j}^TCD_\nu\ell_{m}\right)$&$Q_{lqudD^2}^{(1)}$&$\varepsilon^{\alpha\beta\gamma} \varepsilon^{ij}\left(u_{R\alpha }^TCd_{R\beta }\right)\left(D_\mu Q_{\gamma i}^TCD^\mu\ell_{j}\right)$\\
$Q_{lqudD^2}^{(2)}$&$\varepsilon^{\alpha\beta\gamma} \varepsilon^{ij}\left(u_{R\alpha }^TC\sigma^{\mu\nu}d_{\beta }\right)\left(D_\mu Q_{\gamma i}^TCD_\nu\ell_{j}\right)$&$Q_{eq^2uD^2}^{(1)}$&$\varepsilon^{\alpha\beta\gamma} \varepsilon^{ij}\left(Q_{\alpha i}^TCQ_{\beta j}\right)\left(D_\mu u_{R\gamma }^TCD^\mu e_{R}\right)$\\
$Q_{eq^2uD^2}^{(2)}$&$\varepsilon^{\alpha\beta\gamma} \varepsilon^{ij}\left(Q_{\alpha i}^TC\sigma^{\mu\nu}Q_{\beta j}\right)\left(D_\mu u_{R\gamma }^TCD_\nu e_{R}\right)$&$Q_{eu^2dD^2}^{(1)}$&$\varepsilon^{\alpha\beta\gamma} \left(u_{R\alpha }^TCd_{R\beta}\right)\left(D_\mu u_{R\gamma }^TCD^\mu e_{R}\right)$\\
$Q_{eu^2dD^2}^{(2)}$&$\varepsilon^{\alpha\beta\gamma} \left(u_{R\alpha }^TC\sigma^{\mu\nu}d_{R\beta}\right)\left(D_\mu u_{R\gamma }^TCD_\nu e_{R}\right)$&$Q_{lq^3W}^{(1)}$&$\varepsilon^{\alpha\beta\gamma}\varepsilon^{ik}\varepsilon^{jn}(\tau^I)^m_n\left(Q_{\alpha j}^TC\sigma^{\mu\nu}Q_{\beta k}\right)\left(Q^T_{\gamma i}C\ell_{ m}\right)W_{\mu\nu}^I$\\
$Q_{lq^3W}^{(2)}$&$\varepsilon^{\alpha\beta\gamma}\varepsilon^{km}\varepsilon^{jn}(\tau^I)^i_n\left(Q_{\alpha j}^TC\sigma^{\mu\nu}Q_{\beta k}\right)\left(Q^T_{\gamma i}C\ell_{ m}\right)W_{\mu\nu}^I$&$Q_{lq^3W}^{(3)}$&$\varepsilon^{\alpha\beta\gamma}\varepsilon^{ik}\varepsilon^{jn}(\tau^I)^m_n\left(Q_{\alpha j}^TCQ_{\beta k}\right)\left(Q^T_{\gamma i}C\sigma^{\mu\nu}\ell_{ m}\right)W_{\mu\nu}^I$\\
$Q_{lq^3W}^{(4)}$&$\varepsilon^{\alpha\beta\gamma}\varepsilon^{km}\varepsilon^{jn}(\tau^I)^i_n\left(Q_{\alpha j}^TCQ_{\beta k}\right)\left(Q^T_{\gamma i}C\sigma^{\mu\nu}\ell_{ m}\right)W_{\mu\nu}^I$&$Q_{lq^3B}^{(1)}$&$\varepsilon^{\alpha\beta\gamma}\varepsilon^{ik}\varepsilon^{jm}\left(Q_{\alpha j}^TC\sigma^{\mu\nu}Q_{\beta k}\right)\left(Q^T_{\gamma i}C\ell_{ m}\right)B_{\mu\nu}$\\
$Q_{lq^3B}^{(2)}$&$\varepsilon^{\alpha\beta\gamma}\varepsilon^{ik}\varepsilon^{jm}\left(Q_{\alpha j}^TCQ_{\beta k}\right)\left(Q^T_{\gamma i}C\sigma^{\mu\nu}\ell_{ m}\right)B_{\mu\nu}$&$Q_{lqudW}^{(1)}$&$\varepsilon^{\alpha\beta\gamma}\varepsilon^{jk}(\tau)^i_k\left(u_{R\alpha}^TC\sigma^{\mu\nu}d_{R\beta}\right)\left(Q^T_{\gamma i}C\ell_{ j}\right)W_{\mu\nu}^I$\\
$Q_{lqudW}^{(2)}$&$\varepsilon^{\alpha\beta\gamma}\varepsilon^{jk}(\tau)^i_k\left(u_{R\alpha}^TCd_{R\beta}\right)\left(Q^T_{\gamma i}C\sigma^{\mu\nu}\ell_{ j}\right)W_{\mu\nu}^I$&$Q_{lqudB}^{(1)}$&$\varepsilon^{\alpha\beta\gamma}\varepsilon^{ij}\left(u_{R\alpha}^TC\sigma^{\mu\nu}d_{R\beta}\right)\left(Q^T_{\gamma i}C\ell_{ j}\right)B_{\mu\nu}$\\
$Q_{lqudB}^{(2)}$&$\varepsilon^{\alpha\beta\gamma}\varepsilon^{ij}\left(u_{R\alpha}^TCd_{R\beta}\right)\left(Q^T_{\gamma i}C\sigma^{\mu\nu}\ell_{ j}\right)B_{\mu\nu}$&$Q_{eq^2uW}^{(1)}$&$\varepsilon^{\alpha\beta\gamma}\varepsilon^{jk}(\tau^I)^i_k\left(Q^T_{\alpha i}C\sigma^{\mu\nu}Q_{\beta j}\right)\left(u_{R\gamma}^TCe_R\right)W_{\mu\nu}^I$\\
$Q_{eq^2uW}^{(2)}$&$\varepsilon^{\alpha\beta\gamma}\varepsilon^{jk}(\tau^I)^i_k\left(Q^T_{\alpha i}CQ_{\beta j}\right)\left(u_{R\gamma}^TC\sigma^{\mu\nu}e_R\right)W_{\mu\nu}^I$&$Q_{eq^2uB}^{(1)}$&$\varepsilon^{\alpha\beta\gamma}\varepsilon^{ij}\left(Q^T_{\alpha i}C\sigma^{\mu\nu}Q_{\beta j}\right)\left(u_{R\gamma}^TCe_R\right)B_{\mu\nu}$\\
$Q_{eq^2uB}^{(2)}$&$\varepsilon^{\alpha\beta\gamma}\varepsilon^{ij}\left(Q^T_{\alpha i}CQ_{\beta j}\right)\left(u_{R\gamma}^TC\sigma^{\mu\nu}e_R\right)B_{\mu\nu}$&$Q_{eu^2dB}^{(1)}$&$\varepsilon^{\alpha\beta\gamma}\left(u^T_{R\alpha}C\sigma^{\mu\nu}d_{R\beta}\right)\left(u_{R\gamma}^TCe_R\right)B_{\mu\nu}$\\
$Q_{eu^2dB}^{(2)}$&$\varepsilon^{\alpha\beta\gamma}\left(u^T_{R\alpha}Cd_{R\beta}\right)\left(u_{R\gamma}^TC\sigma^{\mu\nu}e_R\right)B_{\mu\nu}$&&\\
\hline
\multicolumn{4}{|c|}{Dimension-9}\\
\hline
$\mathcal{Q}_{q^3\bar eH^{\dagger 3}}^{(1)}$
&$\varepsilon ^{\alpha\beta\gamma} \left(\bar e_RQ_{\alpha i}\right) \left(Q_{\beta j}^TCQ_{\gamma k}\right) H^{\dagger}{}^{i}H^{\dagger}{}^{j}H^{\dagger}{}^{k}$
&$\mathcal{Q}_{q^3\bar lH^{\dagger2}D}^{(1)}$
&$\varepsilon^{\alpha\beta\gamma}\left(\bar \ell^k \gamma^\mu Q_{\alpha i}\right) \left(Q_{\beta j}^TCD_\mu Q_{\gamma k}\right) H^{\dagger}{}^{i}H^{\dagger}{}^{j}$\\
$\mathcal{Q}_{q^3\bar lH^{\dagger2}D}^{(2)}$&$\varepsilon^{\alpha\beta\gamma}\left(\bar \ell^i \gamma^\mu Q_{\alpha i}\right) \left(Q_{\beta j}^TCD_\mu Q_{\gamma k}\right) H^{\dagger}{}^{j}H^{\dagger}{}^{k}$&$\mathcal{Q}_{q^3\bar lH^{\dagger2}D}^{(3)}$&$\varepsilon^{\alpha\beta\gamma}\left(\bar \ell^j \gamma^\mu Q_{\alpha i}\right) \left(Q_{\beta j}^TCD_\mu Q_{\gamma k}\right) H^{\dagger}{}^{i}H^{\dagger}{}^{k}$\\
$\mathcal{Q}_{q^2d\bar eH^{\dagger2}D}$
&$\varepsilon^{\alpha\beta\gamma}\left(\bar e_R \gamma^\mu d_{R\alpha }\right) \left(Q_{\beta i}^TCD_\mu Q_{\gamma j}\right) H^{\dagger}{}^{i}H^{\dagger}{}^{j}$
&$\mathcal{Q}_{q^2d\bar lH^\dagger D^2}^{(1)}$
&$\varepsilon^{\alpha\beta\gamma}\left(D^\mu\bar\ell^i D_\mu d_{R\alpha} \right) \left(Q_{\beta j}^TC Q_{\gamma i}\right) H^{\dagger}{}^{j}$\\
$\mathcal{Q}_{q^2d\bar lH^\dagger D^2}^{(2)}$
&$\varepsilon^{\alpha\beta\gamma}\left(D_\mu\bar\ell^i D_\nu d_{R\alpha} \right) \left(Q_{\beta j}^TC\sigma^{\mu\nu} Q_{\gamma i}\right) H^{\dagger}{}^{j}$
&$\mathcal{Q}_{ud^2\bar lH^\dagger D^2}^{(1)}$
&$\varepsilon^{\alpha\beta\gamma}\varepsilon_{ij}\left(D^\mu\bar\ell^i D_\mu u_{R\alpha} \right) \left(d_{R\beta}^TC d_{R\gamma }\right) H^{\dagger}{}^{j}$\\
$\mathcal{Q}_{ud^2\bar lH^\dagger D^2}^{(2)}$&$\varepsilon^{\alpha\beta\gamma}\varepsilon_{ij}\left(D_\mu\bar\ell^i D_\nu u_{R\alpha} \right) \left(d_{R\beta}^TC\sigma^{\mu\nu} d_{R\gamma }\right) H^{\dagger}{}^{j}$
&$\mathcal{Q}_{d^3\bar lHD^2}^{(1)}$
&$\varepsilon^{\alpha\beta\gamma}\left(D^\mu\bar\ell^i D_\mu d_{R\alpha} \right) \left(d_{R\beta}^TC d_{R\gamma }\right) H_{i}$\\
$\mathcal{Q}_{d^3\bar lHD^2}^{(2)}$
&$\varepsilon^{\alpha\beta\gamma}\left(D_\mu\bar\ell^i D_\nu d_{R\alpha} \right) \left(d_{R\beta}^TC\sigma^{\mu\nu} d_{R\gamma }\right) H_{i}$
&$\mathcal{Q}_{qd^2\bar eH^\dagger D^2}^{(1)}$
&$\varepsilon^{\alpha\beta\gamma}\left(D^\mu\bar e_R D_\mu Q_{\alpha i} \right) \left(d_{R\beta}^TC d_{R\gamma }\right) H^{\dagger i}$\\
$\mathcal{Q}_{qd^2\bar eH^\dagger D^2}^{(2)}$
&$\varepsilon^{\alpha\beta\gamma}\left(D^\mu\bar e_R D_\nu Q_{\alpha i} \right) \left(d_{R\beta}^TC\sigma^{\mu\nu} d_{R\gamma }\right) H^{\dagger i}$
&$\mathcal{Q}_{Bq^2d\bar lH^\dagger}^{(1)}$
&$\varepsilon^{\alpha\beta\gamma}\left(\bar\ell^jd_{R\alpha} \right) \left(Q_{\beta i}^TC\sigma^{\mu\nu} Q_{\gamma j}\right)B_{\mu\nu}H^{\dagger i}$\\
$\mathcal{Q}_{Bq^2d\bar lH^\dagger}^{(2)}$
&$\varepsilon^{\alpha\beta\gamma}\left(\bar\ell^j\sigma^{\mu\nu}d_{R\alpha} \right) \left(Q_{\beta i}^TC Q_{\gamma j}\right)B_{\mu\nu}H^{\dagger i}$
&$\mathcal{Q}_{Wq^2d\bar lH^\dagger}^{(1)}$
&$\varepsilon^{\alpha\beta\gamma}(\tau^I)^j_k\left(\bar\ell^id_{R\alpha} \right) \left(Q_{\beta i}^TC\sigma^{\mu\nu} Q_{\gamma j}\right)W_{\mu\nu}^IH^{\dagger k}$\\
$\mathcal{Q}_{Wq^2d\bar lH^\dagger}^{(2)}$
&$\varepsilon^{\alpha\beta\gamma}(\tau^I)^j_m\left(\bar\ell^md_{R\alpha} \right) \left(Q_{\beta i}^TC\sigma^{\mu\nu} Q_{\gamma j}\right)W_{\mu\nu}^IH^{\dagger i}$
&$\mathcal{Q}_{Wq^2d\bar lH^\dagger}^{(3)}$
&$\varepsilon^{\alpha\beta\gamma}(\tau^I)^m_j\left(\bar\ell^i\sigma^{\mu\nu}d_{R\alpha} \right) \left(Q_{\beta m}^TC Q_{\gamma i}\right)W_{\mu\nu}^IH^{\dagger j}$\\
$\mathcal{Q}_{Wq^2d\bar lH^\dagger}^{(4)}$
&$\varepsilon^{\alpha\beta\gamma}(\tau^I)^m_i\left(\bar\ell^i\sigma^{\mu\nu}d_{R\alpha} \right) \left(Q_{\beta m}^TC Q_{\gamma j}\right)W_{\mu\nu}^IH^{\dagger j}$
&$\mathcal{Q}_{Bd^3\bar lH}$
&$\varepsilon^{\alpha\beta\gamma}\left(\bar\ell^id_{R\alpha} \right) \left(d_{R\beta}^TC\sigma^{\mu\nu} d_{R\gamma }\right)B_{\mu\nu}H_{i}$\\
$\mathcal{Q}_{Wd^3\bar lH}$
&$\varepsilon^{\alpha\beta\gamma}(\tau)^j_i\left(\bar\ell^id_{R\alpha} \right) \left(d_{R\beta}^TC\sigma^{\mu\nu} d_{R\gamma }\right)W_{\mu\nu}^IH_{j}$
&$\mathcal{Q}_{Bqd^2\bar eH^\dagger}^{(1)}$
&$\varepsilon^{\alpha\beta\gamma}\left(\bar e_RQ_{\alpha i} \right) \left(d_{R\beta}^TC\sigma^{\mu\nu} d_{R\gamma }\right)B_{\mu\nu}H^{\dagger i}$\\
$\mathcal{Q}_{Bqd^2\bar eH^\dagger}^{(2)}$
&$\varepsilon^{\alpha\beta\gamma}\left(\bar e_R\sigma^{\mu\nu}Q_{\alpha i} \right) \left(d_{R\beta}^TC d_{R\gamma }\right)B_{\mu\nu}H^{\dagger i}$
&$\mathcal{Q}_{Wqd^2\bar eH^\dagger}^{(1)}$
&$\varepsilon^{\alpha\beta\gamma}(\tau^I)^i_j\left(\bar e_RQ_{\alpha i} \right) \left(d_{R\beta}^TC\sigma^{\mu\nu} d_{R\gamma }\right)B_{\mu\nu}H^{\dagger j}$\\
$\mathcal{Q}_{Wqd^2\bar eH^\dagger}^{(2)}$
&$\varepsilon^{\alpha\beta\gamma}(\tau^I)^i_j\left(\bar e_R\sigma^{\mu\nu}Q_{\alpha i} \right) \left(d_{R\beta}^TC d_{R\gamma }\right)W_{\mu\nu}^IH^{\dagger j}$
&
&\\
\hline
\end{tabular}}
\caption{Relevant SMEFT operators for BNV processes.}
\label{tab:SMEFT}
\end{table}

\subsection{UV Completions}

The BSM UV models contribute to the SMEFT operators and have properties of the $SU(3)_C\times SU(2)_L\times U(1)_Y$ groups. We derive the complete tree-level UV models and give the minimal Lagrangian of these models for BNV processes. Here, we we select the UV models which can not contribute to the lower dimension operators at higher dimensional level. For example, if one UV particle contribute to both of the dimension 6 and dimension 7 SMEFT operators, its contribution of dimension 7 operators will not be considered which is suppressed with dimension 6 operators. We use the notation of Ref.~\cite{Li:2023cwy,Li:2023pfw} and the $(S\,,~F\,,~V\,,~T)$ correspond to the scalar particles, spin 1/2 fermion particles, vector particles and spin 2 tensor particles. In addition, the massive spin 2 particles are considered as  Fierz-Pauli~\cite{Fierz:1939ix} fields. Furthermore, the representation of the SM group is given as $(SU(3)_C,SU(2)_L)_{U(1)_Y}$ and the UV completions are listed in Tab.~\ref{tab:UV}.

\begin{table}[H]
    \centering
    \resizebox{0.8\textwidth}{!}{
    \begin{tabular}{c|c}
    \hline
    \hline
    \multicolumn{2}{c}{Dimension 6}\\
    \hline
        $\mathcal{O}_{duq}$ & $S_{10}(\mathbf{3},\mathbf{1})_{-\frac{1}{3}}\,,\quad V_7(\mathbf{3},\mathbf{2})_{-\frac{5}{6}}\,,\quad V_8(\mathbf{3},\mathbf{2})_{-\frac{1}{6}}$ \\
        \hline
        $\mathcal{O}_{qqu}$ &$S_{10}(\mathbf{3},\mathbf{1})_{-\frac{1}{3}}\,,\quad V_8(\mathbf{3},\mathbf{2})_{-\frac{1}{6}}$\\ 
        \hline
        $\mathcal{O}_{duu}$ &$S_{9}(\mathbf{3},\mathbf{1})_{-\frac{4}{3}}\,,\quad S_{10}(\mathbf{3},\mathbf{1})_{-\frac{1}{3}}$\\ 
        \hline
        $\mathcal{O}_{qqq}$ &$S_{10}(\mathbf{3},\mathbf{1})_{-\frac{1}{3}}\,,\quad S_{14}(\mathbf{3},\mathbf{3})_{-\frac{1}{3}}\,,$\\ 
        \hline
        \multicolumn{2}{c}{Dimension 7}\\
        \hline
        $\mathcal{O}_{LdddH}$ &$\{S_{11}(\mathbf{3},\mathbf{1})_{-\frac{4}{3}}\,,\,S_{12}(\mathbf{3},\mathbf{2})_{\frac{1}{6}}\}\,,\{S_{11}(\mathbf{3},\mathbf{1})_{-\frac{4}{3}}\,,\,F_{2}(\mathbf{1},\mathbf{1})_{1}\}\,,\{S_{11}(\mathbf{3},\mathbf{1})_{-\frac{4}{3}}\,,\,F_{11}(\mathbf{3},\mathbf{2})_{\frac{1}{6}}\}$\\ 
        \hline
        \multirow{2}{*}{$\mathcal{O}_{LdudH}$}&$\{S_{11}(\mathbf{3},\mathbf{1})_{-\frac{4}{3}}\,,\,S_{13}(\mathbf{3},\mathbf{2})_{\frac{7}{6}}\}\,,\quad\{S_{11}(\mathbf{3},\mathbf{1})_{-\frac{4}{3}}\,,\,F_{1}(\mathbf{1},\mathbf{1})_{0}\}$\\ 
        &$\{S_{11}(\mathbf{3},\mathbf{1})_{-\frac{4}{3}}\,,\,F_{11}(\mathbf{3},\mathbf{2})_{\frac{1}{6}}\}\,,\quad\{S_{13}(\mathbf{3},\mathbf{1})_{\frac{7}{6}}\,,\,F_{10}(\mathbf{3},\mathbf{2})_{-\frac{5}{6}}\}$\\
        \hline
        \multirow{2}{*}{$\mathcal{O}_{eQddH}$} &$\{S_{11}(\mathbf{3},\mathbf{1})_{-\frac{4}{3}}\,,\,S_{13}(\mathbf{3},\mathbf{2})_{\frac{7}{6}}\}\,,\quad\{S_{11}(\mathbf{3},\mathbf{1})_{-\frac{4}{3}}\,,\,F_{3}(\mathbf{1},\mathbf{2})_{\frac{1}{2}}\}$\\ 
        &$\{S_{11}(\mathbf{3},\mathbf{1})_{-\frac{4}{3}}\,,\,F_{8}(\mathbf{3},\mathbf{1})_{-\frac{1}{3}}\}\,,\quad\{S_{13}(\mathbf{3},\mathbf{2})_{\frac{7}{6}}\,,\,F_{10}(\mathbf{3},\mathbf{2})_{-\frac{5}{6}}\}$\\
        \hline
        \multicolumn{2}{c}{Dimension 8}\\
        \hline
        $\mathcal{O}_{eq^2uD^2}$ &$T_{9}(\mathbf{3},\mathbf{2})_{-\frac{5}{6}}$\\ 
        \hline
        $\mathcal{O}_{lqudD^2}^{(1)}$ &$T_{9}(\mathbf{3},\mathbf{2})_{-\frac{5}{6}}\,,\quad T_{10}(\mathbf{3},\mathbf{2})_{\frac{1}{6}}$\\ 
        \hline
        $\mathcal{O}_{lqudD^2}^{(2)}$ &$T_{9}(\mathbf{3},\mathbf{2})_{-\frac{5}{6}}\,,\quad T_{10}(\mathbf{3},\mathbf{2})_{\frac{1}{6}}$\\ 
        \hline
        \hline
    \end{tabular}}
    \caption{The UV completions for SMEFT  operators that contribute to BNV processes.}
    \label{tab:UV}
\end{table}

Moreover, we also obtain the corresponding gauge invariant UV Lagrangian for these particles. The kinematic terms of the particles are given by
\begin{align}
    \mathcal{L}^{\rm UV}_{\rm kin}=&\bigg\{-S_{9}^{\dagger \alpha}(D^2+M_{S_{9}}^2)S_{9\alpha}-S_{10}^{\dagger \alpha}(D^2+M_{S_{10}}^2)S_{10 a}-S_{11}^{\dagger \alpha }(D^2+M_{S_{11}}^2)S_{11 \alpha}\notag\\
    &\quad-S_{12}^{\dagger \alpha i}(D^2+M_{S_{12}}^2)S_{12 \alpha i}-S_{13}^{\dagger \alpha i}(D^2+M_{S_{13}}^2)S_{13 \alpha i}-S_{14}^{\dagger \alpha I}(D^2+M_{S_{14}}^2)S_{14 \alpha}^I\notag\\
    &\quad+\bar F_1(i{\slashed D}-M_{F_1})F_{1}+\bar F_2(i{\slashed D}-M_{F_2})F_{2}+\bar F_{3}^{i}(i{\slashed D}-M_{F_{3}})F_{3i}\notag\\
    &\quad+\bar F_8^\alpha(i{\slashed D}-M_{F_8})F_{8\alpha}+\bar F_{10}^{\alpha i}(i{\slashed D}-M_{F_{10}})F_{10\alpha i}+\bar F_{11}^{\alpha i}(i{\slashed D}-M_{F_{11}})F_{11\alpha i}\notag\\
    &\quad+V_7^{\dagger\mu \alpha i}(g_{\mu\nu}D^2-D_\nu D_\mu+g_{\mu\nu} M_{V_7}^2)V_{7\alpha i}^{\nu}+V_8^{\dagger\mu \alpha i}(g_{\mu\nu}D^2-D_\nu D_\mu+g_{\mu\nu} M_{V_8^2})V_{8\alpha i}^\nu\notag\\
    &\quad-\frac{1}{2}\left\{g^{\mu\lambda}g^{\nu\rho}+g^{\nu\lambda}g^{\mu\rho}-\frac{2}{3}g^{\mu\nu}g^{\rho\lambda}\right\}^{-1}T_{9\alpha}^{\mu\nu}(\square-M_{T_9})T_{9\alpha}^{\rho\lambda}\notag\\
    &\quad-\frac{1}{2}\left\{g^{\mu\lambda}g^{\nu\rho}+g^{\nu\lambda}g^{\mu\rho}-\frac{2}{3}g^{\mu\nu}g^{\rho\lambda}\right\}^{-1}T_{10\alpha}^{\mu\nu}(\square-M_{T_{10}})T_{10\alpha}^{\rho\lambda}\bigg\}\,,
\end{align}
and the corresponding interactions of the UV Lagrangian are
\begin{align}
    \mathcal{L}^{\rm UV}_{\rm int}=&\bigg\{ \mathcal{D}_{{d^\dagger e^\dagger S_{9}}}(\bar{d}_{R}^{{\alpha}}{C}\bar{e}_R^T){S}_{{9\alpha}}+\mathcal{D}_{{uuS_{9}}}\varepsilon ^{{\alpha\beta\gamma}}({u}_{{Rb}}^T{C}{u}_{{Rc}}){S}_{{9\alpha}}+\mathcal{D}_{{du S_{10}}}\varepsilon ^{{\alpha\beta\gamma}}({d}_{{R\beta}}^T{C}{u}_{{R\gamma}}){S}_{{10a}}\notag\\
    &\quad\quad+\mathcal{D}_{{e^\dagger u^\dagger S_{10}}}(\bar{e}_{{R}}{C}\bar{u}_{R}^{{\alpha}T}){S}_{{10\alpha}}+\mathcal{D}_{{L^\dagger Q^\dagger S_{10} }}\varepsilon _{{ij}}  (\bar{\ell}^{{i}}{C}\bar{Q}^{{\alpha j}T}){S}_{{10\alpha}} + \mathcal{D}_{{QQS_{10}}}\varepsilon ^{{\alpha\beta\gamma}} \varepsilon ^{{ij}}(Q_{{\beta i}}^T{C}Q_{{\gamma j}}){S}_{{10\alpha}}\notag\\
    &\quad\quad+\mathcal{D}_{{L^\dagger Q^\dagger S_{14}}}\varepsilon_{{kj}} (\tau ^{{I}})_{{i}}^{{k}}(\bar\ell^{{i}}{C}\bar Q^{{\alpha j}T}){S}_{{14\alpha}}^{{I}} +{\mathcal{D}_{{QQS_{14}}}}\varepsilon ^{{\alpha\beta\gamma}} \varepsilon ^{{jk}} (\tau ^{{I}})_{{k}}^{{i}}(Q_{{\beta i}}^T{C}Q_{{\gamma j}}){S}_{{14\alpha}}^{{I}}\notag\\
    &\quad\quad+{\mathcal{D}_{{QQS_{14}}}}\varepsilon ^{{\alpha\beta\gamma}} \varepsilon ^{{ik}} (\tau ^{{I}})_{{k}}^{{j}}(Q_{{\beta i}}^T{C}Q_{{\gamma j}}){S}_{{14a}}^{{I}}+  \mathcal{D}_{{d^\dagger L^\dagger V_{7}}}(\bar{d}_{{R}}^{{\alpha}}\gamma _{\mu }{C}\bar{\ell}^{{i}T}){V}_{{7\alpha i}}^{\mu }  +  \mathcal{D}_{{Q^\dagger e^\dagger V_{7}}} (\bar Q^{{\alpha i}}\gamma _{\mu }C\bar{e}_R^T){V}_{{7\alpha i}}^{{\mu }}\notag\\
    &\quad\quad+ \mathcal{D}_{{QuV_{7}}}\varepsilon ^{{\alpha\beta\gamma}} \varepsilon ^{{ij}}(Q_{{\beta j}}^T{C}\gamma _{\mu }{u}_{R\gamma}){V}_{{7\alpha i}}^{\mu }+ \mathcal{D}_{{QdV_{8}}}\varepsilon ^{{\alpha\beta\gamma}} \varepsilon ^{{ij}}(Q_{{\beta j}}^T{C}\gamma _{\mu }{d}_{{R\gamma}}){V}_{{8\alpha i}}^{\mu }+ \mathcal{D}_{{L^\dagger u^\dagger V_{8}}} (\bar{u}_{R}^{{\alpha}}\gamma _{\mu }{C}\bar{\ell}^{{i}T}){V}_{{8\alpha i}}^{\mu }\notag\\
    &\quad\quad+ \mathcal{D}_{{d d S_{11} }}\epsilon ^{{\alpha\beta\gamma}}(d_{R\beta}^T{C}d_{{R\gamma}}){S}_{{11\alpha}} 
    + \mathcal{D}_{{d^\dagger LS_{12}}}\epsilon ^{{ji}} (\bar{d}_{{R}}^{{\alpha}}{\ell}_{{j}}){S}_{{12\alpha i}}+ \mathcal{D}_{{ Q^\dagger eS_{13} }} (\bar{Q}^{\alpha i}e_R){S}_{{13\alpha i}}\notag\\
    &\quad\quad+ \mathcal{D}_{{u^\dagger LS_{13}}}\epsilon ^{{ji}} (\bar{u}_{R}^\alpha{\ell}_j){S}_{{13\alpha i}}+ \mathcal{D}_{{F_{1}LH}}\epsilon ^{{ji}}({F}_{{1}}^T{C}{\ell}_{{i}}){H}_{{j}} 
    + \mathcal{D}_{{F_{2} LH^\dagger}} ({F}_{{2}}^T{C}{\ell}_{{i}}){H}^{{\dagger i}}\notag\\
    &\quad\quad+ \mathcal{D}_{{ F_{3}e H }}\epsilon^{{ij}}(F_{3i}^TCe_R){H}_j+  \mathcal{D}_{{Q^\dagger F_8H}} (\bar Q^{\alpha i}F_{{8\alpha}}){H}_{{i}} 
    + \mathcal{D}_{{d^\dagger F_{10L}H}}\epsilon ^{{ij}}  (\bar{d}_R^{{\alpha}}{F}_{{10\alpha i}}){H}_{{j}}\notag\\
    &\quad\quad+ \mathcal{D}_{{d^\dagger F_{11}H^\dagger }}  (\bar{d}_{{R}}^{{\alpha}}{F}_{{11\alpha i}}){H}^{{\dagger i}} 
    + \mathcal{D}_{{u^\dagger F_{11L}H}}\epsilon ^{{ij}}  (\bar{u}_{{R}}^{{\alpha}}{F}_{{11\alpha i}}){H}_{{j}}+\rm h.c.\bigg\}\notag\\
    &+\bigg\{ \mathcal{C}_{{HS_{11}^\dagger S_{12}}}\epsilon ^{{ji}} {H}_{{j}}{S}_{{11}}^{\dagger \alpha}{S}_{{12\alpha i}} +  \mathcal{C}_{{HS_{11}S_{13}^\dagger }} {H}_{{i}}{S}_{{11\alpha}}{S}_{{13}}^{\dagger \alpha i}+ \mathcal{D}_{{u^\dagger F_{1}S_{11} }}(\bar{u}_{{R}}^{{\alpha}}{F}_{{1}}){S}_{{11\alpha}}\notag\\
    &\quad\quad+ \mathcal{D}_{{d^\dagger F_{2}^\dagger S_{11}}}(\bar{d}_{{R}}^{{\alpha}}C\bar{F}_{{2}}^T){S}_{{11\alpha}} + \mathcal{D}_{{F_{3}QS_{11}^\dagger }}\epsilon ^{{ij}} ({F}_{{3i}}^T{C}{Q}_{{\alpha j}}){S}_{{11}}^{\dagger \alpha}+ \mathcal{D}_{{e^\dagger F_{8L}S_{11}^\dagger }}(\bar{e}_{{R}}{F}_{{8\alpha}}){S}_{{11}}^{\dagger \alpha}\notag\\
    &\quad\quad+  \mathcal{D}_{{F_{11}^\dagger LS_{11}}}(\bar{F}_{{11}}^{{\alpha i}}{\ell}_{{i}}){S}_{{11a}}+ \mathcal{D}_{{d F_{10R} S_{13} }}\epsilon^{{ij}} \epsilon ^{{\alpha\beta\gamma}}({d}_{{Rc}}^TCF_{{10\alpha i}}){S}_{{13bj}}+\rm h.c.\bigg\}\notag\\
    &+\bigg\{g_{T_9qu}\varepsilon^{\alpha\beta\gamma}\epsilon^{ij}( {Q^{T}}_{\alpha i}C\gamma_\mu\overleftrightarrow{\partial}_\nu u_{R\beta})T_{9\gamma j}^{\mu\nu}+g_{T_9qe}( {Q^{T}}_{\alpha i}C\gamma_\mu\overleftrightarrow{\partial}_\nu e_{R})T_{9}^{\mu\nu\dagger \alpha i}\notag\\
    &\quad\quad+g_{T_{10}qd}\varepsilon^{\alpha\beta\gamma}\epsilon^{ij}( {Q^{T}}_{\alpha i}C\gamma_\mu\overleftrightarrow{\partial}_\nu d_{R\beta})T_{10\gamma j}^{\mu\nu}+g_{T_{10}\ell u}( {\ell^{T}}_{ i}C\gamma_\mu\overleftrightarrow{\partial}_\nu u_{R\alpha})T_{10}^{\mu\nu\dagger \alpha i}+\rm h.c.\bigg\}\,,
\end{align}
where the coupling constants \(\mathcal{C}\) have mass dimension, the coupling constants \(\mathcal{D}\) are dimensionless, and the coupling constants \(g_{T...}\) have dimension of inverse mass (i.e., \([\text{mass}]^{-1}\)).

\begin{figure}[H]
    \centering
    \resizebox{0.7\textwidth}{!}{
    \begin{minipage}[t]{0.4\textwidth}
        \centering
        \includegraphics[width=\linewidth]{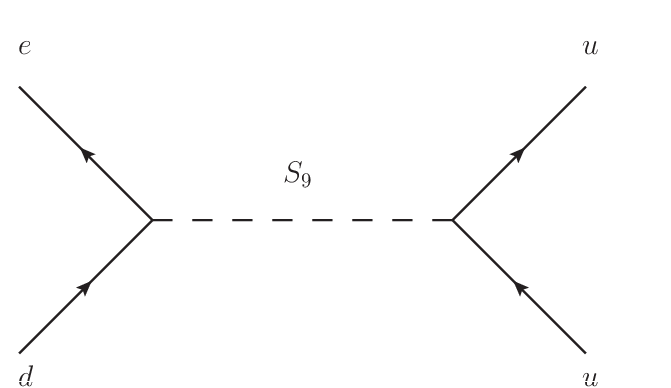}
    \end{minipage}
    \begin{minipage}[t]{0.48\textwidth}
        \centering
        \includegraphics[width=\linewidth]{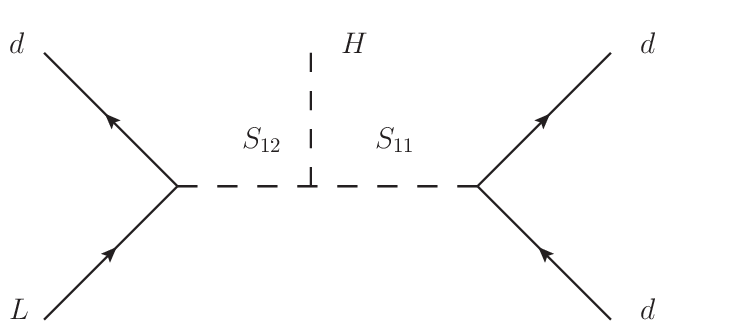}
    \end{minipage}}
    \centering
    \caption{Contributions to $\mathcal{O}_{duu}$ and $\mathcal{O}_{LdddH}$ from the UV completions with $S_{9}$ and $(S_{11},S_{12})$ particles.}
    \label{Dimensionex}
\end{figure}

The matching between the UV models and SMEFT operators is straightforward. We take the UV particles $S_9$ and $(S_{11},S_{12})$ as the example. The diagrams of the matching for $S_9$ and $(S_{11},S_{12})$ are shown in Fig.\ref{Dimensionex}. After integrating out the UV particles, the matching results for the SMEFT operators become
\begin{align}
 {Q_{duu}} =
\frac{ \mathcal{D}_{{d^\dagger e^\dagger S_{9}}} \mathcal{D}_{{ uu S_9 }}^*}{{M_{S_{9}}^2}}\,,\quad\mathcal{O}_{LdddH}=&\frac{ \mathcal{D}^*_{{d dS_{11} }} \mathcal{D}_{{d^\dagger LS_{12}}} \mathcal{C}_{{HS_{11}^\dagger S_{12}}}}{{M_{S_{11}}^2} {M_{S_{12}}^2}}\,.
\end{align}

\subsection{Grand Unified Theory Scenarios}

The idea behind grand unification is that, at some higher energy, the gauge group factors of $\mathcal{G}_{\rm SM}=SU(3)_C\times SU(2)_L\times U(1)_Y$ unify and we have only one gauge group $\mathcal{G}_{\rm GUT}$. Since our observations are mostly in agreement with a model based on the gauge group $\mathcal{G}_{\rm SM}$, we require that $\mathcal{G}_{\rm GUT}$ has $\mathcal{G}_{\rm SM}$ as a subgroup. Similarly to the Higgs mechanism, we assume that $\mathcal{G}_{\rm GUT}$ spontaneously breaks to $\mathcal{G}_{\rm SM}$ at some high energy scale.

There are numerous GUT models such as $SU(5)$~\cite{Georgi:1974sy}, Pati-Salam~\cite{Pati:1973uk,Pati:1974yy} and $SO(10)$~\cite{Fritzsch:1974nn} models. Since the original $SU(5)$ model has been excluded, we consider the $SU(5)$ flipped~\cite{Barr:1981qv,Derendinger:1983aj} models. In addition, the $SO(10)$ group is the larger group which contain the $SU(5)$ flipped $\mathcal{G}_{51}=SU(5)_C\times U(1)_X$ and Pati-Salam $\mathcal{G}_{\rm PS}=SU(4)_C\times SU(2)_L\times SU(2)_R$ group. Here, we give the breaking chains of $SO(10)$ for the UV completions through flipped $SU(5)$ method in Tab.~\ref{tab:flipped} and Pati-Salam method in Tab.~\ref{tab:PS}.

\begin{table}[H]
    \centering
    \begin{tabular}{|c|c|c|c|c|c|c|c|c|c|}
    \hline
        $SO(10)$ &$\rightarrow$& $\mathcal{G}_{51}$&$\rightarrow$&$\mathcal{G}_{\rm SM}$ &$SO(10)$ &$\rightarrow$& $\mathcal{G}_{51}$&$\rightarrow$&$\mathcal{G}_{\rm SM}$ \\
        \hline
        $\mathbf{45}_V$ &$\rightarrow$ &$\mathbf{24}_0$&$\rightarrow$&$V_7(\mathbf{3},\mathbf{2})_{-\frac{5}{6}}$&$\mathbf{45}_F$ &$\rightarrow$ &$\mathbf{24}_0$&$\rightarrow$&$F_1(\mathbf{1},\mathbf{1})_0$\\
        $\mathbf{16}_V$ &$\rightarrow$ &$\mathbf{10}_{-1}$&$\rightarrow$&$V_8(\mathbf{3},\mathbf{2})_{\frac{1}{6}}$&${\mathbf{16}_F}$ &$\rightarrow$ &${\overline{10}_{-1}}$&$\rightarrow$&${F_2(\mathbf{1},\mathbf{1})_1}$\\
        $\mathbf{120}_S$ &$\rightarrow$ &$\overline{\mathbf{45}}_{-2}$&$\rightarrow$&$S_9(\mathbf{3},\mathbf{1})_{-\frac{4}{3}}$&${\mathbf{10}_F}$ &$\rightarrow$ &${\mathbf{5}_2}$&$\rightarrow$&${F_3(\mathbf{1},\mathbf{2})_{\frac{1}{2}}}$\\
        $\mathbf{10}_S$ &$\rightarrow$ &$\mathbf{5}_2$&$\rightarrow$&$S_{10}(\mathbf{3},\mathbf{1})_{-\frac{1}{3}}$&${\mathbf{10}_F}$ &$\rightarrow$ &${\mathbf{5}_2}$&$\rightarrow$&${F_8(\mathbf{3},\mathbf{1})_{-\frac{1}{3}}}$\\
        ${\mathbf{16}_S}$ &$\rightarrow$ &${\mathbf{5}_2}$&$\rightarrow$&${S_{11}(\mathbf{3},\mathbf{1})_{-\frac{1}{3}}}$&$\mathbf{45}_F$ &$\rightarrow$ &$\mathbf{24}_0$&$\rightarrow$&$F_{10}(\mathbf{3},\mathbf{2})_{-\frac{5}{6}}$\\
        ${\mathbf{16}_S}$ &$\rightarrow$ &${\mathbf{10}_{-1}}$&$\rightarrow$&${S_{12}(\mathbf{3},\mathbf{2})_\frac{1}{6}}$&${\mathbf{16}_F}$ &$\rightarrow$ &${\mathbf{10}_{-1}}$&$\rightarrow$&$F_{11}(\mathbf{3},\mathbf{2})_{\frac{1}{6}}$\\
        $\overline{\mathbf{126}}_S$ &$\rightarrow$ &$\overline{\mathbf{45}}_{-2}$&$\rightarrow$&${S_{13}(\mathbf{3},\mathbf{2})_\frac{7}{6}}$&$\mathbf{120}_S$ &$\rightarrow$ &$\mathbf{45}_2$&$\rightarrow$&$S_{14}(\mathbf{3},\mathbf{3})_{-\frac{1}{3}}$\\
         \hline
    \end{tabular}
    \caption{The $SO(10)$ breaking chain for UV completions in $SU(5)$ flipped method.}
    \label{tab:flipped}
\end{table}

\begin{table}[H]
    \centering
    \begin{tabular}{|c|c|c|c|c|c|c|c|c|c|}
    \hline
        $SO(10)$ &$\rightarrow$& $\mathcal{G}_{PS}$&$\rightarrow$&$\mathcal{G}_{\rm SM}$&$SO(10)$ &$\rightarrow$& $\mathcal{G}_{PS}$&$\rightarrow$&$\mathcal{G}_{\rm SM}$ \\
        \hline
        $\mathbf{45}_V$ &$\rightarrow$ &$(\mathbf{6},\mathbf{2},\mathbf{2})$&$\rightarrow$&$V_7(\mathbf{3},\mathbf{2})_{-\frac{5}{6}}$&${\mathbf{45}_F}$ &$\rightarrow$ &${(\mathbf{1},\mathbf{1},\mathbf{3})}$&$\rightarrow$&${F_1(\mathbf{1},\mathbf{1})_0}$\\
        $\mathbf{16}_V$ &$\rightarrow$ &$(\mathbf{4},\mathbf{2},\mathbf{1})$&$\rightarrow$&$V_8(\mathbf{3},\mathbf{2})_{\frac{1}{6}}$&${\mathbf{16}_F}$ &$\rightarrow$ &${(\overline{\mathbf{4}},\mathbf{1},\mathbf{2})}$&$\rightarrow$&${F_2(\mathbf{1},\mathbf{1})_1}$\\
        $\mathbf{120}_S$ &$\rightarrow$ &$(\mathbf{6},\mathbf{1},\mathbf{3})$&$\rightarrow$&$S_9(\mathbf{3},\mathbf{1})_{-\frac{4}{3}}$&${\mathbf{10}_F}$ &$\rightarrow$ &${(\mathbf{1},\mathbf{2},\mathbf{2})}$&$\rightarrow$&${F_3(\mathbf{1},\mathbf{2})_{\frac{1}{2}}}$\\
        $\mathbf{10}_S$ &$\rightarrow$ &$(\mathbf{6},\mathbf{1},\mathbf{1})$&$\rightarrow$&$S_{10}(\mathbf{3},\mathbf{1})_{-\frac{1}{3}}$&${\mathbf{10}_F}$ &$\rightarrow$ &${(\mathbf{6},\mathbf{1},\mathbf{1})}$&$\rightarrow$&${F_8(\mathbf{3},\mathbf{1})_{-\frac{1}{3}}}$\\
        ${\mathbf{16}_S}$ &$\rightarrow$ &${(\mathbf{6},\mathbf{1},\mathbf{1})}$&$\rightarrow$&${S_{11}(\mathbf{3},\mathbf{1})_{-\frac{1}{3}}}$&$\mathbf{45}_F$ &$\rightarrow$ &$(\mathbf{6},\mathbf{2},\mathbf{2})$&$\rightarrow$&$F_{10}(\mathbf{3},\mathbf{2})_{-\frac{5}{6}}$\\
        ${\mathbf{16}_S}$ &$\rightarrow$ &${(\mathbf{4},\mathbf{2},\mathbf{1})}$&$\rightarrow$&${S_{12}(\mathbf{3},\mathbf{2})_\frac{1}{6}}$&${\mathbf{16}_F}$&$\rightarrow$&$(\mathbf{4},\mathbf{2},\mathbf{1})$&$\rightarrow$&$F_{11}(\mathbf{3},\mathbf{2})_{\frac{1}{6}}$\\
        $\overline{\mathbf{126}}_S$ &$\rightarrow$ &${(\mathbf{15},\mathbf{2},\mathbf{2})}$&$\rightarrow$&${S_{13}(\mathbf{3},\mathbf{2})_\frac{7}{6}}$&$\mathbf{120}_S$ &$\rightarrow$ &$(\mathbf{6},\mathbf{3},\mathbf{1})$&$\rightarrow$&$S_{14}(\mathbf{3},\mathbf{3})_{-\frac{1}{3}}$\\
         \hline
    \end{tabular}
    \caption{The $SO(10)$ breaking chain for UV completions in Pati-Salam flipped method.}
    \label{tab:PS}
\end{table}

\newpage

\section{Conclusion}
\label{sec:con}
The BNV processes are of profound theoretical and experimental significance, serving as a unique window into NP at scales far beyond the reach of current collider experiments. They are generic predictions of GUTs and a crucial ingredient for explaining the observed baryon asymmetry in the universe. Experimental searches for BNV signals, such as proton decay, have been a central focus of particle physics for decades. With next-generation detectors like DUNE, Hyper-K, and JUNO poised to significantly improve sensitivity, there is an urgent need for robust theoretical frameworks to interpret future experimental results. The EFT approach provides a model-independent tool for parameterizing BNV processes. A rigorous mapping from the quark operators to the chiral operators is essential to translate high-energy operator effects into low-energy hadronic observables.

In this work, we have performed a comprehensive, model-independent analysis of BNV baryon decays within the EFT framework, with the primary focus on the computation of BNV baryon decay rates. A defining feature of our analysis is the detailed and systematic mapping from the LEFT to the chiral Lagrangian, which we have developed explicitly for the complete set of dimension 8 LEFT operators. This mapping includes a rigorous treatment of chiral symmetry breaking, the embedding of baryon fields within the SU(3) chiral flavor group, the renormalization group evolution between the LEFT scale and the chiral scale, and the construction of chiral Lagrangian operators that correctly reproduce the BNV effects encoded in the LEFT operators. Our approach is centered on the systematic incorporation of higher-dimensional operators, with a focus on the complete dimension-8 operator basis in the LEFT and the extension of the SMEFT to LEFT matching up to dimension 9. By explicitly accounting for the UV completion of these operators, we have been able to ensure consistency between our EFT calculations and the underlying high-energy dynamics. Using an automated framework, we efficiently navigate the complex operator space to obtain the decay rates for BNV baryon decay channels from contributions of generic UV models to the leading dimension-6 and dimension-7 BNV operators. This is made possible by our detailed LEFT-to-chiral-Lagrangian mapping, which is critical for reliably translating operator effects into decay rate predictions. Complementing this EFT analysis, we have also presented a discussion of BNV processes in GUT frameworks, connecting our model-independent decay rate predictions to explicit, well-motivated high-scale new physics scenarios.

Our key findings highlight the critical role of higher-dimensional operators in shaping the phenomenology of BNV baryon decay rates, as well as the importance of a rigorous LEFT-to-chiral-Lagrangian mapping for accurate decay rate calculations. We find that the complete set of dimension-8 LEFT operators allows us to include the full chiral representations for $\Delta B=1$ processes, while the conventional dimension 6 operators only yield the $(\bar{\mathbf{3}},\mathbf{3})$ and $(\mathbf{8},\mathbf{1})$ representations in the chiral Lagrangian. Accordingly, we systematically derive the corresponding chiral Lagrangians for the full three quark representations $(\bar{\mathbf{3}},\mathbf{3})$, $(\mathbf{8},\mathbf{1})$, $(\bar{\mathbf{6}},\mathbf{3})$, and $(\mathbf{10},\mathbf{1})$, i.e., the representations realized within the chiral Lagrangian. Furthermore, extending the matching between SMEFT and LEFT up to dimension 9 reveals that the higher-dimensional operators allow us to accommodate both $(B-L)$-conserving and $(B-L)$-violating baryon decay processes. In addition, they enable the inclusion of photon contributions and more general many-body final states, resulting in a much richer set of decay channels and a wider variety of UV completions. We obtain the complete UV completions for dimension 6 and dimension 7 SMEFT operators, while only representative UV completions are considered for higher-dimensional operators.
The physical content of these extended channels can only be reliably extracted through a careful mapping to the chiral Lagrangian that incorporates low-energy QCD dynamics. Within bottom-up EFT framework, we obtain the complete UV completions for dimension 6 and dimension 7 SMEFT operators, while only representative UV completions are considered for higher-dimensional operators
Against this background, our complementary GUT analysis further demonstrates how these reconstructed UV structures can be linked to concrete high-scale models, offering a consistent framework to interpret our results in well-motivated NP scenarios.

The results presented in this work provide a robust theoretical foundation for the study of BNV baryon decays, with the primary outcome being a comprehensive set of decay rate calculations for the considered BNV operators—enabled by our detailed treatment of the LEFT-to-chiral-Lagrangian mapping. This mapping fills a critical gap in existing literature by providing an explicit, systematic connection between higher-dimensional LEFT operators and their chiral Lagrangian, ensuring that BNV effects are correctly translated into low-energy observables. While we do not compare our results directly with experimental data, the framework we have developed can be straightforwardly applied to interpret future experimental searches for BNV processes, aiding in the identification of potential NP signals. Our automated analysis tools, detailed LEFT-to-chiral-Lagrangian mapping, and decay rate calculations are designed to be useful for the broader community, facilitating further studies of BNV in both EFT and GUT frameworks.

Looking ahead, our analysis opens several avenues for future research. The systematic framework developed in this work can be readily extended to other BNV processes, including neutron-antineutron oscillations~\cite{Baldo-Ceolin:1994hzw,Phillips:2014fgb,Buchoff:2015qwa,Bijnens:2017xrz,Rinaldi:2018osy,Haidenbauer:2019fyd}, and applied to other EFT such as axion-like particle EFT\cite{Song:2023jqm,Song:2023lxf}, sterile neutrino EFT\cite{Li:2021tsq,Helo:2018bgb} for BNV processes. Furthermore, the LECs in our chiral Lagrangian can be further evaluated using lattice QCD calculations in future work. This will refine our theoretical predictions and strengthen the connection between our framework and experimental observations, laying a more solid foundation for future phenomenological studies and experimental interpretation of BNV processes.

\section*{Acknowledgments}

We would like to thank Xiao-Dong Ma for helpful discussion. This work is supported by the National Science Foundation of China under Grants No. 12347105, No. 12375099 and No. 12047503, and the National Key Research and Development Program of China Grant No. 2020YFC2201501, No. 2021YFA0718304.

\appendix

\section{Naive Dimension Analysis}
\label{app:nda-matching}

The matching between the LEFT operators and the chiral Lagrangian is not a one-to-one procedure, which means that every LEFT operator can be matched to several $\chi$PT operators, and each $\chi$PT operator receives contributions from several LEFT operators (will be clarified in the subsequent examples). Thus, we need a method to assess all these contributions systematically. In detail, we want to distinguish which one is more dominant among the LEFT operators matching to the same $\chi$PT operator, and which one is more dominant among the $\chi$PT operators matched from the same LEFT operator. In this section, we use the NDA~\cite{Manohar:1983md,Gavela:2016bzc,Jenkins:2013sda} of the effective operators to organize them, an approach consistent with that adopted in Ref.~\cite{Song:2025snz,Li:2026mco}.

The LEFT Lagrangian is organized as an expansion in powers of the electroweak scale inverse $1/\Lambda_{\text{EW}}$. The NDA master formula~\cite{Manohar:1983md,Gavela:2016bzc} in the 4-dimensional spacetime states that the operators in the Lagrangian are normalized as
\begin{equation}
\label{eq:NDA_1}
\frac{\Lambda_{\text{EW}}^4}{16\pi^2} {\left[\frac{\partial}{\Lambda_{\text{EW}}}\right]}^{N_p}{\left[\frac{4\pi\phi}{\Lambda_{\text{EW}}}\right]}^{N_\phi}{\left[\frac{4\pi A}{\Lambda_{\text{EW}}}\right]}^{N_A}{\left[\frac{4\pi\psi}{\Lambda_{\text{EW}}^{3/2}}\right]}^{N_\psi}{\left[\frac{g}{4\pi}\right]}^{N_g}{\left[\frac{y}{4\pi}\right]}^{N_y}{\left[\frac{\lambda}{16\pi^2}\right]}^{N_\lambda}\,.
\end{equation}
For convenience, we express the master formula by replacing the vector field $A$ with the field-strength tensor $F\sim \partial A$ and omitting the renormalizable coupling constants $g,y,\lambda$,
\begin{equation}
\text{LEFT:\quad }
    \frac{\Lambda_{\text{EW}}^4}{16\pi^2}\left[\frac{\partial}{\Lambda_{\text{EW}}}\right]^{N_p} \left[\frac{4\pi F}{\Lambda_{\text{EW}}^2}\right]^{N_F} \left[\frac{4\pi \psi}{\Lambda_{\text{EW}}^{3/2}}\right]^{N_\psi}\,,
\end{equation}
which implies that the 4-fermion operators of dimension-6, 7, and 8 are normalized as
\begin{equation}
    \frac{{(4\pi)}^2}{\Lambda_{\text{EW}}^2}\psi^4\,,\quad\frac{{(4\pi)}^2}{\Lambda_{\text{EW}}^2}\frac{\partial}{\Lambda_{\text{EW}}}\psi^4\,,\quad\frac{{(4\pi)}^2}{\Lambda_{\text{EW}}^2}\frac{\partial^2}{\Lambda_{\text{EW}}^2}\psi^4\,.
\end{equation}
In particular, the spurions are dimensionless in the NDA.
On the other hand, the NDA master formula for the $\chi$PT is 
\begin{equation}
    \text{$\chi$PT:\quad }f^2\Lambda_\chi^2 \left[\frac{\partial}{\Lambda_\chi}\right]^{N_p} \left[\frac{\psi}{f\sqrt{\Lambda_\chi}}\right]^{N_\psi} \left[\frac{F}{\Lambda_\chi f}\right]^{N_A} \,,  \label{eq:nda_chiral}
\end{equation}
where $4\pi f\sim \Lambda_\chi$. To relate these two NDA formulae we first replace all occurrences of the scale $\Lambda_{\text{EW}}$ in Eq.~\eqref{eq:NDA_1} by the scale $\Lambda_\chi$,
\begin{align}
    & \frac{\Lambda_{\text{EW}}^4}{16\pi^2} \left[\frac{\Lambda_\chi}{\Lambda_{\text{EW}}}\right]^{N_p+2N_F + \frac{3}{2}N_\psi} \left[\frac{\partial}{\Lambda_\chi}\right]^{N_p} \left[\frac{F}{\Lambda_\chi^2}\right]^{N_F} \left[\frac{\psi}{\sqrt{\Lambda_\chi}f}\right]^{N_\psi} \notag \\
    =& \left[\frac{\Lambda_\chi}{\Lambda_{\text{EW}}}\right]^{\mathcal{D}} \left(f^2\Lambda_\chi^2 \left[\frac{\partial}{\Lambda_\chi}\right]^{N_p} \left[\frac{F}{\Lambda_\chi^2}\right]^{N_F} \left[\frac{\psi}{\sqrt{\Lambda_\chi}f}\right]^{N_\psi}\right)\,,
\end{align}
where $\mathcal{D}=N_p+2N_F + \frac{3}{2}N_\psi-4$. The expression inside the parentheses is similar to the NDA formula in the $\chi$PT, so we replace it by the $\chi$PT NDA formula in Eq.~\eqref{eq:nda_chiral} and obtain the NDA formula for the matching,
\begin{align}
    \text{matching:\quad }& \frac{\Lambda_{\text{EW}}^4}{16\pi^2}\left[\frac{\partial}{\Lambda_{\text{EW}}}\right]^{N_p} \left[\frac{4\pi F}{\Lambda_{\text{EW}}^2}\right]^{N_F} \left[\frac{4\pi \psi}{\Lambda_{\text{EW}}^{3/2}}\right]^{N_\psi} \notag \\
    \sim & \left[\frac{\Lambda_\chi}{\Lambda_{\text{EW}}}\right]^{\mathcal{D}} \left(f^2\Lambda_\chi^2 \left[\frac{\partial}{\Lambda_\chi}\right]^{N_p} \left[\frac{\psi}{f\sqrt{\Lambda_\chi}}\right]^{N_\psi} \left[\frac{F}{\Lambda_\chi f}\right]^{N_A} \right)\,. \label{eq:NDA_2}
\end{align}

\section{Expansion of Chiral Operators in Hadron Field}
\label{app:chiral-expansion}

Using $u=e^{X}$ in Eq.~\eqref{eq:CCWZ}, with $X\equiv \tfrac{i}{2f}\Pi$ and $\Pi$ defined in Eq.~\eqref{eq:NGBPi}, one has $X^\dagger=-X$ and thus $u^\dagger=e^{-X}$. The chiral building blocks introduced in Eq.~\eqref{eq:buildblock} take the form
\begin{align}
  uu_\mu u^\dagger&=
  i\left(- e^{X}\partial_\mu e^{-X}-e^{2X}\partial_\mu e^{-X}e^{-X}\right)\sim \frac{1}{f}\partial_\mu \Pi\,, \nonumber\\
  u^\dagger u_\mu u&=
  i\left(e^{-2X}\partial_\mu e^{X}e^X+ e^{-X} \partial_\mu e^{X}\right)\sim -\frac{1}{f}\partial_\mu\Pi\,, \nonumber\\
  uu_\mu u&=
  i\left(\partial_\mu e^{X}e^X+e^{2X} \partial_\mu e^X\right)\sim -\frac{1}{f}\partial_\mu\Pi\,, \nonumber\\
  u^\dagger u_\mu u^\dagger&=
  i\left(-e^{-2X}\partial_\mu e^{-X}-\partial_\mu e^{-X}e^{-X}\right)\sim \frac{1}{f}\partial_\mu \Pi\,, \nonumber\\
  uBu &= e^{X}B e^{X} \sim B+\frac{i}{2f}\Pi B+\frac{i}{2f}B\Pi\,, \nonumber\\
  u^\dagger Bu^\dagger &= e^{-X}B e^{-X}\sim B-\frac{i}{2f}\Pi B-\frac{i}{2f}B\Pi\,, \nonumber\\
  uBu^\dagger &= e^{X}B e^{-X}\sim B+\frac{i}{2f}\Pi B-\frac{i}{2f}B\Pi\,, \nonumber\\
  u^\dagger Bu &= e^{-X}B e^{X}\sim B-\frac{i}{2f}\Pi B+\frac{i}{2f}B\Pi\,,
\end{align}
where $B$ is the octet baryon defined in Eq.~\eqref{eq:Bdefine} and this building blocks are $3\times 3$ matrices.

Then the chiral operators can be expanded in hadron field. For example, the LO chiral operator
\begin{align}
{\rm LO}:\quad\alpha[\hat{\mathcal{C}}_{1}^{(6)}]_{abc}\varepsilon_{abd}(\lambda^A)^d_c(T_L)^a(T_L)^b(T_L)^c[(u^\dagger{B_L^TC}u)^Ae_L]\,,
\end{align}
can be matched from the LEFT operator $[\hat{\mathcal{C}}_{1}^{(6)}]_{121}\varepsilon^{\alpha\beta\gamma}(u^T_{L\alpha}Cd_{L\beta})(u_{L\gamma}^TCe_L)$. Thus, the $(abc)$ takes the component $(121)$ and $(\lambda^A)^d_c(u^\dagger{B_L^TC}u)^A=\sqrt{2}(u^\dagger{B_L^TC}u)^d_c$. In this work, only the two-body baryon decay processes are considered and the corresponding expansion becomes  \begin{align}
    &\alpha\varepsilon_{12d}(\lambda^A)^d_1(T_L)^1(T_L)^2(T_L)^1[(u^\dagger{B_L^TC}u)^Ae_L]\notag\\
    =&\alpha\varepsilon_{123}(\lambda^A)^3_1(T_L)^1(T_L)^2(T_L)^1[(u^\dagger{B_L^TC}u)^Ae_L]\notag\notag\\
    =&\alpha(p_L^TCe_L)+\frac{i\alpha}{2f}\bigg({\sqrt{3}} K^+ \Lambda - \sqrt{2} n \pi^+ - {\sqrt{3}} p \eta - p \pi^0+ K^+ \Sigma^0 + \sqrt{2} K^0 \Sigma^+ \bigg)\,.
\end{align}

\section{Operator Matching: From SMEFT to LEFT}
\label{app:MatchingSM2L}

In this appendix, we give the matching between SMEFT and LEFT. The BNV matching becomes directly and there are only two types $|\Delta B-\Delta L|=0$ and $|\Delta B-\Delta L|=2$ in the matching. In the SMEFT operators, the dimension 6 operators are $|\Delta B-\Delta L|=0$, the dimension 7 operators are $|\Delta B-\Delta L|=2$, the dimension 8 operators are $|\Delta B-\Delta L|=0$ and the dimension 9 operators are $|\Delta B-\Delta L|=2$. Here, we only consider the dominate contribution. For example, the dimension 6 $|\Delta B-\Delta L|=0$ LEFT operators only received the contribution of SMEFT dimension 6 operators and the higher-dimensional operators are suppressed, thus are neglected. Moreover, both of the flavor structures for SMEFT and LEFT operators are simplified, we only consider the direct contribution without the flavor mixing.

\subsection*{Dimension-6 Operators}
\begin{align}
\C_1^{(6)} &= C_{qqq}\,, & \C_3^{(6)} &= C_{duq}\,, & \C_5^{(6)} &= C_{qqu}\,, \\
\C_7^{(6)} &= C_{duu}\,, & \C_9^{(6)} &= -C_{qqq}\,, & \C_{11}^{(6)} &= -C_{duq}\,, \\
\C_2^{(6)} &= \frac{v}{\sqrt{2}\Lambda}C_{LdQQH}\,, &
\C_4^{(6)} &= \frac{v^3}{2\sqrt{2}\Lambda^3}C_{q^3\bar eH^{\dagger 3}}\,, \\
\C_6^{(6)} &= \frac{v}{\sqrt{2}\Lambda}\bigl[C_{LdudH} + C_{LdddH}\bigr]\,, &
\C_8^{(6)} &= \frac{v}{\sqrt{2}\Lambda}C_{eQddH}\,, \\
\C_{10}^{(6)} &= \frac{v}{\sqrt{2}\Lambda}C_{LdQQH}\,, &
\C_{12}^{(6)} &= \frac{v}{\sqrt{2}\Lambda}\bigl[C_{LdudH} + C_{LdddH}\bigr]\,.
\end{align}

\subsection*{Dimension-7 Operators}
\begin{align}
\C_1^{(7)} &= C_{lq^2uHD,lud^2HD}\,, & \C_3^{(7)} &= C_{lu^2dHD}\,, \\
\C_5^{(7)} &= C_{eq^3HD}\,, & \C_7^{(7)} &= C_{equ^2HD}\,, \\
\C_9^{(7)} &= C_{lq^2dHD}\,, & \C_{11}^{(7)} &= C_{lu^2dHD}\,, \\
\C_2^{(7)} &= \frac{v^2}{2\Lambda^2}C_{q^3\bar lH^{\dagger2}D}\,, &
\C_4^{(7)} &= C_{LQddD}\,, \\
\C_6^{(7)} &= \frac{v^2}{2\Lambda^2}C_{q^2d\bar eH^{\dagger2}D}\,, &
\C_8^{(7)} &= C_{edddD}\,, \\
\C_{10}^{(7)} &= \frac{v^2}{2\Lambda^2}C_{q^3\bar lH^{\dagger2}D}\,, &
\C_{12}^{(7)} &= C_{LQddD}\,.
\end{align}

\subsection*{Dimension-8 Operators ($D^2$ Sector)}
\begin{align}
\C_1^{(8)} &= C_{lq^3D^2}^{(1)}\,, & \C_3^{(8)} &= C_{lqudD^2}^{(1)}\,, & \C_5^{(8)} &= C_{eq^2uD^2}^{(1)}\,, \\
\C_7^{(8)} &= C_{eu^2dD^2}^{(1)}\,, & \C_{9}^{(8)} &= C_{lq^3D^2}^{(1)}\,, & \C_{11}^{(8)} &= C_{lqudD^2}^{(1)}\,, \\
\C_2^{(8)} &= \frac{v}{\sqrt{2}\Lambda}C_{q^2d\bar lH^\dagger D^2}^{(1)}\,, \\
\C_4^{(8)} &= \frac{v}{\sqrt{2}\Lambda}\bigl[C_{ud^2\bar lH^\dagger D^2}^{(1)} + C_{d^3\bar lH D^2}^{(1)}\bigr]\,, \\
\C_8^{(8)} &= \frac{v}{\sqrt{2}\Lambda}C_{qd^2\bar eH^\dagger D^2}^{(1)}\,, &
\C_{10}^{(8)} &= \frac{v}{\sqrt{2}\Lambda}C_{q^2d\bar lH^\dagger D^2}^{(1)}\,, \\
\C_{12}^{(8)} &= \frac{v}{\sqrt{2}\Lambda}\bigl[C_{ud^2\bar lH^\dagger D^2}^{(1)} + C_{d^3\bar lH D^2}^{(1)}\bigr]\,, &
\C_{13}^{(8)} &= C_{lq^3D^2}^{(2)}\,, \\
\C_{15}^{(8)} &= C_{lqudD^2}^{(2)}\,, & \C_{17}^{(8)} &= C_{eq^2uD^2}^{(2)}\,, & \C_{19}^{(8)} &= C_{eu^2dD^2}^{(2)}\,, \\
\C_{21}^{(8)} &= C_{lq^3D^2}^{(2)}\,, & \C_{23}^{(8)} &= C_{lqudD^2}^{(2)}\,, \\
\C_{14}^{(8)} &= \frac{v}{\sqrt{2}\Lambda}C_{q^2d\bar lH^\dagger D^2}^{(2)}\,, \\
\C_{16}^{(8)} &= \frac{v}{\sqrt{2}\Lambda}\bigl[C_{ud^2\bar lH^\dagger D^2}^{(2)} + C_{d^3\bar lH D^2}^{(2)}\bigr]\,, \\
\C_{20}^{(8)} &= \frac{v}{\sqrt{2}\Lambda}C_{qd^2\bar eH^\dagger D^2}^{(2)}\,, \\
\C_{22}^{(8)} &= \frac{v}{\sqrt{2}\Lambda}C_{q^2d\bar lH^\dagger D^2}^{(2)}\,, \\
\C_{24}^{(8)} &= \frac{v}{\sqrt{2}\Lambda}\bigl[C_{ud^2\bar lH^\dagger D^2}^{(2)} + C_{d^3\bar lH D^2}^{(2)}\bigr]\,.
\end{align}

\subsection*{Dimension-8 Operators ($W/B$ Sector)}
\begin{align}
\C_{25}^{(8)} &= \frac{1}{2}\bigl[C_{lq^3W}^{(1)} + C_{lq^3B}^{(1)}\bigr]\,, \\
\C_{27}^{(8)} &= \frac{1}{2}\bigl[C_{lqudB}^{(1)} + C_{lqudW}^{(1)}\bigr]\,, \\
\C_{29}^{(8)} &= \frac{1}{2}\bigl[C_{eq^2uW}^{(1)} + C_{eq^2uB}^{(1)}\bigr]\,, \\
\C_{31}^{(8)} &= \frac{1}{2}C_{eu^2dB}^{(1)}\,, \\
\C_{33}^{(8)} &= \frac{1}{2}\bigl[C_{lqudB}^{(1)} + C_{lqudW}^{(1)}\bigr]\,, \\
\C_{35}^{(8)} &= \frac{1}{2}\bigl[C_{lq^3W}^{(1)} + C_{lq^3B}^{(1)}\bigr]\,, \\
\C_{26}^{(8)} &= \frac{v}{2\sqrt{2}\Lambda}\bigl[C_{Bq^2d\bar lH^\dagger}^{(1)} + C_{Wq^2d\bar lH^\dagger }^{(1)}\bigr]\,, \\
\C_{28}^{(8)} &= \frac{v}{2\sqrt{2}\Lambda}\bigl[C_{Bud^2\bar lH^\dagger }^{(1)} + C_{Wud^2\bar lH^\dagger}^{(1)} + C_{Wud^2\bar lH^\dagger}^{(2)} + C_{Bd^3\bar lH } + C_{Wd^3\bar lH}\bigr]\,, \\
\C_{32}^{(8)} &= \frac{v}{2\sqrt{2}\Lambda}\bigl[C_{Bqd^2\bar eH^\dagger }^{(1)} + C_{Wqd^2\bar eH^\dagger }^{(1)}\bigr]\,, \\
\C_{34}^{(8)} &= \frac{-v}{2\sqrt{2}\Lambda}\bigl[C_{Bq^2d\bar lH^\dagger }^{(1)} + C_{Wq^2d\bar lH^\dagger }^{(1)}\bigr]\,, \\
\C_{36}^{(8)} &= \frac{-v}{2\sqrt{2}\Lambda}\bigl[C_{Bud^2\bar lH^\dagger }^{(1)} + C_{Wud^2\bar lH^\dagger}^{(1)} + C_{Wud^2\bar lH^\dagger}^{(2)} + C_{Bd^3\bar lH } + C_{Wd^3\bar lH}\bigr]\,, \\
\C_{37}^{(8)} &= \frac{1}{2}\bigl[C_{lq^3W}^{(2)} + C_{lq^3B}^{(2)}\bigr]\,, \\
\C_{39}^{(8)} &= \frac{1}{2}\bigl[C_{lqudB}^{(2)} + C_{lqudW}^{(2)}\bigr]\,, \\
\C_{41}^{(8)} &= \frac{1}{2}\bigl[C_{eq^2uW}^{(2)} + C_{eq^2uB}^{(2)}\bigr]\,, \\
\C_{43}^{(8)} &= \frac{1}{2}C_{eu^2dB}^{(2)}\,, \\
\C_{45}^{(8)} &= \frac{1}{2}\bigl[C_{lqudB}^{(2)} + C_{lqudW}^{(2)}\bigr]\,, \\
\C_{47}^{(8)} &= \frac{1}{2}\bigl[C_{lq^3W}^{(2)} + C_{lq^3B}^{(2)}\bigr]\,, \\
\C_{38}^{(8)} &= \frac{v}{2\sqrt{2}\Lambda}\bigl[C_{Bq^2d\bar lH^\dagger }^{(2)} + C_{Wq^2d\bar lH^\dagger }^{(2)}\bigr]\,, \\
\C_{40}^{(8)} &= \frac{v}{2\sqrt{2}\Lambda}\bigl[C_{Bud^2\bar lH^\dagger }^{(2)} + C_{Wud^2\bar lH^\dagger}^{(3)} + C_{Wud^2\bar lH^\dagger}^{(4)}\bigr]\,, \\
\C_{44}^{(8)} &= \frac{v}{2\sqrt{2}\Lambda}\bigl[C_{Bqd^2\bar eH^\dagger }^{(2)} + C_{Wqd^2\bar eH^\dagger }^{(2)}\bigr]\,, \\
\C_{46}^{(8)} &= \frac{-v}{2\sqrt{2}\Lambda}\bigl[C_{Bq^2d\bar lH^\dagger }^{(2)} + C_{Wq^2d\bar lH^\dagger }^{(2)}\bigr]\,, \\
\C_{48}^{(8)} &= \frac{-v}{2\sqrt{2}\Lambda}\bigl[C_{Bud^2\bar lH^\dagger }^{(2)} + C_{Wud^2\bar lH^\dagger}^{(3)} + C_{Wud^2\bar lH^\dagger}^{(4)}\bigr]\,.
\end{align}

\bibliography{ref}

@article{Goldstone:1961eq,
    author = "Goldstone, J.",
    title = "{Field Theories with Superconductor Solutions}",
    doi = "10.1007/BF02812722",
    journal = "Nuovo Cim.",
    volume = "19",
    pages = "154--164",
    year = "1961"
}

@article{Goldstone:1962es,
    author = "Goldstone, Jeffrey and Salam, Abdus and Weinberg, Steven",
    title = "{Broken Symmetries}",
    doi = "10.1103/PhysRev.127.965",
    journal = "Phys. Rev.",
    volume = "127",
    pages = "965--970",
    year = "1962"
}

@article{Weinberg:1978kz,
    author = "Weinberg, Steven",
    editor = "Deser, S.",
    title = "{Phenomenological Lagrangians}",
    reportNumber = "HUTP-78-A051A",
    doi = "10.1016/0378-4371(79)90223-1",
    journal = "Physica A",
    volume = "96",
    number = "1-2",
    pages = "327--340",
    year = "1979"
}

@article{Weinberg:1968de,
    author = "Weinberg, Steven",
    title = "{Nonlinear realizations of chiral symmetry}",
    doi = "10.1103/PhysRev.166.1568",
    journal = "Phys. Rev.",
    volume = "166",
    pages = "1568--1577",
    year = "1968"
}

@article{Gasser:1983yg,
    author = "Gasser, J. and Leutwyler, H.",
    title = "{Chiral Perturbation Theory to One Loop}",
    reportNumber = "CERN-TH-3689",
    doi = "10.1016/0003-4916(84)90242-2",
    journal = "Annals Phys.",
    volume = "158",
    pages = "142",
    year = "1984"
}

@article{Gasser:1984gg,
    author = "Gasser, J. and Leutwyler, H.",
    title = "{Chiral Perturbation Theory: Expansions in the Mass of the Strange Quark}",
    reportNumber = "CERN-TH-3798",
    doi = "10.1016/0550-3213(85)90492-4",
    journal = "Nucl. Phys. B",
    volume = "250",
    pages = "465--516",
    year = "1985"
}

@article{Gasser:1987rb,
    author = "Gasser, J. and Sainio, M. E. and Svarc, A.",
    title = "{Nucleons with chiral loops}",
    reportNumber = "BUTP-87-17",
    doi = "10.1016/0550-3213(88)90108-3",
    journal = "Nucl. Phys. B",
    volume = "307",
    pages = "779--853",
    year = "1988"
}

@article{Li:2026mco,
    author = "Li, Gang and Song, Chuan-Qiang and Tang, Feng-Jie and Yu, Jiang-Hao",
    title = "{A Comprehensive Effective Field Theory Framework for Coherent Elastic Neutrino-Nucleus Scattering}",
    eprint = "2601.19883",
    archivePrefix = "arXiv",
    primaryClass = "hep-ph",
    month = "1",
    year = "2026"
}

@article{Bernard:1992qa,
    author = "Bernard, Veronique and Kaiser, Norbert and Kambor, Joachim and Meissner, Ulf G.",
    title = "{Chiral structure of the nucleon}",
    reportNumber = "BUTP-92-15, CRN-92-24, TUM-T31-28-92",
    doi = "10.1016/0550-3213(92)90615-I",
    journal = "Nucl. Phys. B",
    volume = "388",
    pages = "315--345",
    year = "1992"
}

@article{Coleman:1969sm,
    author = "Coleman, Sidney R. and Wess, J. and Zumino, Bruno",
    title = "{Structure of phenomenological Lagrangians. 1.}",
    doi = "10.1103/PhysRev.177.2239",
    journal = "Phys. Rev.",
    volume = "177",
    pages = "2239--2247",
    year = "1969"
}

@article{Callan:1969sn,
    author = "Callan, Jr., Curtis G. and Coleman, Sidney R. and Wess, J. and Zumino, Bruno",
    title = "{Structure of phenomenological Lagrangians. 2.}",
    doi = "10.1103/PhysRev.177.2247",
    journal = "Phys. Rev.",
    volume = "177",
    pages = "2247--2250",
    year = "1969"
}

@article{Weinberg:1979sa,
    author = "Weinberg, Steven",
    title = "{Baryon and Lepton Nonconserving Processes}",
    reportNumber = "HUTP-79-A050",
    doi = "10.1103/PhysRevLett.43.1566",
    journal = "Phys. Rev. Lett.",
    volume = "43",
    pages = "1566--1570",
    year = "1979"
}

@article{Georgi:1974sy,
    author = "Georgi, H. and Glashow, S. L.",
    title = "{Unity of All Elementary Particle Forces}",
    doi = "10.1103/PhysRevLett.32.438",
    journal = "Phys. Rev. Lett.",
    volume = "32",
    pages = "438--441",
    year = "1974"
}

@article{Sakharov:1967dj,
    author = "Sakharov, A. D.",
    title = "{Violation of CP Invariance, C asymmetry, and baryon asymmetry of the universe}",
    doi = "10.1070/PU1991v034n05ABEH002497",
    journal = "Pisma Zh. Eksp. Teor. Fiz.",
    volume = "5",
    pages = "32--35",
    year = "1967"
}

@inproceedings{Takhistov:2016eqm,
    author = "Takhistov, Volodymyr",
    collaboration = "Super-Kamiokande",
    title = "{Review of Nucleon Decay Searches at Super-Kamiokande}",
    booktitle = "{51st Rencontres de Moriond on EW Interactions and Unified Theories}",
    eprint = "1605.03235",
    archivePrefix = "arXiv",
    primaryClass = "hep-ex",
    reportNumber = "UCI-TR-2016-11",
    pages = "437--444",
    year = "2016"
}

@article{Hyper-Kamiokande:2018ofw,
    author = "Abe, K. and others",
    collaboration = "Hyper-Kamiokande",
    title = "{Hyper-Kamiokande Design Report}",
    eprint = "1805.04163",
    archivePrefix = "arXiv",
    primaryClass = "physics.ins-det",
    month = "5",
    year = "2018"
}

@article{DUNE:2020ypp,
    author = "Abi, Babak and others",
    collaboration = "DUNE",
    title = "{Deep Underground Neutrino Experiment (DUNE), Far Detector Technical Design Report, Volume II: DUNE Physics}",
    eprint = "2002.03005",
    archivePrefix = "arXiv",
    primaryClass = "hep-ex",
    reportNumber = "FERMILAB-PUB-20-025-ND, FERMILAB-DESIGN-2020-02",
    month = "2",
    year = "2020"
}

@article{JUNO:2015zny,
    author = "An, Fengpeng and others",
    collaboration = "JUNO",
    title = "{Neutrino Physics with JUNO}",
    eprint = "1507.05613",
    archivePrefix = "arXiv",
    primaryClass = "physics.ins-det",
    doi = "10.1088/0954-3899/43/3/030401",
    journal = "J. Phys. G",
    volume = "43",
    number = "3",
    pages = "030401",
    year = "2016"
}

@article{Lazarides:1980nt,
    author = "Lazarides, George and Shafi, Q. and Wetterich, C.",
    title = "{Proton Lifetime and Fermion Masses in an SO(10) Model}",
    reportNumber = "FREIBURG-THEP-80-2",
    doi = "10.1016/0550-3213(81)90354-0",
    journal = "Nucl. Phys. B",
    volume = "181",
    pages = "287--300",
    year = "1981"
}

@article{Inoue:1982pi,
    author = "Inoue, Kenzo and Kakuto, Akira and Komatsu, Hiromasa and Takeshita, Seiichiro",
    title = "{Aspects of Grand Unified Models with Softly Broken Supersymmetry}",
    reportNumber = "KYUSHU-82-HE-5",
    doi = "10.1143/PTP.68.927",
    journal = "Prog. Theor. Phys.",
    volume = "68",
    pages = "927",
    year = "1982",
    note = "[Erratum: Prog.Theor.Phys. 70, 330 (1983)]"
}

@article{Barger:1992ac,
    author = "Barger, Vernon D. and Berger, M. S. and Ohmann, P.",
    title = "{Supersymmetric grand unified theories: Two loop evolution of gauge and Yukawa couplings}",
    eprint = "hep-ph/9209232",
    archivePrefix = "arXiv",
    reportNumber = "MAD-PH-711",
    doi = "10.1103/PhysRevD.47.1093",
    journal = "Phys. Rev. D",
    volume = "47",
    pages = "1093--1113",
    year = "1993"
}

@article{Nath:2006ut,
    author = "Nath, Pran and Fileviez Perez, Pavel",
    title = "{Proton stability in grand unified theories, in strings and in branes}",
    eprint = "hep-ph/0601023",
    archivePrefix = "arXiv",
    doi = "10.1016/j.physrep.2007.02.010",
    journal = "Phys. Rept.",
    volume = "441",
    pages = "191--317",
    year = "2007"
}

@article{Hisano:1992jj,
    author = "Hisano, J. and Murayama, H. and Yanagida, T.",
    title = "{Nucleon decay in the minimal supersymmetric SU(5) grand unification}",
    eprint = "hep-ph/9207279",
    archivePrefix = "arXiv",
    reportNumber = "TU-400",
    doi = "10.1016/0550-3213(93)90636-4",
    journal = "Nucl. Phys. B",
    volume = "402",
    pages = "46--84",
    year = "1993"
}

@article{Carena:1993ag,
    author = "Carena, Marcela and Pokorski, S. and Wagner, C. E. M.",
    title = "{On the unification of couplings in the minimal supersymmetric Standard Model}",
    eprint = "hep-ph/9303202",
    archivePrefix = "arXiv",
    reportNumber = "MPI-PH-93-10",
    doi = "10.1016/0550-3213(93)90161-H",
    journal = "Nucl. Phys. B",
    volume = "406",
    pages = "59--89",
    year = "1993"
}

@article{Ellis:1978xg,
    author = "Ellis, John R. and Gaillard, Mary K. and Nanopoulos, Dimitri V.",
    title = "{Baryon Number Generation in Grand Unified Theories}",
    reportNumber = "CERN-TH-2596",
    doi = "10.1016/0370-2693(79)91190-0",
    journal = "Phys. Lett. B",
    volume = "80",
    pages = "360",
    year = "1979",
    note = "[Erratum: Phys.Lett.B 82, 464 (1979)]"
}

@article{Chang:1984qr,
    author = "Chang, D. and Mohapatra, R. N. and Gipson, J. and Marshak, R. E. and Parida, M. K.",
    title = "{Experimental Tests of New SO(10) Grand Unification}",
    reportNumber = "VPI-HEP-84/9",
    doi = "10.1103/PhysRevD.31.1718",
    journal = "Phys. Rev. D",
    volume = "31",
    pages = "1718",
    year = "1985"
}

@article{Babu:2016bmy,
    author = "Babu, K. S. and Bajc, Borut and Saad, Shaikh",
    title = "{Yukawa Sector of Minimal SO(10) Unification}",
    eprint = "1612.04329",
    archivePrefix = "arXiv",
    primaryClass = "hep-ph",
    reportNumber = "OSU-HEP-16-08",
    doi = "10.1007/JHEP02(2017)136",
    journal = "JHEP",
    volume = "02",
    pages = "136",
    year = "2017"
}

@article{Wilczek:1979hc,
    author = "Wilczek, Frank and Zee, A.",
    title = "{Operator Analysis of Nucleon Decay}",
    reportNumber = "Print-79-0709 (PRINCETON)",
    doi = "10.1103/PhysRevLett.43.1571",
    journal = "Phys. Rev. Lett.",
    volume = "43",
    pages = "1571--1573",
    year = "1979"
}

@article{Abbott:1980zj,
    author = "Abbott, L. F. and Wise, Mark B.",
    title = "{The Effective Hamiltonian for Nucleon Decay}",
    reportNumber = "SLAC-PUB-2487",
    doi = "10.1103/PhysRevD.22.2208",
    journal = "Phys. Rev. D",
    volume = "22",
    pages = "2208",
    year = "1980"
}

@article{Buchmuller:1985jz,
    author = "Buchmuller, W. and Wyler, D.",
    title = "{Effective Lagrangian Analysis of New Interactions and Flavor Conservation}",
    reportNumber = "CERN-TH-4254/85",
    doi = "10.1016/0550-3213(86)90262-2",
    journal = "Nucl. Phys. B",
    volume = "268",
    pages = "621--653",
    year = "1986"
}

@article{Rinaldi:2018osy,
    author = "Rinaldi, Enrico and Syritsyn, Sergey and Wagman, Michael L. and Buchoff, Michael I. and Schroeder, Chris and Wasem, Joseph",
    title = "{Neutron-antineutron oscillations from lattice QCD}",
    eprint = "1809.00246",
    archivePrefix = "arXiv",
    primaryClass = "hep-lat",
    reportNumber = "RBRC-1290, MIT-CTP/5051, LLNL-JRNL-757017",
    doi = "10.1103/PhysRevLett.122.162001",
    journal = "Phys. Rev. Lett.",
    volume = "122",
    number = "16",
    pages = "162001",
    year = "2019"
}

@article{Buchoff:2015qwa,
    author = "Buchoff, Michael I. and Wagman, Michael",
    title = "{Perturbative Renormalization of Neutron-Antineutron Operators}",
    eprint = "1506.00647",
    archivePrefix = "arXiv",
    primaryClass = "hep-ph",
    doi = "10.1103/PhysRevD.93.016005",
    journal = "Phys. Rev. D",
    volume = "93",
    number = "1",
    pages = "016005",
    year = "2016",
    note = "[Erratum: Phys.Rev.D 98, 079901 (2018)]"
}

@article{Phillips:2014fgb,
    author = "Phillips, II, D. G. and others",
    title = "{Neutron-Antineutron Oscillations: Theoretical Status and Experimental Prospects}",
    eprint = "1410.1100",
    archivePrefix = "arXiv",
    primaryClass = "hep-ex",
    reportNumber = "FERMILAB-PUB-14-263-T",
    doi = "10.1016/j.physrep.2015.11.001",
    journal = "Phys. Rept.",
    volume = "612",
    pages = "1--45",
    year = "2016"
}

@article{Li:2021tsq,
    author = "Li, Hao-Lin and Ren, Zhe and Xiao, Ming-Lei and Yu, Jiang-Hao and Zheng, Yu-Hui",
    title = "{Operator bases in effective field theories with sterile neutrinos: d {\ensuremath{\leq}} 9}",
    eprint = "2105.09329",
    archivePrefix = "arXiv",
    primaryClass = "hep-ph",
    doi = "10.1007/JHEP11(2021)003",
    journal = "JHEP",
    volume = "11",
    pages = "003",
    year = "2021"
}

@article{Song:2023jqm,
    author = "Song, Huayang and Sun, Hao and Yu, Jiang-Hao",
    title = "{Complete EFT operator bases for dark matter and weakly-interacting light particle}",
    eprint = "2306.05999",
    archivePrefix = "arXiv",
    primaryClass = "hep-ph",
    doi = "10.1007/JHEP05(2024)103",
    journal = "JHEP",
    volume = "05",
    pages = "103",
    year = "2024"
}

@article{Song:2023lxf,
    author = "Song, Huayang and Sun, Hao and Yu, Jiang-Hao",
    title = "{Effective field theories of axion, ALP and dark photon}",
    eprint = "2305.16770",
    archivePrefix = "arXiv",
    primaryClass = "hep-ph",
    doi = "10.1007/JHEP01(2024)161",
    journal = "JHEP",
    volume = "01",
    pages = "161",
    year = "2024"
}

@article{Helo:2018bgb,
    author = "Helo, Juan C. and Hirsch, Martin and Ota, Toshihiko",
    title = "{Proton decay and light sterile neutrinos}",
    eprint = "1803.00035",
    archivePrefix = "arXiv",
    primaryClass = "hep-ph",
    doi = "10.1007/JHEP06(2018)047",
    journal = "JHEP",
    volume = "06",
    pages = "047",
    year = "2018"
}

@article{Baldo-Ceolin:1994hzw,
    author = "Baldo-Ceolin, M. and others",
    title = "{A New experimental limit on neutron - anti-neutron oscillations}",
    reportNumber = "DFPD-94-EP-13",
    doi = "10.1007/BF01580321",
    journal = "Z. Phys. C",
    volume = "63",
    pages = "409--416",
    year = "1994"
}

@article{Grzadkowski:2010es,
    author = "Grzadkowski, B. and Iskrzynski, M. and Misiak, M. and Rosiek, J.",
    title = "{Dimension-Six Terms in the Standard Model Lagrangian}",
    eprint = "1008.4884",
    archivePrefix = "arXiv",
    primaryClass = "hep-ph",
    reportNumber = "IFT-9-2010, TTP10-35",
    doi = "10.1007/JHEP10(2010)085",
    journal = "JHEP",
    volume = "10",
    pages = "085",
    year = "2010"
}

@article{Lehman:2014jma,
    author = "Lehman, Landon",
    title = "{Extending the Standard Model Effective Field Theory with the Complete Set of Dimension-7 Operators}",
    eprint = "1410.4193",
    archivePrefix = "arXiv",
    primaryClass = "hep-ph",
    doi = "10.1103/PhysRevD.90.125023",
    journal = "Phys. Rev. D",
    volume = "90",
    number = "12",
    pages = "125023",
    year = "2014"
}

@article{Li:2020gnx,
    author = "Li, Hao-Lin and Ren, Zhe and Shu, Jing and Xiao, Ming-Lei and Yu, Jiang-Hao and Zheng, Yu-Hui",
    title = "{Complete set of dimension-eight operators in the standard model effective field theory}",
    eprint = "2005.00008",
    archivePrefix = "arXiv",
    primaryClass = "hep-ph",
    doi = "10.1103/PhysRevD.104.015026",
    journal = "Phys. Rev. D",
    volume = "104",
    number = "1",
    pages = "015026",
    year = "2021"
}

@article{Murphy:2020rsh,
    author = "Murphy, Christopher W.",
    title = "{Dimension-8 operators in the Standard Model Effective Field Theory}",
    eprint = "2005.00059",
    archivePrefix = "arXiv",
    primaryClass = "hep-ph",
    doi = "10.1007/JHEP10(2020)174",
    journal = "JHEP",
    volume = "10",
    pages = "174",
    year = "2020"
}

@article{Li:2020xlh,
    author = "Li, Hao-Lin and Ren, Zhe and Xiao, Ming-Lei and Yu, Jiang-Hao and Zheng, Yu-Hui",
    title = "{Complete set of dimension-nine operators in the standard model effective field theory}",
    eprint = "2007.07899",
    archivePrefix = "arXiv",
    primaryClass = "hep-ph",
    doi = "10.1103/PhysRevD.104.015025",
    journal = "Phys. Rev. D",
    volume = "104",
    number = "1",
    pages = "015025",
    year = "2021"
}

@article{Jenkins:2017jig,
    author = "Jenkins, Elizabeth E. and Manohar, Aneesh V. and Stoffer, Peter",
    title = "{Low-Energy Effective Field Theory below the Electroweak Scale: Operators and Matching}",
    eprint = "1709.04486",
    archivePrefix = "arXiv",
    primaryClass = "hep-ph",
    doi = "10.1007/JHEP03(2018)016",
    journal = "JHEP",
    volume = "03",
    pages = "016",
    year = "2018",
    note = "[Erratum: JHEP 12, 043 (2023)]"
}

@article{Jenkins:2017dyc,
    author = "Jenkins, Elizabeth E. and Manohar, Aneesh V. and Stoffer, Peter",
    title = "{Low-Energy Effective Field Theory below the Electroweak Scale: Anomalous Dimensions}",
    eprint = "1711.05270",
    archivePrefix = "arXiv",
    primaryClass = "hep-ph",
    doi = "10.1007/JHEP01(2018)084",
    journal = "JHEP",
    volume = "01",
    pages = "084",
    year = "2018",
    note = "[Erratum: JHEP 12, 042 (2023)]"
}

@article{Liao:2020zyx,
    author = "Liao, Yi and Ma, Xiao-Dong and Wang, Quan-Yu",
    title = "{Extending low energy effective field theory with a complete set of dimension-7 operators}",
    eprint = "2005.08013",
    archivePrefix = "arXiv",
    primaryClass = "hep-ph",
    doi = "10.1007/JHEP08(2020)162",
    journal = "JHEP",
    volume = "08",
    pages = "162",
    year = "2020"
}

@article{Liao:2020jmn,
    author = "Liao, Yi and Ma, Xiao-Dong",
    title = "{An explicit construction of the dimension-9 operator basis in the standard model effective field theory}",
    eprint = "2007.08125",
    archivePrefix = "arXiv",
    primaryClass = "hep-ph",
    doi = "10.1007/JHEP11(2020)152",
    journal = "JHEP",
    volume = "11",
    pages = "152",
    year = "2020"
}

@article{Li:2020tsi,
    author = "Li, Hao-Lin and Ren, Zhe and Xiao, Ming-Lei and Yu, Jiang-Hao and Zheng, Yu-Hui",
    title = "{Low energy effective field theory operator basis at d {\ensuremath{\leq}} 9}",
    eprint = "2012.09188",
    archivePrefix = "arXiv",
    primaryClass = "hep-ph",
    doi = "10.1007/JHEP06(2021)138",
    journal = "JHEP",
    volume = "06",
    pages = "138",
    year = "2021"
}

@article{Murphy:2020cly,
    author = "Murphy, Christopher W.",
    title = "{Low-Energy Effective Field Theory below the Electroweak Scale: Dimension-8 Operators}",
    eprint = "2012.13291",
    archivePrefix = "arXiv",
    primaryClass = "hep-ph",
    doi = "10.1007/JHEP04(2021)101",
    journal = "JHEP",
    volume = "04",
    pages = "101",
    year = "2021"
}

@article{Claudson:1981gh,
    author = "Claudson, Mark and Wise, Mark B. and Hall, Lawrence J.",
    title = "{Chiral Lagrangian for Deep Mine Physics}",
    reportNumber = "HUTP-81/A036",
    doi = "10.1016/0550-3213(82)90401-1",
    journal = "Nucl. Phys. B",
    volume = "195",
    pages = "297--307",
    year = "1982"
}

@article{JLQCD:1999dld,
    author = "Aoki, S. and others",
    collaboration = "JLQCD",
    title = "{Nucleon decay matrix elements from lattice QCD}",
    eprint = "hep-lat/9911026",
    archivePrefix = "arXiv",
    reportNumber = "HUPD-9919",
    doi = "10.1103/PhysRevD.62.014506",
    journal = "Phys. Rev. D",
    volume = "62",
    pages = "014506",
    year = "2000"
}

@article{Aoki:2006ib,
    author = "Aoki, Y. and Dawson, C. and Noaki, J. and Soni, A.",
    title = "{Proton decay matrix elements with domain-wall fermions}",
    eprint = "hep-lat/0607002",
    archivePrefix = "arXiv",
    reportNumber = "BNL-HET-6-5, RBRC-606",
    doi = "10.1103/PhysRevD.75.014507",
    journal = "Phys. Rev. D",
    volume = "75",
    pages = "014507",
    year = "2007"
}

@article{Aoki:2008ku,
    author = "Aoki, Y. and Boyle, P. and Cooney, P. and Del Debbio, L. and Kenway, R. and Maynard, C. M. and Soni, A. and Tweedie, R.",
    collaboration = "RBC-UKQCD",
    title = "{Proton lifetime bounds from chirally symmetric lattice QCD}",
    eprint = "0806.1031",
    archivePrefix = "arXiv",
    primaryClass = "hep-lat",
    reportNumber = "EDINBURGH-2008-19, RBRC-728, BNL-HET-08-13",
    doi = "10.1103/PhysRevD.78.054505",
    journal = "Phys. Rev. D",
    volume = "78",
    pages = "054505",
    year = "2008"
}

@article{QCDSF:2008qtn,
    author = "Braun, Vladimir M. and others",
    collaboration = "QCDSF",
    title = "{Nucleon distribution amplitudes and proton decay matrix elements on the lattice}",
    eprint = "0811.2712",
    archivePrefix = "arXiv",
    primaryClass = "hep-lat",
    reportNumber = "DESY-08-166, EDINBURGH-2008-45",
    doi = "10.1103/PhysRevD.79.034504",
    journal = "Phys. Rev. D",
    volume = "79",
    pages = "034504",
    year = "2009"
}

@article{Aoki:2017puj,
    author = "Aoki, Yasumichi and Izubuchi, Taku and Shintani, Eigo and Soni, Amarjit",
    title = "{Improved lattice computation of proton decay matrix elements}",
    eprint = "1705.01338",
    archivePrefix = "arXiv",
    primaryClass = "hep-lat",
    reportNumber = "RBRC-1235, KEK-CP-358",
    doi = "10.1103/PhysRevD.96.014506",
    journal = "Phys. Rev. D",
    volume = "96",
    number = "1",
    pages = "014506",
    year = "2017"
}

@article{Yoo:2021gql,
    author = "Yoo, Jun-Sik and Aoki, Yasumichi and Boyle, Peter and Izubuchi, Taku and Soni, Amarjit and Syritsyn, Sergey",
    title = "{Proton decay matrix elements on the lattice at physical pion mass}",
    eprint = "2111.01608",
    archivePrefix = "arXiv",
    primaryClass = "hep-lat",
    reportNumber = "RBRC-1333, KEK-CP-0385",
    doi = "10.1103/PhysRevD.105.074501",
    journal = "Phys. Rev. D",
    volume = "105",
    number = "7",
    pages = "074501",
    year = "2022"
}

@article{Liao:2025vlj,
    author = "Liao, Yi and Ma, Xiao-Dong and Wang, Hao-Lin",
    title = "{New Chiral Structures for Baryon Number Violating Nucleon Decays}",
    eprint = "2504.14855",
    archivePrefix = "arXiv",
    primaryClass = "hep-ph",
    doi = "10.1103/d8m7-5xxx",
    journal = "Phys. Rev. Lett.",
    volume = "135",
    number = "16",
    pages = "161801",
    year = "2025"
}

@article{Hambye:2017qix,
    author = "Hambye, Thomas and Heeck, Julian",
    title = "{Proton decay into charged leptons}",
    eprint = "1712.04871",
    archivePrefix = "arXiv",
    primaryClass = "hep-ph",
    reportNumber = "ULB-TH-17-22",
    doi = "10.1103/PhysRevLett.120.171801",
    journal = "Phys. Rev. Lett.",
    volume = "120",
    number = "17",
    pages = "171801",
    year = "2018"
}

@article{Heeck:2019kgr,
    author = "Heeck, Julian and Takhistov, Volodymyr",
    title = "{Inclusive Nucleon Decay Searches as a Frontier of Baryon Number Violation}",
    eprint = "1910.07647",
    archivePrefix = "arXiv",
    primaryClass = "hep-ph",
    reportNumber = "UCI-TR-2019-24",
    doi = "10.1103/PhysRevD.101.015005",
    journal = "Phys. Rev. D",
    volume = "101",
    number = "1",
    pages = "015005",
    year = "2020"
}

@article{He:2021mrt,
    author = "He, Xiao-Gang and Ma, Xiao-Dong",
    title = "{An EFT toolbox for baryon and lepton number violating dinucleon to dilepton decays}",
    eprint = "2102.02562",
    archivePrefix = "arXiv",
    primaryClass = "hep-ph",
    doi = "10.1007/JHEP06(2021)047",
    journal = "JHEP",
    volume = "06",
    pages = "047",
    year = "2021"
}

@article{He:2021sbl,
    author = "He, Xiao-Gang and Ma, Xiao-Dong",
    title = "{$\Delta B=2$ neutron decay into antiproton mode $n\to \bar pe^+\nu(\bar\nu)$}",
    eprint = "2101.01405",
    archivePrefix = "arXiv",
    primaryClass = "hep-ph",
    doi = "10.1016/j.physletb.2021.136298",
    journal = "Phys. Lett. B",
    volume = "817",
    pages = "136298",
    year = "2021"
}

@article{Fan:2024gzc,
    author = "Fan, Wei-Qi and Liao, Yi and Ma, Xiao-Dong and Wang, Hao-Lin",
    title = "{Baryon number violating hydrogen decay}",
    eprint = "2412.20774",
    archivePrefix = "arXiv",
    primaryClass = "hep-ph",
    doi = "10.1016/j.physletb.2025.139335",
    journal = "Phys. Lett. B",
    volume = "862",
    pages = "139335",
    year = "2025"
}

@article{Gargalionis:2024nij,
    author = "Gargalionis, John and Herrero-Garc{\'\i}a, Juan and Schmidt, Michael A.",
    title = "{Model-independent estimates for loop-induced baryon-number-violating nucleon decays}",
    eprint = "2401.04768",
    archivePrefix = "arXiv",
    primaryClass = "hep-ph",
    reportNumber = "CPPC-2024-02",
    doi = "10.1007/JHEP06(2024)182",
    journal = "JHEP",
    volume = "06",
    pages = "182",
    year = "2024"
}

@article{Beneito:2023xbk,
    author = "Beneito, I, Arnau Bas and Gargalionis, John and Herrero-Garcia, Juan and Santamaria, Arcadi and Schmidt, Michael A.",
    title = "{An EFT approach to baryon number violation: lower limits on the new physics scale and correlations between nucleon decay modes}",
    eprint = "2312.13361",
    archivePrefix = "arXiv",
    primaryClass = "hep-ph",
    reportNumber = "IFIC/23-52, CPPC-2023-12",
    doi = "10.1007/JHEP07(2024)004",
    journal = "JHEP",
    volume = "07",
    pages = "004",
    year = "2024",
    note = "[Erratum: JHEP 02, 065 (2026)]"
}

@article{Li:2024liy,
    author = "Li, Tong and Schmidt, Michael A. and Yao, Chang-Yuan",
    title = "{Baryon-number-violating nucleon decays in ALP effective field theories}",
    eprint = "2406.11382",
    archivePrefix = "arXiv",
    primaryClass = "hep-ph",
    reportNumber = "CPPC-2024-05, DESY-24-082",
    doi = "10.1007/JHEP08(2024)221",
    journal = "JHEP",
    volume = "08",
    pages = "221",
    year = "2024"
}

@article{Li:2025slp,
    author = "Li, Tong and Schmidt, Michael A. and Yao, Chang-Yuan",
    title = "{Baryon-number-violating nucleon decays in sterile neutrino effective field theories}",
    eprint = "2502.14303",
    archivePrefix = "arXiv",
    primaryClass = "hep-ph",
    reportNumber = "CPPC-2025-02",
    doi = "10.1007/JHEP06(2025)077",
    journal = "JHEP",
    volume = "06",
    pages = "077",
    year = "2025"
}

@article{Fridell:2023tpb,
    author = "Fridell, K{\r{a}}re and Hati, Chandan and Takhistov, Volodymyr",
    title = "{Noncanonical nucleon decays as window into light new physics}",
    eprint = "2312.13740",
    archivePrefix = "arXiv",
    primaryClass = "hep-ph",
    reportNumber = "KEK-TH-2588, KEK-Cosmo-0336, KEK-QUP-2023-0038, IPMU23-0051,
  IFIC/23-54, IPMU23-0051, ULB-TH/23-18, IFIC/23-54",
    doi = "10.1103/PhysRevD.110.L031701",
    journal = "Phys. Rev. D",
    volume = "110",
    number = "3",
    pages = "L031701",
    year = "2024"
}

@article{Fan:2025xhi,
    author = "Fan, Wei-Qi and Liao, Yi and Ma, Xiao-Dong and Wang, Hao-Lin",
    title = "{Comprehensive investigation on baryon number violating nucleon decays involving an axion-like particle}",
    eprint = "2507.11844",
    archivePrefix = "arXiv",
    primaryClass = "hep-ph",
    month = "7",
    year = "2025"
}

@article{Ma:2025mjy,
    author = "Ma, Xiao-Dong and Schmidt, Michael A. and Zhang, Weihang",
    title = "{Baryon-number-violating nucleon decays in SMEFT extended with a light scalar}",
    eprint = "2511.02169",
    archivePrefix = "arXiv",
    primaryClass = "hep-ph",
    doi = "10.1007/JHEP02(2026)175",
    journal = "JHEP",
    volume = "02",
    pages = "175",
    year = "2026"
}

@article{Song:2025snz,
    author = "Song, Chuan-Qiang and Sun, Hao and Yu, Jiang-Hao",
    title = "{Systematic Spurion Matching between Low Energy EFT and Chiral Lagrangian}",
    eprint = "2501.09787",
    archivePrefix = "arXiv",
    primaryClass = "hep-ph",
    month = "1",
    year = "2025"
}

@article{Li:2025xmq,
    author = "Li, Gang and Song, Chuan-Qiang and Yu, Jiang-Hao",
    title = "{Matching from quark to hadronic operators: external source vs spurion methods}",
    eprint = "2507.02538",
    archivePrefix = "arXiv",
    primaryClass = "hep-ph",
    month = "7",
    year = "2025"
}

@article{Becher:1999he,
    author = "Becher, Thomas and Leutwyler, H.",
    title = "{Baryon chiral perturbation theory in manifestly Lorentz invariant form}",
    eprint = "hep-ph/9901384",
    archivePrefix = "arXiv",
    reportNumber = "BUTP-99-1",
    doi = "10.1007/PL00021673",
    journal = "Eur. Phys. J.",
    volume = "9",
    number = "4",
    pages = "643--671",
    year = "1999"
}

@article{Jenkins:1990jv,
    author = "Jenkins, Elizabeth Ellen and Manohar, Aneesh V.",
    title = "{Baryon chiral perturbation theory using a heavy fermion Lagrangian}",
    reportNumber = "UCSD-PTH-90-23",
    doi = "10.1016/0370-2693(91)90266-S",
    journal = "Phys. Lett. B",
    volume = "255",
    pages = "558--562",
    year = "1991"
}

@article{Krause:1990xc,
    author = "Krause, A.",
    title = "{Baryon Matrix Elements of the Vector Current in Chiral Perturbation Theory}",
    doi = "10.5169/seals-116214",
    journal = "Helv. Phys. Acta",
    volume = "63",
    pages = "3--70",
    year = "1990"
}

@article{Oller:2006yh,
    author = "Oller, Jose Antonio and Verbeni, Michela and Prades, Joaquim",
    title = "{Meson-baryon effective chiral lagrangians to O(q**3)}",
    eprint = "hep-ph/0608204",
    archivePrefix = "arXiv",
    reportNumber = "CAFPE-69-06, UGFT-199-06",
    doi = "10.1088/1126-6708/2006/09/079",
    journal = "JHEP",
    volume = "09",
    pages = "079",
    year = "2006"
}

@article{Song:2024fae,
    author = "Song, Chuan-Qiang and Sun, Hao and Yu, Jiang-Hao",
    title = "{Complete CP-eigen bases of meson-baryon chiral lagrangian up to p$^{5}$-order}",
    eprint = "2404.15047",
    archivePrefix = "arXiv",
    primaryClass = "hep-ph",
    doi = "10.1007/JHEP09(2024)171",
    journal = "JHEP",
    volume = "09",
    pages = "171",
    year = "2024"
}

@article{Bijnens:2017xrz,
    author = "Bijnens, Johan and Kofoed, Erik",
    title = "{Chiral perturbation theory for neutron{\textendash}antineutron oscillations}",
    eprint = "1710.04383",
    archivePrefix = "arXiv",
    primaryClass = "hep-ph",
    reportNumber = "LU-TP-17-30",
    doi = "10.1140/epjc/s10052-017-5411-7",
    journal = "Eur. Phys. J. C",
    volume = "77",
    number = "12",
    pages = "867",
    year = "2017"
}

@article{Kleiss:1985gy,
    author = "Kleiss, R. and Stirling, W. James and Ellis, S. D.",
    title = "{A New Monte Carlo Treatment of Multiparticle Phase Space at High-energies}",
    reportNumber = "CERN-TH-4299/85",
    doi = "10.1016/0010-4655(86)90119-0",
    journal = "Comput. Phys. Commun.",
    volume = "40",
    pages = "359",
    year = "1986"
}

@article{Manohar:1983md,
    author = "Manohar, Aneesh and Georgi, Howard",
    title = "{Chiral Quarks and the Nonrelativistic Quark Model}",
    reportNumber = "HUTP-83/A042a",
    doi = "10.1016/0550-3213(84)90231-1",
    journal = "Nucl. Phys. B",
    volume = "234",
    pages = "189--212",
    year = "1984"
}

@article{Gavela:2016bzc,
    author = "Gavela, B. M. and Jenkins, E. E. and Manohar, A. V. and Merlo, L.",
    title = "{Analysis of General Power Counting Rules in Effective Field Theory}",
    eprint = "1601.07551",
    archivePrefix = "arXiv",
    primaryClass = "hep-ph",
    reportNumber = "CERN-TH-2016-015, FTUAM-16-2, IFT-UAM-CSIC-16-006",
    doi = "10.1140/epjc/s10052-016-4332-1",
    journal = "Eur. Phys. J. C",
    volume = "76",
    number = "9",
    pages = "485",
    year = "2016"
}

@article{Jenkins:2013sda,
    author = "Jenkins, Elizabeth E. and Manohar, Aneesh V. and Trott, Michael",
    title = "{Naive Dimensional Analysis Counting of Gauge Theory Amplitudes and Anomalous Dimensions}",
    eprint = "1309.0819",
    archivePrefix = "arXiv",
    primaryClass = "hep-ph",
    reportNumber = "CERN-PH-TH-2013-213",
    doi = "10.1016/j.physletb.2013.09.020",
    journal = "Phys. Lett. B",
    volume = "726",
    pages = "697--702",
    year = "2013"
}

@article{Pati:1974yy,
    author = "Pati, Jogesh C. and Salam, Abdus",
    title = "{Lepton Number as the Fourth Color}",
    reportNumber = "IC-74-7",
    doi = "10.1103/PhysRevD.10.275",
    journal = "Phys. Rev. D",
    volume = "10",
    pages = "275--289",
    year = "1974",
    note = "[Erratum: Phys.Rev.D 11, 703--703 (1975)]"
}

@article{Pati:1973uk,
    author = "Pati, Jogesh C. and Salam, Abdus",
    title = "{Unified Lepton-Hadron Symmetry and a Gauge Theory of the Basic Interactions}",
    reportNumber = "IC-73-41-INT-REP",
    doi = "10.1103/PhysRevD.8.1240",
    journal = "Phys. Rev. D",
    volume = "8",
    pages = "1240--1251",
    year = "1973"
}

@article{Fritzsch:1974nn,
    author = "Fritzsch, Harald and Minkowski, Peter",
    title = "{Unified Interactions of Leptons and Hadrons}",
    reportNumber = "CALT-68-467",
    doi = "10.1016/0003-4916(75)90211-0",
    journal = "Annals Phys.",
    volume = "93",
    pages = "193--266",
    year = "1975"
}

@article{Barr:1981qv,
    author = "Barr, Stephen M.",
    title = "{A New Symmetry Breaking Pattern for SO(10) and Proton Decay}",
    reportNumber = "Print-81-0903 (WASH.U., SEATTLE), 40048-82-PT2",
    doi = "10.1016/0370-2693(82)90966-2",
    journal = "Phys. Lett. B",
    volume = "112",
    pages = "219--222",
    year = "1982"
}

@article{Derendinger:1983aj,
    author = "Derendinger, J. P. and Kim, Jihn E. and Nanopoulos, Dimitri V.",
    title = "{Anti-SU(5)}",
    reportNumber = "CERN-TH-3770",
    doi = "10.1016/0370-2693(84)91238-3",
    journal = "Phys. Lett. B",
    volume = "139",
    pages = "170--176",
    year = "1984"
}

@article{Li:2023cwy,
    author = "Li, Xu-Xiang and Ren, Zhe and Yub, Jiang-Hao",
    title = "{Complete tree-level dictionary between simplified BSM models and SMEFT d{\ensuremath{\leq}}7 operators}",
    eprint = "2307.10380",
    archivePrefix = "arXiv",
    primaryClass = "hep-ph",
    doi = "10.1103/PhysRevD.109.095041",
    journal = "Phys. Rev. D",
    volume = "109",
    number = "9",
    pages = "095041",
    year = "2024"
}

@article{Li:2023pfw,
    author = "Li, Hao-Lin and Ni, Yu-Han and Xiao, Ming-Lei and Yu, Jiang-Hao",
    title = "{Complete UV resonances of the dimension-8 SMEFT operators}",
    eprint = "2309.15933",
    archivePrefix = "arXiv",
    primaryClass = "hep-ph",
    doi = "10.1007/JHEP05(2024)238",
    journal = "JHEP",
    volume = "05",
    pages = "238",
    year = "2024"
}

@article{Jenkins:2013zja,
    author = "Jenkins, Elizabeth E. and Manohar, Aneesh V. and Trott, Michael",
    title = "{Renormalization Group Evolution of the Standard Model Dimension Six Operators I: Formalism and lambda Dependence}",
    eprint = "1308.2627",
    archivePrefix = "arXiv",
    primaryClass = "hep-ph",
    doi = "10.1007/JHEP10(2013)087",
    journal = "JHEP",
    volume = "10",
    pages = "087",
    year = "2013"
}

@article{Jenkins:2013wua,
    author = "Jenkins, Elizabeth E. and Manohar, Aneesh V. and Trott, Michael",
    title = "{Renormalization Group Evolution of the Standard Model Dimension Six Operators II: Yukawa Dependence}",
    eprint = "1310.4838",
    archivePrefix = "arXiv",
    primaryClass = "hep-ph",
    reportNumber = "CERN-PH-TH/2015-247",
    doi = "10.1007/JHEP01(2014)035",
    journal = "JHEP",
    volume = "01",
    pages = "035",
    year = "2014"
}

@article{Alonso:2013hga,
    author = "Alonso, Rodrigo and Jenkins, Elizabeth E. and Manohar, Aneesh V. and Trott, Michael",
    title = "{Renormalization Group Evolution of the Standard Model Dimension Six Operators III: Gauge Coupling Dependence and Phenomenology}",
    eprint = "1312.2014",
    archivePrefix = "arXiv",
    primaryClass = "hep-ph",
    reportNumber = "CERN-PH-TH-2013-305, CERN-PH-TH/2013-305",
    doi = "10.1007/JHEP04(2014)159",
    journal = "JHEP",
    volume = "04",
    pages = "159",
    year = "2014"
}

@article{Fierz:1939ix,
    author = "Fierz, M. and Pauli, W.",
    title = "{On relativistic wave equations for particles of arbitrary spin in an electromagnetic field}",
    doi = "10.1098/rspa.1939.0140",
    journal = "Proc. Roy. Soc. Lond. A",
    volume = "173",
    pages = "211--232",
    year = "1939"
}

@article{Haidenbauer:2019fyd,
    author = "Haidenbauer, Johann and Mei{\ss}ner, Ulf-G",
    title = "{Neutron-antineutron oscillations in the deuteron studied with $NN$ and $\bar NN$ interactions based on chiral effective field theory}",
    eprint = "1910.14423",
    archivePrefix = "arXiv",
    primaryClass = "hep-ph",
    doi = "10.1088/1674-1137/44/3/033101",
    journal = "Chin. Phys. C",
    volume = "44",
    number = "3",
    pages = "033101",
    year = "2020"
}

@article{Li:2022tec,
    author = "Li, Hao-Lin and Ren, Zhe and Xiao, Ming-Lei and Yu, Jiang-Hao and Zheng, Yu-Hui",
    title = "{Operators for generic effective field theory at any dimension: on-shell amplitude basis construction}",
    eprint = "2201.04639",
    archivePrefix = "arXiv",
    primaryClass = "hep-ph",
    doi = "10.1007/JHEP04(2022)140",
    journal = "JHEP",
    volume = "04",
    pages = "140",
    year = "2022"
}

@article{Li:2024ghg,
    author = "Li, Xuan-He and Sun, Hao and Tang, Feng-Jie and Yu, Jiang-Hao",
    title = "{Complete CP eigen-bases of mesonic chiral Lagrangian up to p$^{8}$-order}",
    eprint = "2404.14152",
    archivePrefix = "arXiv",
    primaryClass = "hep-ph",
    doi = "10.1007/JHEP08(2024)189",
    journal = "JHEP",
    volume = "08",
    pages = "189",
    year = "2024"
}

@article{Low:2022iim,
    author = "Low, Ian and Shu, Jing and Xiao, Ming-Lei and Zheng, Yu-Hui",
    title = "{Amplitude/operator basis in chiral perturbation theory}",
    eprint = "2209.00198",
    archivePrefix = "arXiv",
    primaryClass = "hep-ph",
    doi = "10.1007/JHEP01(2023)031",
    journal = "JHEP",
    volume = "01",
    pages = "031",
    year = "2023"
}

@article{Liao:2012uj,
    author = "Liao, Yi and Liu, Ji-Yuan",
    title = "{Generalized Fierz Identities and Applications to Spin-3/2 Particles}",
    eprint = "1206.5141",
    archivePrefix = "arXiv",
    primaryClass = "hep-ph",
    doi = "10.1140/epjp/i2012-12121-0",
    journal = "Eur. Phys. J. Plus",
    volume = "127",
    pages = "121",
    year = "2012"
}

@article{Nieves:2003in,
    author = "Nieves, Jose F. and Pal, Palash B.",
    title = "{Generalized Fierz identities}",
    eprint = "hep-ph/0306087",
    archivePrefix = "arXiv",
    reportNumber = "SINP-TNP-03-16",
    doi = "10.1119/1.1757445",
    journal = "Am. J. Phys.",
    volume = "72",
    pages = "1100--1108",
    year = "2004"
}

@article{Bali:2022qja,
    author = {Bali, Gunnar S. and Collins, Sara and S{\"o}ldner, Wolfgang and Weish{\"a}upl, Simon},
    collaboration = "RQCD",
    title = "{Leading order mesonic and baryonic SU(3) low energy constants from Nf=3 lattice QCD}",
    eprint = "2201.05591",
    archivePrefix = "arXiv",
    primaryClass = "hep-lat",
    doi = "10.1103/PhysRevD.105.054516",
    journal = "Phys. Rev. D",
    volume = "105",
    number = "5",
    pages = "054516",
    year = "2022"
}

\end{document}